\documentclass[
twocolumn,
notitlepage,
nofootinbib,
english,
aps,
pra,
superscriptaddress,
10pt
]{revtex4-2}

\usepackage[utf8]{inputenc}
\usepackage{babel}

\usepackage{amsmath}
\usepackage{amssymb}
\usepackage{bbm}
\usepackage{braket}

\usepackage{enumitem}           %Control layout of itemize, enumerate, description
\usepackage{etaremune}          %Reverse-counting enumerate environment
\usepackage{fontawesome5}
\usepackage{soul}
\usepackage[dvipsnames]{xcolor}
\pdfoutput=1
\usepackage[frozencache]{minted}
\usepackage{multirow}
\usepackage{tabularray}
\usepackage[many]{tcolorbox}

\usepackage[labelfont=bf,justification=raggedright]{caption}
\usepackage[labelfont=bf,textfont=normalfont,font=small,justification=raggedright]{subcaption}

\usepackage[frozencache]{minted}
\definecolor{mintbg}{gray}{0.95}
\setminted{
    bgcolor=mintbg, 
    frame=lines, 
    rulecolor=\color{gray!50}, 
    numbers=left
}

\newcommand\mydots{\makebox[1em][c]{.\hfil.\hfil.}}
\newcommand{\tr}{\textrm{Tr}}
\newcommand{\kvac}{\ket{\textrm{vac}}}

\newcommand{\etal}{et.\ al.\ }

\makeatletter
\def\l@subsubsection#1#2{}
\makeatother

\renewcommand{\footnoterule}{\kern -1ex \hspace{-3mm}\rule{\linewidth}{0.5pt}\\\vspace{1ex}}
\newcommand{\fnt}[1]{%
    \footnote{%
        \begin{tcolorbox}[
            size=fbox,
            width=\dimexpr\linewidth-1.5em\relax, % Subtract marker width
            colframe=white, 
            colback=black!10,
            before skip=-3mm
        ]
        \textcolor{black}{#1}
        \end{tcolorbox}%
    }
}

\usepackage[linktocpage]{hyperref}
\hypersetup{
    colorlinks=true,
    urlcolor=blue,
    citecolor=blue,
    linkcolor=blue
}

\begin{document}

\title{Foundations of Quantum Optics for Quantum Information: Crash Course on Nonclassical States and Quantum Correlations}

\author{Jhoan Eusse}
\affiliation{Universidad de Antioquia, Instituto de Física,  Calle 70 No. 52-21, Medellín, Colombia.}

\author{Esteban Vasquez}
\affiliation{TU Wien, Atominstitut \& VCQ,  Stadionallee 2, 1020 Vienna, Austria}

\author{Tom Rivlin}
\affiliation{TU Wien, Atominstitut \& VCQ,  Stadionallee 2, 1020 Vienna, Austria}

\author{Elizabeth Agudelo}
\affiliation{TU Wien, Atominstitut \& VCQ,  Stadionallee 2, 1020 Vienna, Austria}

\date{\today}

\begin{abstract}
    Nonclassical states of light and their correlations lie at the heart of quantum optics, serving as fundamental resources that underpin both the exploration of quantum phenomena and the realisation of quantum information protocols.
    These lecture notes provide an accessible yet rigorous introduction to the foundations of quantum optics, emphasising their relevance to quantum information science and technology.
    Starting from the quantisation of the electromagnetic field and the bosonic formalism of Fock space, the notes develop a unified framework for describing and analysing quantum states of light.
    Key families of states---thermal, coherent, and squeezed---are introduced as paradigmatic examples illustrating the transition from classical to nonclassical behaviour.
    The concepts of convexity, classicality, and quasiprobability representations are presented as complementary tools for characterising quantumness and defining operational notions such as $P$-nonclassicality.
    The discussion extends naturally to Gaussian states, composite systems, and continuous-variable entanglement, highlighting how nonclassicality serves as a resource for generating and quantifying quantum correlations.
    Theoretical developments are complemented by computational and experimental perspectives, including simulations of optical states using the Python library Strawberry Fields and data analysis from simulated data.
    Together, these notes aim to bridge the foundational concepts of quantum optics and modern quantum information, offering both conceptual insight and practical tools for students and researchers entering the field.
\end{abstract}
\maketitle

\tableofcontents

% DAY 1
\section{Introduction}

Quantum theory is one of the most successful scientific frameworks ever developed: it has been confirmed to extraordinary precision in countless experiments.
It not only explains phenomena that classical physics cannot—like superconductivity, lasers, and the structure of matter—but also underpins technologies that shape everyday life, including transistors, medical imaging, and emerging quantum computers. 
Its predictive power and practical impact make it both a cornerstone of modern science and a driver of future innovation.
Fittingly, 2025 was labeled the \textit{International Year of Quantum Science and Technology} \cite{unesco_2025}, as part of a global initiative to celebrate these achievements and highlight the transformative role that quantum physics is expected to play in shaping the rest of this century.

Quantum technologies have rapidly advanced over the past two decades, transitioning from theoretical concepts to experimental demonstrations and early commercial applications \cite{Mckinsey_2025}. 
Today, they encompass a broad range of platforms, including superconducting circuits \cite{BarendsKellyMegrantEtAl_2014, Jiang_2025}, trapped ions \cite{Foss-FeigPaganoPotterYao_2024}, photonics \cite{Ramakrishnan_2023}, and spin-based systems \cite{Burkard_2023}, each with distinct strengths in correlation times, scalability and controllability.
Quantum sensing and metrology have reached remarkable precision, exploiting quantum phenomena such as entanglement and squeezing to surpass classical limits in detecting magnetic fields and gravitational waves, and in setting time standards \cite{GiovannettiLloydMaccone_2011, TothApellaniz_2014, Demkowicz-DobrzanskiJarzynaKolodynski_2015, DegenReinhardCappellaro_2017, MontenegroMukhopadhyayYousefjaniEtAl_2025}. 
Quantum communication, particularly through quantum key distribution, has moved toward secure, real-world networks, with satellite-based implementations  already demonstrating global reach \cite{BedingtonArrazolaLing2017,HenrietBeguinSignolesEtAl2020,deForgesdeParnyAlibartDebaudEtAl2023,LiCaiRenEtAl2025,AbuGhanem2025}.
The pursuit of quantum computing has seen significant milestones, with small-scale universal quantum processors achieving tens to hundreds to thousands of qubits \cite{EveredBluvsteinKalinowskiEtAl2023,ManetschNomuraBatailleEtAl2025,JiangDengFanEtAl2025,MohseniSchererJohnsonEtAl2025}. 
While error rates and decoherence remain challenges, algorithms for quantum simulation, optimization, and cryptography \cite{GeorgescuAshhabNori_2014, MitraJanaBhattacharya_2017,MartonAsboth2023,AbbasAmbainisAugustino_2024, AcharyaAbaninAghababaie-BeniEtAl2025,TripathiKowsariSauravEtAl2025} are being actively developed. 
The concept of quantum advantage --- where a quantum device performs a task beyond the practical reach of classical computers --- has been experimentally demonstrated in specialized scenarios, such as random circuit sampling \cite{BoulandFeffermanNirkheVazirani_2019} and boson sampling \cite{Oh_2025}. 
However, fully general-purpose quantum advantage, especially for applications of widespread practical significance, is still far on the horizon, as fault-tolerant architectures and error-corrected qubits will be required for reliable, large-scale computation, and these remain out of reach for now \cite{Monroe_2024}.
Overall, the field is at a transformative stage: experimental capabilities are rapidly maturing, and theoretical progress continues to identify problems where quantum resources can provide meaningful speedups or precision gains. 
The convergence of hardware development, algorithmic innovation, and applications in secure communication, simulation, and sensing positions quantum technologies to potentially revolutionize computation, information processing, and measurement science, although significant technical and engineering challenges remain before broad quantum advantage is realized.

Quantum information theory provides the fundamental framework that underpins quantum technologies. 
Formalising how information is encoded, processed, and transmitted using quantum systems defines the principles behind quantum computing, quantum communication, and quantum sensing. 
Concepts such as superposition, qubits, and entanglement are directly derived from this theory, guiding the design of algorithms, protocols, and devices that exploit uniquely quantum effects for quantum advantages. 
In essence, quantum information theory acts as both the roadmap and the rulebook for turning abstract quantum phenomena into practical technologies.

Quantum information theory relies on reinterpreting the principles of quantum mechanics through an information-theoretic lens \cite{Wilde2013,Hayashi2017,Watrous2018}.
Tasks such as storing, processing, and transmitting information are inherently tied to the physical properties of the systems in which the information is encoded. 
As a result, physics, information, and computation are deeply interconnected. 
The initial extension of classical information science into the quantum domain naturally focused on qubits—the quantum analogues of classical bits. 
Early on, it became evident that uniquely quantum effects, such as superposition and entanglement, provide significant advantages for information processing, enabling tasks to be performed faster, more efficiently, and with higher precision \cite{Landauer1961,Bennett1982,BennettBrassard1984,BennettEtAl1993,BennettDiVincenzoSmolinWootters1996,Shor1997}.

Light, in particular, offers remarkable versatility as a quantum information carrier. 
It allows the implementation of both finite, discrete-variable (DV) systems, such as qubits encoded in polarization or photon number, and infinite, continuous-variable (CV) systems, such as quadrature amplitudes in optical fields. 
This dual capability makes photonics a highly flexible platform, able to address a wide range of quantum protocols and applications \cite{WalmsleyRaymer2005,PolzikShapiro2007,KokMunroNemotoEtAl2007,OBrienFurusawaVuckovic_2009,PanChenLuEtAl2012,FlaminiSpagnoloSciarrino2018,SlussarenkoPryde2019}. 
Being able to manipulate both discrete and continuous degrees of freedom expands the possibilities for encoding, transmitting, and processing quantum information, further highlighting the central role of optics in advancing quantum technologies.
Moreover, photons can encode quantum information in multiple CV and DV degrees of freedom --- such as polarization, phase, time bins, or quadratures --- providing flexible platforms for qubits and higher-dimensional quantum systems.
And finally, photons are inherently low-noise and weakly interacting, which allows them to travel long distances without significant decoherence. 
This all makes optical systems ideal for quantum communication, enabling secure transmission of information through quantum key distribution and the development of quantum networks. 

DV and CV systems have very different properties that make them useful in different information processing tasks --- using either type of system comes with different constraints and trade-offs.
For instance, for a DV setting such as a pair of photons, one can only rarely engineer entangled states, since doing so is conditioned on infrequent measurement outcomes (coinciding counts in the detector), but any entanglement that is detected is guaranteed to be maximal \cite{AnwarPerumangattSteinlechnerEtAl2021}.
By contrast, for a CV setting such an EM field mode, entanglement generation is more efficient, since there is almost always entanglement between modes and detecting it is not dependent on conditioning on detector outcomes, but any entanglement one can detect is imperfect and non-maximal \cite{OuPereiraKimblePeng1992,DuanGiedkeCiracZoller2003,WasilewskiFernholzJensenEtAl2009}.
Though it should be noted that it is known that one can, in principle, reach universal computation and efficient quantum information processing using electromagnetic fields, and optical elements \cite{KnillLaflammeMilburn_2001}.  
In fact, many protocols originally developed for DV qubits have been generalized to the CV setting.
For example, very shortly after the teleportation protocol was proposed for qubits in 1993 by Bennett \etal \cite{BennettEtAl1993}, it was proposed for fields in 1994 by Vaidman \cite{Vaidman_1994}.
In particular, entangled states are ``easily" prepared, (unitarily) transformed, and measured in quantum optics laboratories \cite{AnwarPerumangattSteinlechnerJenneweinLing_2021}.

Broadly speaking, light is an exceptionally versatile and controllable quantum system, making it a fascinating platform for the study of quantum information \cite{BraunsteinVanLoock_2005,KokMunroNemotoEtAl2007,OBrienFurusawaVuckovic2009,AndersenGehringMarquardtLeuchs2016,FlaminiSpagnoloSciarrino2018,SlussarenkoPryde2019}. 
And since this is the fundamental framework that underpins quantum technologies, it is clear that understanding quantum optics is key to understanding the emerging quantum technological advances of recent years.

Another reason optics is exciting for quantum technologies is the wide range of available experimental tools and techniques. 
For decades, laser sources, beam splitters, interferometers, and single-photon detectors have enabled the precise preparation, manipulation, and measurement of quantum states of light in relatively inexpensive, easily constructed laboratory settings \cite{leonhardt_1997}. 
These tools enable fundamental experiments that explore the boundary between classical and quantum physics, such as tests of superposition \cite{PfleegorMandel1967}, contextuality \cite{LapkiewiczLiSchaeffEtAl2011}, nonlocality \cite{AspectDalibardRoger1982},  entanglement \cite{KwiatMattleWeinfurterEtAl1995}, and other quantum correlations \cite{KohnkeAgudeloSchunemannSchlettweinVogelSperlingHage_2021}. 
Photonics also integrates naturally with emerging technologies like integrated photonic circuits, offering scalability and miniaturization that are crucial for building practical quantum devices \cite{Silberhorn2007,PolitiCryanRarityEtAl2008,EcksteinChristMosleySilberhorn2011,CarolanHarroldSparrowEtAl2015,WangSciarrinoLaingThompson2020}.

Finally, optics allows direct visualization and experimental control of quantum phenomena, which is both scientifically exciting and technologically promising. 
Optical systems can simulate complex quantum systems \cite{AspuruGuzikWalther2012}, test quantum algorithms \cite{ZhangZhanLiaoEtAl2021}, and explore quantum-enhanced sensing and metrology with unprecedented precision \cite{Barbieri2022}. 
This combination of fundamental insight, practical applicability, and elegant experimental accessibility makes optics a particularly attractive and ``cool'' platform for advancing quantum technologies and pushing the boundaries of what quantum systems can achieve.

The cornerstone of quantum optics --- and the fundamental source of quantum effects for quantum technologies --- is the quantum states of light and their correlations. 
Throughout this document, we introduce and discuss these states, providing a crash course on the foundations of quantum optics for quantum information, with a focus on nonclassical states and quantum correlations.
There exists an extensive collection of excellent textbooks on quantum optics and quantum information where these topics are discussed in great detail. 
In these notes, we have curated a selection of concepts and results with the aim of providing a self-contained introduction to optical nonclassicality and its connection to quantum correlations.

In the following, we first build the foundations by introducing the quantization of the electromagnetic field and its description in terms of harmonic oscillator modes (Sec. \ref{sec:em_qho}), and by reviewing the bosonic formalism of Fock space and second quantization (Sec. \ref{sec:Fock}). 
We then turn to specific families of quantum states, beginning with the coherent states often regarded as the “most classical” quantum states (Sec. \ref{sec:coherent}) and continuing with squeezed states (Sec. \ref{sec:squeezed}) and their role as paradigmatic examples of nonclassical light. 
The discussion is then broadened to include mixed states (Sec. \ref{sec:squeezed}), convexity and criteria for classicality (Sec. \ref{sec:convexCL}), and the characterization of states through quasiprobability distributions (Sec. \ref{sec:quasip}). 
We also examine the notions of Gaussianity (Sec. \ref{sec:gaussianiry}), composite systems (Sec. \ref{sec:BS_transf}), and entanglement (Sec. \ref{sec:entanglement}), before comparing entanglement-based and P-function-based perspectives on nonclassicality (Sec. \ref{sec:NCLvsENT}).
In addition, the theoretical discussions are complemented by computational viewpoints, featuring the analysis of data obtained through balanced homodyne detection and simulations of optical states using the Strawberry Fields Python library (Sec. \ref{sec:labs}).
Throughout the document, several proposed exercises are provided in the form of footnotes to help students practice, consolidate their knowledge, and gain a deeper understanding of the concepts and definitions presented here.
Together, these lecture notes provide a comprehensive overview of the conceptual and mathematical tools used to analyze classicality and quantumness in bosonic systems, with an emphasis on their relevance for quantum optics and quantum technologies.
\section{Electromagnetic field modes and the quantum harmonic oscillator}
\label{sec:em_qho}

To describe the quantum properties of light, it is necessary to quantize the electromagnetic (EM) field. 
A particularly enlightening approach is to start with the simplest scenario: a single mode of the field confined within a one-dimensional cavity of length $L$ with perfectly conducting walls. 
The conducting boundary conditions require the electric field to vanish at the walls, leading to discrete allowed wavevectors $k_n = n \pi / L$ with corresponding angular frequencies $\omega_n = c k_n$. 
In classical field theory, each value of $n$ corresponds to a different `normal mode' of the field. 
Let us select one of them and focus on its quantization.

Classically, the Hamiltonian of the EM field can be written in terms of the field energy stored in its electric and magnetic components. 
For a single mode, the field variables reduce effectively to two conjugate degrees of freedom, analogous to the position and momentum of a mechanical harmonic oscillator. 
More explicitly, the Hamiltonian for the single mode takes the form
\begin{equation}
    H = \frac{1}{2}\left( p^2 + \omega^2 x^2 \right),
\end{equation}
where $x$ and $p$ are canonical coordinates, and $\omega$ is the frequency of the cavity mode. 
This formal identity establishes a direct correspondence between the quantized EM mode and the quantum harmonic oscillator.

Constructing the quantum version of a system like this is called \textit{quantizing} it. 
To do so (specifically to do a kind of quantization called ``second quantization'', which will be defined in more detail soon), one promotes the variables $x$ and $p$ to operators satisfying the canonical commutation relation%
\fnt{Considering that $[\hat x,\hat p]=i\hbar$, prove $[\hat a,\hat a^\dagger]=1$.}
\begin{equation}\label{eq:xp_comm}
    [\hat x, \hat p] = i \hbar.
\end{equation}
$\hat{x}$ and $\hat{p}$ are often called field quadratures, since they are the two orthogonal components of a harmonic oscillator mode, analogous to in-phase and quadrature-phase components of a classical oscillation.
Next, it is convenient to introduce the dimensionless, non-Hermitian operators:
\begin{gather}
    \hat a = \frac{\omega \hat x + i \hat p}{\sqrt{2\hbar \omega}},
    \label{eq:annihilation}
    \\
    \hat a^\dagger = \frac{\omega \hat x - i \hat p}{\sqrt{2\hbar \omega}},
    \label{eq:creation}
\end{gather}
which satisfy the bosonic commutation relation $[\hat a, \hat a^\dagger] = 1$. 

In terms of these operators, the Hamiltonian reads
\begin{equation}
    \hat H = \hbar \omega \left( \hat a^\dagger \hat a + \frac{1}{2} \right).
\end{equation}
The eigenstates of the number operator $\hat N = \hat a^\dagger \hat a$ are the \textit{photon number states} or \textit{Fock states}, denoted $\ket{n}$, with eigenvalues%
\fnt{Show that $n$ must be a nonnegative integer.}
\begin{equation}
    \hat N \ket{n} = n \ket{n}, \qquad n = 0, 1, 2, \ldots
    \label{eq:number_op}
\end{equation}
As we can see, the Hamiltonian is a linear function of the number operator, i.e.
$\hat H = a\,\hat N + b\hat I$, with \(a,b\in\mathbb{R}\).
If \(\{\ket{n}\}\) is the (orthonormal) eigenbasis of $\hat N$, then acting with \(\hat H\) on \(\ket{n}\) gives
$\hat H\ket{n} = (a\hat N + b\hat I)\ket{n} = (a n + b)\ket{n}$.
Which is to say, both operators can be diagonalized simultaneously, and the eigenvalue equation for $\hat H$ is
\begin{equation*}
    \hat H \ket{n}=\hbar\omega\left(n+\frac{1}{2}\right)\ket{n}.
\end{equation*}
This implies that the energy eigenvalues are given by $E_n= \hbar\omega\left(n+\frac{1}{2}\right)$.
Notice that $[\hat N,\hat a]=-\hat a$ and $[\hat N,\hat a^\dagger]=\hat a^\dagger$, so $\hat N \hat a\ket{n}= (n-1)\,   \hat a^\dagger\ket{n}$  and
$\hat N \hat a^\dagger\ket{n} 
= (n+1)\,   \hat a^\dagger\ket{n}$.
Hence, $\hat a^\dagger\ket{n}$ (and also $\hat a\ket{n}$) is an eigenstate of the number operator $\hat N$ with eigenvalue $n+1$ ($n-1$), that is, increased (decreased) by one ``unit of quantum energy" $\hbar\omega$.
Hence, we commonly call $\hat a^\dagger$ and $\hat a$ the \textit{creation} and \textit{annihilation} operators respectively. 

The states $\ket{n}$ describe discrete excitations of the cavity mode, with the interpretation that the field contains exactly $n$ photons, each carrying an energy $\hbar \omega$. 
Notice that the smallest possible value of $n$ is zero, and so the lowest state, $\ket{0}$, is the vacuum with no photons present. However, interestingly, the field still possesses non-vanishing energy, known as the zero-point energy: $E_0=\frac{1}{2}\hbar\omega$.

The quantized electric field operator for the single mode can be expressed as
\begin{equation}
    \hat{E}(t) = i \sqrt{\frac{\hbar \omega}{2 \epsilon_0 V}} \left( \hat a e^{-i \omega t} -\hat a^\dagger e^{i \omega t} \right),
\end{equation}
where $V$ is the effective mode volume (cf. e.g.\ \cite{GerryKnight_2005}). 
Here we see how the corresponding magnetic field operator has an analogous structure.
The annihilation operator $\hat a$ is related to the amplitude and phase of the field.
The expectation value $\braket{\hat a}$ is a complex number.
A nonzero $\braket{\hat a}$ would signal that the state has a well-defined average phase and amplitude. 
The magnitude and phase of $\braket{\hat a}$ are respectively related to the average amplitude and phase of the field.
A nonzero $\braket{\hat a}$ implies the state has a preferred direction in phase space (defined by amplitude and phase), i.e.\ the field “points” in a certain direction in the complex plane.
When the system is in a number state $\ket{n}$, the expectation value of the field operator vanishes,
\begin{equation}
    \bra{n} \hat{E}(t) \ket{n} = 0,
\end{equation}
reflecting the fact that number states have an exact photon number but a completely undefined phase (this is a manifestation of the number–phase uncertainty relation, cf.\ e.g., Ch. 3 Ref. \cite{VogelWelsch_2006}).
However, the fluctuations of the field are nonzero. 
We can see that the variance of the electric field in such a state is
\begin{equation}
    (\Delta \hat{E})^2 = \bra{n} \hat{E}^2 \ket{n} - \bra{n} \hat{E} \ket{n}^2 
    \propto (2n + 1).
\end{equation}
Thus, even in the vacuum state $\ket{0}$, field fluctuations persist arising due to the non-commutativity of $\hat x$ and $\hat p$, Eq.~\eqref{eq:xp_comm}. 
These fluctuations are a hallmark of the quantum nature of the EM field and underlie many nonclassical effects (e.g.\ \cite{LambRetherford1947, Bethe1947, Welton1948, Sanchez-MondragonNarozhnyEberly1983, Agarwal1985}).

Ultimately, the quantization of the EM field reveals a picture in which each normal mode of the cavity is equivalent to a quantum harmonic oscillator. 
The photon number states provide a natural basis for describing the energy content of a mode, and they exhibit strong quantum fluctuations in the field observables. 
These fluctuations become central in phenomena such as spontaneous emission and the generation of nonclassical states of light. 
This harmonic oscillator analogy thus provides the foundation for understanding both the discrete- and continuous-variable properties of the quantized fields.
\section{Bosons: Fock space and second quantization}
\label{sec:Fock}

In non-relativistic quantum mechanics, one typically begins by studying systems with a fixed number of particles, an approach often referred to as first quantization. 
For instance, the Hilbert space of a single particle is $\mathcal{H}$, while for $N$ distinguishable particles the corresponding state space is the tensor product $\mathcal{H}^{\otimes N}$. 
In this framework the particle number is specified from the outset and remains fixed.
However, this approach has intrinsic limitations: it cannot account for phenomena such as the experimentally observed states that involve coherent superpositions of different particle numbers, nor can it naturally incorporate quantization of fields like the EM field.
In such quantum systems one may also encounter interference effects between states with different particle numbers. 
A paradigmatic example is provided by the laser, which ideally produces a %coherent 
superposition of photon number states with contributions from every $n \in \mathbb{N}$. 
Within the framework of first quantization, however, superpositions of states belonging to Hilbert spaces with different, fixed particle numbers are not well defined. 

To overcome these restrictions, one introduces the formalism of second quantization. 
This approach not only resolves the conceptual issues left open by first quantization but also provides the modern language of quantum theory. 
It forms the mathematical backbone of relativistic quantum field theory, underlies the Standard Model of particle physics, and serves as the natural framework for quantum optics and its applications in quantum technologies.

The relation between first and second quantization is twofold. 
Starting from first quantization, one can systematically construct the formalism of second quantization by extending the Hilbert space to include arbitrary particle numbers, encoded in the Fock space \cite{Fock1932}. 
Conversely, first quantization can be recovered as the special case of second quantization when the particle number is fixed. 

In the context of the quantum harmonic oscillator, as discussed earlier, the equally spaced energy levels can be interpreted as states with varying numbers of bosonic excitations, which are created and annihilated by ladder-like operators. 
Bosons are quantum particles that obey Bose–Einstein statistics, allowing multiple identical particles to occupy the same quantum state simultaneously. 
For photons---the bosons of the EM field---these excitations correspond to discrete light quanta, so that each harmonic oscillator mode of the field is populated by an integer number of photons. 
Since bosons are indistinguishable, the many-body wavefunction must be symmetric under the exchange of any two particles. 
Consequently, the physical Hilbert space is not the full tensor product but rather its symmetrized subspace, which contains only states with the correct bosonic symmetry.
Thus, the physical Hilbert space is the symmetrized subspace of the $N$-fold tensor product,
\begin{equation}
    \mathcal{H}_N^{\text{Boson}} = \mathrm{Sym} \left( \mathcal{H}^{\otimes N} \right).
\end{equation}
Each element of $\mathcal{H}_N^{\text{Boson}}$ describes an $N$-boson system, such as $N$ photons populating a set of modes.
To consistently describe situations where interference between states with different number of particles play a role, one introduces the \textit{Fock space}, defined as the direct sum of all $N$-boson Hilbert spaces:
\begin{equation}
    \mathcal{H}_\text{Fock} = \bigoplus_{N=0}^\infty \mathcal{H}_N^{\text{Boson}}.    
\end{equation}
The case with $N=0$ corresponds to the vacuum state $\ket{0}$ with $\mathcal{H}^{\otimes 0} = \mathbb{C}$. Then $N=1$ contains all single-particle states, $N=2$ the two-boson states, and so on. 
Thanks to this construction as a direct sum, the contributions with definite particle number can be conveniently arranged into a single vector \cite{SperlingAgudelo_2023}
	\begin{equation}
	\begin{aligned}
    &|\Psi\rangle_\mathrm{Fock}
		=\bigoplus_{n\in\mathbb N}|\psi^{(n)}\rangle=
		\begin{bmatrix}
			|\psi^{(0)}\rangle
			\\
			|\psi^{(1)}\rangle
			\\
			|\psi^{(2)}\rangle
			\\
			\vdots
		\end{bmatrix}\\
		&=
		\begin{bmatrix}
			|\psi^{(0)}\rangle
			\\
			0
			\\
			0
			\\
			\vdots
		\end{bmatrix}
		{+}
		\begin{bmatrix}
			0
			\\
			|\psi^{(1)}\rangle
			\\
			0
			\\
			\vdots
		\end{bmatrix}
		{+}
		\begin{bmatrix}
			0
			\\
			0
			\\
			|\psi^{(2)}\rangle
			\\
			\vdots
		\end{bmatrix}
		{+}\cdots
	\end{aligned}
	\end{equation}
enabling, in a natural way, the possibility of forming superpositions that mix states with different particle counts.   

Notice that the normalization works as expected:
\begin{align}
    \braket{\Psi|\Psi}_\mathrm{Fock} &=
		\begin{bmatrix}
			|\psi^{(0)}\rangle
			&
			|\psi^{(1)}\rangle
			&
			|\psi^{(2)}\rangle
			&
			\cdots
		\end{bmatrix}
		\begin{bmatrix}
			|\psi^{(0)}\rangle
			\\
			|\psi^{(1)}\rangle
			\\
			|\psi^{(2)}\rangle
			\\
			\vdots
		\end{bmatrix}
        \\ &
        = \sum_{n=0}^{\infty} \braket{\psi^{(n)}|\psi^{(n)}}
        =1,
\end{align}
so the squared norm of each $n$-particle component can be interpreted as the probability of finding $n$ particles.
The vacuum, $|\psi^{(0)}\rangle=1$ {when} normalized, has as Fock-state representation 
\begin{equation}
\kvac_\mathrm{Fock}=
\begin{bmatrix}
    1
    \\
    0
    \\
    0
    \\
    \vdots
\end{bmatrix},
\end{equation}
and to move between states with different particle numbers, we have the creation and annihilation operators introduced previously. 
For bosons, the creation operator $\hat a_j^\dagger$ adds one particle to mode $j$:
\begin{align}
    \hat a_j^\dagger \ket{n_0, \mydots, n_j, \mydots} 
    = \sqrt{n_j+1}\, \ket{n_0, \mydots, n_j+1, \mydots}.
\end{align}
The annihilation operator $a_j$ removes one particle from mode $j$:
\begin{align}
    \hat a_j \ket{n_0, \mydots, n_j, \mydots} 
    = \sqrt{n_j}\, \ket{n_0, \mydots, n_j-1, \mydots}.
\end{align}
And they obey the bosonic commutation relations
\begin{align}
    [\hat a_j, \hat a_k^\dagger] = \delta_{jk},
\end{align}
generalizing $[\hat a, \hat a^\dagger] = 1$ to the multimode scenario.%
\fnt{Consider the expectation value of the operator $\hat{a}_i^{\dagger} \hat{a}_j^{\dagger} \hat{a}_l \hat{a}_k$ in the Fock state $\left|\Phi_N\right\rangle=\left|n_1, n_2, \mydots, n_i, \mydots\right\rangle$:
$\left\langle\Phi_N\right| \hat{a}_i^{\dagger} \hat{a}_j^{\dagger} \hat{a}_l \hat{a}_k\left|\Phi_N\right\rangle$.
Assume that $i=j=k=l$, and prove the following relationship for bosons:
$
\left\langle\Phi_N\right| \hat{a}_i^{\dagger} \hat{a}_i^{\dagger} \hat{a}_i \hat{a}_i\left|\Phi_N\right\rangle=n_i\left(n_i-1\right) .$}
The number operator for mode $j$ is defined as $\hat{n}_j = \hat{a}_j^\dagger \hat{a}_j$, with eigenvalues
\begin{align}
    \hat{n}_j \ket{n_0, n_1, \mydots} = n_j \ket{n_0, n_1, \mydots}.
\end{align}
This makes the link between algebraic operators and measurable quantities explicit: $n_j$ is the number of particles detected in mode $j$.%
\fnt{We are still working with the expectation value of $\hat{a}_i^{\dagger} \hat{a}_j^{\dagger} \hat{a}_l \hat{a}_k$ in the Fock state $\left.\left|\Phi_N\right\rangle=\mid n_1, n_2, \mydots, n_i, \mydots\right\rangle:$
$
\left\langle\Phi_N\right| \hat{a}_i^{\dagger} \hat{a}_j^{\dagger} \hat{a}_l \hat{a}_k\left|\Phi_N\right\rangle
$.
Assume that $i \neq j$, and that $i=k$ and $j=l$. Prove the following relationship: 
$
\left\langle\Phi_N\right| \hat{a}_i^{\dagger} \hat{a}_j^{\dagger} \hat{a}_j \hat{a}_i\left|\Phi_N\right\rangle=n_i n_j .
$}
To construct a basis of the Fock space, we first choose an orthonormal basis $\{\ket{b_j}\}$ of the single-particle Hilbert space. 
We then label Fock states by their \textit{occupation numbers} $n_j$, which count how many particles occupy each basis state $\ket{b_j}$:
\begin{align}
    \ket{n_0} \otimes \ket{n_1} \otimes \ket{n_2}\otimes \mydots = \ket{n_0, n_1, n_2, \mydots}.
\end{align}
For bosons, the $n_j$ can be any non-negative integer. 
These occupation number states form a natural orthonormal basis of the Fock space and are often called \textit{number states}. 

To emphasize, an important conceptual novelty arose here: in Fock space, one can form quantum superpositions of states with different particle numbers. 
For example, $\ket{\psi} = \alpha \ket{0} + \beta \ket{1}$, describes a system in a superposition of the vacuum and a single-particle state. This is the first time in quantum mechanics that we encounter superpositions between states belonging to Hilbert spaces of different dimensions. 
Fock space unifies these sectors into a single larger Hilbert space, making such superpositions well-defined.
These superpositions play a fundamental role in quantum optics and quantum field theory. 
For instance, as we will discuss in detail later (Sec. \ref{sec:coherent}), coherent states of light, which approximate classical EM waves, are superpositions of all photon number states with Poissonian weights.
\section{Description of statistical mixtures}
\label{sec:Mixtures}

Until now, we have considered (isolated) systems whose state was unequivocally described by a state vector, meaning we have all information about the state that it is possible to have.
But we will now consider the case where our information about the system is not complete. 
In situations such as this, the quantum system of interest is best described not as the single state $\ket{\psi}$ in Hilbert space, but rather as a statistical mixture of multiple states. 
Crucially, this is a \textit{classical} statistical mixture. This means that our lack of knowledge about the possible outcomes of measurements arises due to our limited knowledge of the preparation of the state (as could easily happen for a classical system), as opposed to arising due to a quantum superposition.
When we have a classical statistical mixture like this, we call the resulting state a mixed state, and any state that is not a classical statistical mixture we call a pure state. 
(Of course, whether the state is formed from a quantum superposition or from a classical mixture, it is always a \textit{quantum} state, in that it lives in a Hilbert space and obeys the Schr\"{o}dinger equation. The word `classical' here refers only to the fact that uncertainty about measurement outcomes has a classical analog.)

One common situation in which mixed states arise --- when we lack all possible information about the state --- is when we describe physical systems which are small parts of much larger ones.
In such cases, we will not have complete information about the overall system, and this lack of information will manifest as a classical `mixedness'.

To motivate the definition of the \textit{density operator} $\hat{\rho}$ that we use to describe mixed states, consider an observable $\hat L$. 
Its expectation value with respect to a pure state $\ket{\psi_{k}}$ is given by $\braket{\psi_{k}|\hat L|\psi_{k}}$. 
But now let us assume that the probability of the pure state $\ket{\psi_{k}}$ arising is given by $p_k$. 
Now we need to average over all the $\ket{\psi_{k}}$:
\begin{align*}
    \braket{\hat L} &
    = \sum_k p_k \braket{\psi_k|\hat L|\psi_k}
    \\ &
    = \sum_k p_k\, \tr\left[\hat L\ket{\psi_k} \!\! \bra{\psi_k}\right]
    \\ & 
    = \tr\left[\sum_k p_k \hat L\ket{\psi_k} \!\! \bra{\psi_k}\right] 
    \\ &
    = \tr\left[\hat L \left(\sum_k p_k \ket{\psi_k} \!\! \bra{\psi_k}\right)\right],
\end{align*}
For simplicity, we considered a discrete mixing here, but a direct analog in the CV setting is possible using integration.

The expression in parentheses defines the so-called density, or statistical operator
\begin{equation}
\label{eq:op_den}
    \hat \rho = \sum_k p_k \ket{\psi_k} \!\! \bra{\psi_k},
\end{equation}
with which we can describe an ensemble of pure states that is distributed according to (classical) probabilities $p_k$. 
Consequently, quantum expectation values of any observable $\hat L$ that include statistical mixtures can be written as $\braket{\hat L}=\tr(\hat \rho\hat L)$. 
It is called the density operator because, as a Hermitian, measurable operator, it gives the density of the state at a given point in Hilbert space. 
It encodes ``how much'' of each pure state appears in the ensemble --- a \textit{density} of states.
If we have a deterministic collection of quantum states with $p_k = \delta_{kl}$, then $\hat \rho = \ket{\psi_l}\bra{\psi_l}$ defines a \textit{pure state}.
If $p_k$ includes at least two non-zero contributions, then we call it a mixed state.
The density operator $\hat{\rho}$ can thus represent either pure or mixed states, unifying both under a single formalism.

The difference between a ``density matrix'' type uncertainty and a ``quantum superposition'' of a pure state lies in the ability of quantum amplitudes to interfere, which you can measure by preparing many copies of the same state and then measuring incompatible observables.
Given a state, mixed or pure, you can compute the probability distribution $P(\lambda_k)$ for measuring eigenvalues $\lambda_k$, for any observable you want. 
The difference is the way you combine probabilities; in a quantum superposition, you have complex numbers that can interfere. 
In a classical probability distribution, the coefficients only add positively.

One metric to quantify how pure a quantum state is is called the \textit{purity}, $\mathcal{P}(\hat\rho)=\tr(\hat\rho^2)$.
We have $\mathcal{P}(\hat\rho)=1$ for pure states and $\mathcal{P}(\hat\rho)<1$ for mixed states. 
The upper bound for the purity, $\mathcal{P}(\hat\rho)=1$, arises when the density operator represents a pure state $\hat{\rho}=\ket{\psi}\bra{\psi}$, while the lower bound  for the purity, $\mathcal{P}(\hat\rho)=1/2$, occurs when the density operator is that of the so-called \textit{maximally mixed} state.
This is the state one obtains when uniformly averaging over all states $\ket{\psi}\in \mathbbm{C}^d$:
\begin{equation*}
    \hat \rho_{\rm{MM}} = \sum_\psi \frac{1}{d} \ket{\psi} \!\! \bra{\psi} = \frac{1}{d} \mathbbm{1},
\end{equation*}
and reflects the situation where one has the least possible knowledge about the preparation.

From the construction of $\hat\rho$ we can directly infer several properties of the density operators:
(a) \textit{Unit trace}
\begin{align*}
    \tr(\hat \rho) 
     &= \sum_k p_k \tr(\ket{\psi_k} \!\! \bra{\psi_k}) 
    \\ &= \sum_k p_k =1,
\end{align*}
since $\tr(\ket{\psi}\bra{\psi})=\braket{\psi|\psi}=1$ and $p_k$ is a proper normalized probability distribution.
(b) \textit{Hermiticity}
\begin{align*}
    \hat \rho^\dagger 
    & = \left(\sum_k p_k \ket{\psi_k} \!\! \bra{\psi_k}\right)^\dagger 
    \\ &= \sum_k p_k^* (\ket{\psi_k} \!\! \bra{\psi_k})^\dagger 
    \\ &= \sum_k p_k \ket{\psi_k} \!\! \bra{\psi_k} 
    = \hat \rho,
\end{align*}
since probabilities are real numbers and $\ket{a}\bra{b}^\dagger = \ket{b}\bra{a}$.
(c) \textit{Positive semi-definite} $\hat\rho\ge 0$
\begin{align*}
   \braket{\phi |\hat \rho|\phi} 
   &= \sum_k p_k \braket{\phi|\psi_k} \braket{\psi_k|\phi} 
   \\ &= \sum_k p_k |\braket{\psi_k|\phi} |^2 \ge 0.
\end{align*}
These all hold true $\forall\ket{\psi_k} \in \mathbbm{C}^d$ because $p_k\ge 0$ for every proper probability distribution and $|\braket{\psi_k|\phi} |^2 \ge 0$.%
\fnt{Using an arbitrary matrix $\hat 0 \ne\hat M\in \mathbbm{C}^{d\times d}$, show that $\hat\rho = \hat M^\dagger\hat M/\tr(\hat M^\dagger\hat M)\in \mathbbm{C}^{d\times d}$ is a valid density operator.}

Since $\hat{\rho}$ is a Hermitian operator, it admits a spectral decomposition of the form
\begin{equation*}
    \hat\rho=\sum_{j=0}^{d-1} p_j\ket{\phi_j} \!\! \bra{\phi_j}.
\end{equation*}
Here, $\{\ket{\phi_j}\}$ are orthonormal eigenvectors and the $p_j$ are the corresponding eigenvalues, i.e.\ $\hat\rho\ket{\phi_j}=p_j\ket{\phi_j}$.
The $p_j$ are non-negative because $0 \le \braket{\phi_j |\hat \rho|\phi_j}= \braket{\phi_j |p_j|\phi_j}= p_j\braket{\phi_j |\phi_j}=p_j$ and sum to unity: $\tr\hat\rho =\sum_j p_j =1$.
The eigenvalues of the statistical operator $\hat\rho$ define a probability distribution over eigenvectors, and we can write $p_k=\sum_j p_j\delta_{kl}$. 
This means that the probability distribution associated with a mixed state can be described through its spectral decomposition.
It is interesting to note that multiple different probability distributions can be represented by the same density operator --- they are truly indistinguishable from each other given access only to the $\hat{\rho}$ they permit.

We now turn to an important example of a mixed state that appears frequently in quantum theory, the \textit{thermal state} $\hat{\rho}_{{\rm{th}}}$ (cf.\ Ch.~11 Ref.~\cite{BertlmannFriis2023} for a more detailed description).
Consider a single field mode in a cavity as before, but now let that cavity have some temperature $T$, and let that field mode and the cavity be in thermal equilibrium with each other. 
First, let us define the inverse temperature $\beta=1/k_BT$ using the Boltzmann constant $k_B$. 
Then, if the energy of the mode is $E_n$, the probability to find the system to have energy $E_n$ is given by:
$$
P_n = \frac{e^{-\beta E_n}}{\sum_n e^{-\beta E_n}}.
$$
(This expression requires several additional assumptions we do not need to consider too deeply here, such as having nothing but energy exchanged between the mode and the cavity.)
The sum in the denominator is over all possible modes, and so spans from 0 to $\infty$. 
It is used often enough to be given its own name, the \textit{partition function} $\mathcal{Z}=\sum_n e^{-\beta E_n}$. 
This set of probabilities $\{P_n\}_n$ is called the \textit{Boltzmann distribution}, and is actually entirely classical, despite describing discretized energy levels. 
(In fact it played a key role in Planck's discovery of quantum physics.)
In terms of quantum optics, if we have photons whose energies are classically distributed according to the Boltzmann distribution (called \textit{thermally} distributed), then the resulting state is a thermal state, given in the Fock description by
$$
\hat{\rho}_{{\rm{th}}} = \sum_{n=0}^\infty P_n \ket{n} \!\! \bra{n}.
$$

We can calculate the average photon number $\bar{n}$ of this new state using the number operator $\hat{N}$ defined in Eq. \eqref{eq:number_op}:
$$
\bar{n} = \tr(\hat{N} \hat{\rho}_{th}) = \frac{1}{e^{E_n/k_BT} - 1}.
$$
Therefore
$e^{-E_n/k_BT} = \bar{n}/(1+\bar{n})$.
With this, we can also rewrite our thermal state as:
$$
\hat{\rho}_{{\rm{th}}} = \frac{1}{1+\bar{n}} \sum_{n=0}^\infty \left(\frac{\bar{n}}{\bar{n}+1}\right)^n \ket{n} \!\! \bra{n}.
$$

The density operator formalism allows us to describe quantum states that arise from statistical mixtures, such as the thermal state, while retaining a complete description of all observable predictions. 
With this foundation in place, we now explore coherent states, pure states that sit at the boundary between quantum and classical descriptions.
\section{Coherent states: The ``most classical'' states?}
\label{sec:coherent}

Here we introduce a hugely important type of state in quantum optics, called the coherent state.
One way to motivate it is that it is the state whose behavior is closest to that of classical light, for instance recovering the oscillatory behaviour when position and momentum operator are measured.
It turns out that the way to achieve this is to have a statistical mixture of all possible numbers of photons (all number states $\ket{n}$) with an appropriate distribution. 
Let us explore how this works.

Previously, we introduced the creation and annihilation operators $\hat{a}^{\dagger}$ and $\hat{a}$. 
When applied to a Fock state in a mode, they increase or decrease the number of excitations in that mode by 1.
We previously did not introduce the eigenvalues and eigenvectors of $\hat{a}$. 
Let them be given by $\hat{a}\ket{\alpha}=\alpha\ket{\alpha}$ and $\bra{a}\hat{a}^{\dagger}=\bra{\alpha}\alpha^*$. 
These states $\ket{\alpha}$ turn out to be the coherent states we have been looking for: the ``most classical'' states that behave like classical light.
But these states have many peculiar features. 
First, the eigenvalues $\alpha$ are in general complex, because $\hat{a}$ is not Hermitian.
Next, $\hat{a}^{\dagger}$ has no eigenket (and $\hat{a}$ has no eigenbra), meaning that there is no scalar $r$ satisfying $\hat{a}^{\dagger}\ket{\alpha}=r\ket{\alpha}$.
Their relationship with the number states $\ket{n}$ is also somewhat involved, with the following form%
\fnt{Derive this! In other words, considering $\ket{\alpha}=\sum C_n \ket{n}$ prove that $C_n=\exp[-|\alpha|^2/2]\, \alpha^n/\sqrt{n!}$}
\begin{equation}
\label{eq:coherent}
    \ket{\alpha}=e^{-\frac{|\alpha|^2}{2}}\sum_{n=0}^\infty \frac{\alpha^n}{\sqrt{n!}} \ket{n}.
\end{equation}
So, this is a non-Hermitian operator with complex eigenvalues, given as a superposition over all possible numbers of excitations in a mode. 

How on earth is this the \textit{most} classical state of light?
It turns out that coherent states actually have many properties that match closely with classical analogs. 
When the wavefunction of a coherent state is expressed in terms of position or momentum, it is a simple Gaussian wavepacket.
And crucially, as this wavepacket evolves in time, it does not disperse --- its width in both position and momentum remains the same. 
So a coherent state will oscillate around a point with a specific spatial width and a specific momentum spread forever. 
In fact they have the property that $\braket{\alpha|\hat{x}|\alpha}$ and $\braket{\alpha|\hat{p}|\alpha}$ evolve in time in perfectly sinusoidal ways, exactly as a classical harmonic oscillator would. 
This is why they are called \textit{coherent states} --- they remain coherently put-together throughout their evolution.
As a consequence their quantum fluctuations remain locked in a fixed balance. 
Coherent states saturate the Heisenberg uncertainty limit and they preserve this minimal uncertainty for all time (as we will discuss later).

(Note that the name `coherent state' is unfortunate, as the word `coherence' has another meaning in quantum theory that has the potential to confuse. 
The word can refer to the amount of superposing between two states, often represented by off-diagonal elements of a density matrix. 
Coherence is what powers quantum effects such as entanglement and measurement incompatibility. 
But the coherent state in quantum optics is not known for having an especially large or small amount of quantum coherence --- here the term instead reflects their remarkable dynamical stability and classical behavior.)

The coherent states have the following basic properties:
(a) When expressed as a superposition of the energy eigenstates $\{\ket{n}\}$, the distribution as a function of $n$ squared is Poissonian (with mean number $\braket{\hat n}=|\alpha|^2$),

(b) They are related to the position and momentum eigenstates in the following way (ignoring factors of $\hbar$, $m$, and $\omega$): $\hat{a} = \left(\hat{x}+i\hat{p}\right)/\sqrt{2}$ and $\hat{a}^{\dagger} = \left(\hat{x}-i\hat{p}\right)/\sqrt{2}$. This further means that their eigenvalues are related by (again ignoring factors) $\alpha = \left( x + ip \right)/\sqrt{2}$ and $\alpha^* = \left( x - ip \right)/\sqrt{2}$, respectively.
    
(c) They satisfy the minimum uncertainty relation for $\hat x$ and $\hat p$ at all times: $\Delta\hat x \Delta\hat p=\hbar/2$. 
To put that more formally, recall that any two observables $\hat A$ and $\hat B$ satisfy the Heisenberg-Robertson uncertainty relation \cite{Heisenberg_1927, Robertson_1929}%
\fnt{Prove the Heisenberg-Sch\"odinger uncertainty relation.}
\begin{equation}
\label{eq:RS_unce}
\Delta \hat{A} \Delta \hat{B} \geq \frac{1}{2} \lvert \braket{[\hat{A},\hat{B}]} \rvert.
\end{equation}
For coherent states\fnt{Defining the dimensionless quadrature operators $\hat X = \left(\hat a + \hat a^{\dagger} \right)/2$ and $\hat P = -i\left(\hat a - \hat a^{\dagger} \right)/2$ prove that for coherent states $ \Delta \hat{X}  = \Delta \hat{P} = 1/2$.}, this uncertainty relation yield $\Delta \hat{x} \Delta \hat{p} =\hbar/2$.
    
(d) They can be generated by translating the ground state $\kvac$ by some finite distance in the phase space, i.e.
\begin{equation}
    \ket{\alpha} = e^{\alpha \hat{a}^\dagger -\alpha^* \hat{a}}\kvac,
\end{equation}
where $\hat{D}(\alpha)=e^{\alpha \hat{a}^\dagger -\alpha^* \hat{a}}$ is the so-called \textit{displacement operator}.%
\fnt{Let us define the displaced Fock state as $\hat D(\alpha)\ket{n}$. 
What is the variance of $\hat N$ in such state?}

This final feature requires a bit of unpacking. You may be familiar with classical phase spaces that are defined in terms of $x$ and $p$, and there is a way to write down a quantum equivalent. 
But we have also just seen how to convert from the parameters $x$ and $p$ to $\alpha$ and $\alpha^*$. 
This means we can also define a quantum phase space in terms of ${\rm{Re}}\left(\alpha\right)$ and ${\rm{Im}}\left(\alpha\right)$.
In this phase space, a state is a vector in the ($\text{Re}(\alpha), \text{Im}(\alpha)$) plane, and the vector can be translated. 
The aforementioned operator performs this translation --- it displaces the state by some amount $\alpha$, which can be seen from the following expressions:
\begin{align}    \hat{D}^\dagger(\alpha)\,\hat{a}\,\hat{D}(\alpha) &= \hat{a} + \alpha, \label{eq:displ}\\    \hat{D}^\dagger(\alpha)\,\hat{a}^\dagger\,\hat{D}(\alpha) &= \hat{a}^\dagger + \alpha^*.
\end{align}
Then, since $\hat{D}^\dagger(\alpha) = \hat{D}(-\alpha)$, the Hermitian conjugate of the displacement operator can be interpreted as a displacement of opposite amplitude 
%\ev{maybe magnitude is not the right word here} 
$-\alpha$, i.e.\ $\hat{D}(\alpha) \,\hat{a}\,\hat{D}^\dagger(\alpha) = \hat{a} - \alpha$. 
Note also that the displacement operator is unitary, and therefore satisfies
$\hat{D}(\alpha)\, \hat{D}^\dagger(\alpha) 
= \hat{D}^\dagger(\alpha)\,\hat{D}(\alpha) = \mathbbm{1}$, where $\mathbbm{1}$ denotes the identity operator.  
(We will shortly also see an equivalent definition of the displacement operator in the $\left(x,p\right)$ phase space.)

Now that we have introduced the coherent state and the displacement operator, and understood their basic properties, we can put them to use in some interesting ways.
First, one can define the \textit{characteristic function} of a quantum state $\hat{\rho}$ as simply the expectation value of the displacement operator with respect to the state:
\begin{align}
    \chi_{\hat{\rho}}(\gamma) 
    = \braket{\hat{D}(\gamma)}_{\hat{\rho}} 
    = \tr\!\left[\hat{\rho}\,\hat{D}(\gamma)\right],
    \label{eq:char_function_def}
\end{align}
where $\gamma \in \mathbb{C}$ parametrizes a shift in phase space.
The characteristic function provides a compact way of encoding the properties of the quantum state. 
Since $\hat{D}(\gamma)$ generates finite shifts, or \textit{displacements} in the creation operators, Eq. \eqref{eq:displ}, the quantity $\chi_{\hat{\rho}}(\gamma)$ can be viewed as a probe of how the state $\hat{\rho}$ responds to displacements in phase space. 
In this sense, the characteristic function acts as a generating function for the statistical moments of the field operators, much like a classical characteristic function encodes the moments of a random variable. 
This is exactly the case because $\hat{D}(\gamma)$ serves as the kernel of the traditional two-dimensional Fourier transform. 
To see this, we can express $\hat{D}(\gamma)$ in terms of the quadrature operators $\hat{x}$ and $\hat{p}$: $\hat{D}(\gamma) = e^{\gamma \hat{a}^\dagger - \gamma^* \hat{a}} = e^{i (\mu \hat{x} + \nu \hat{p})}$, where the parameters $\mu$ and $\nu$ are defined as $\mu = \sqrt{2/\hbar \omega} \, {\rm{Re}}(\gamma)$, and $\nu = \sqrt{2 \omega/\hbar} \, {\rm{Im}}(\gamma)$.
Recall that in traditional probability theory, the \textit{characteristic function} of a random variable $X$ is defined as $\varphi_X(t) = \mathbb{E}[e^{i t X}]$, where $\mathbb{E}$ denotes expected value.
This is essentially the one-dimensional version of what we have in the quantum case. 
Analogously, the quantum characteristic function can be written as $\chi_{\hat{\rho}}(\gamma) = \mathbb{E}[e^{i (\mu \hat{x} + \nu \hat{p})}]$.
Thus, $\hat{D}(\gamma)$ naturally plays the role of the Fourier kernel in phase space, just as $e^{i t X}$ does in classical probability.

For a coherent state $\ket{\alpha}$ one finds
\begin{align}
    \chi_{\alpha}(\gamma) 
    = \exp\!\left(-\tfrac{1}{2}|\gamma|^2 + \alpha \gamma^* - \alpha^* \gamma \right),
\end{align}
which clearly shows the shift of the vacuum result by the amplitude $\alpha$. 
This example highlights how displacements are directly reflected in the functional 
form of $\chi_{\hat{\rho}}(\gamma)$.

We end our discussion of coherent states and characteristic functions with a further connection they have with quadrature operators.
It turns out to be useful to introduce a generalized definition of a quadrature operator that has the canonical $\hat{x}$ and $\hat{p}$ operators as specific instances. In a single mode of the electromagnetic field, the definition is
\begin{equation}
\hat X_\varphi = \frac{1}{2}\left(\hat a e^{-i\varphi} + \hat a^\dagger e^{i\varphi}\right),
\end{equation}
where $\varphi$ denotes the phase of the quadrature.  
We recover the canonical position $\hat{x}$ from $\hat X_0$, and the canonical $\hat{p}$ from $\hat X_{\pi/2}$.
Then the eigenstates of the generalized quadrature operator, $|x,\varphi\rangle$, satisfy $\hat X_\varphi |x,\varphi\rangle = x_\varphi |x,\varphi\rangle$, meaning that measuring in this basis yields a definite value of the quadrature.
In quantum optical experiments, the Balanced Homodyne Detection (BHD) detection scheme measures the quadratures of a quantum optical field \cite{YuenShapiro1980, SmitheyBeckRaymerFaridani1993, LvovskyRaymer_2009}. 
By adjusting the measurement phase $\varphi$, one can access different quadratures $\hat X_\varphi$.  
Repeating the measurement many times yields the \textit{probability distribution} $p(x,\varphi)$ for the quadrature at phase $\varphi$.
\begin{align}
p(x,\varphi) = \bra{x,\varphi} \hat{\rho} \ket{x,\varphi} = \tr (\hat{\rho}\ket{x,\varphi}\bra{x,\varphi}).
\label{pdftrace}
\end{align}
Note that
\begin{gather}
\ket{x,\varphi}\bra{x,\varphi} = \hat{\delta} (\hat{X}_{\varphi}-x)\\ 
= \frac{1}{2\pi} \int_\mathbb{R}e^{iy\hat{X}_{\varphi}}e^{-iyx} dy = \frac{1}{2\pi} \int_\mathbb{R}\hat{D}(iye^{-i\varphi})\, e^{-iyx} dy,\nonumber
\label{pdfdispl}
\end{gather}
since $\hat{D}(iye^{-i\varphi}) = e^{iy\hat{a}^\dagger e^{-i\varphi} + iy \hat{a}e^{i\varphi}} = e^{iy \hat{X}_{\varphi}}$.

There exists a plethora of criteria to certify different quantum effects from BHD data, many of which are tailored to the specific state under consideration. 
A useful way to test such criteria is to work with simulated data. 
To simulate the data, we need the characteristic function (cf. Sec. \ref{sec:labs}), which enables us to compute the expected probability distribution, as shown next.
The characteristic function of a one-dimensional probability distribution, $p(x,\varphi)$, is given by
\begin{align*}
    \Psi(y,\varphi) = \int_\mathbb{R} p(x,\varphi) \, e^{iyx} dx.
\end{align*}
Inverting it, we get an expression for $p(x,\varphi)$
\begin{align}
    p(x,\varphi) &= \frac{1}{2\pi}\int_\mathbb{R} \Psi (y,\varphi)\, e^{-iyx} dy \label{pdfchar}\\
    &= \tr \left[\hat{\rho} \frac{1}{2\pi} \int_\mathbb{R} \hat{D}(iye^{-i\varphi})e^{-iyx} dy\right] \nonumber\\
    &= \frac{1}{2\pi} \int_\mathbb{R} \tr \left[\hat{\rho}\hat{D}(iye^{-i\varphi}) \right] e^{-iyx} dy.\nonumber
\end{align}
Here, one can notice that
\begin{align*}
    \Psi(y,\varphi) = \tr \left(\hat{\rho}\hat{D}(iye^{-i\varphi}) \right) = \braket{\hat{D}(iye^{-i\varphi})}.
\end{align*}
The characteristic function $\chi_{\hat{\rho}}(\gamma)$ for our quantum state related to $\Psi(y,\varphi)$ is given by $\chi_{\hat{\rho}}(iye^{-i\varphi}) = \tr (\hat{\rho} \hat{D}(iye^{-i\varphi}))$ $\Rightarrow \Psi(y,\varphi) = \chi_{\hat{\rho}}(iye^{-i\varphi})$.
Replacing in Eq.~(\ref{pdfchar}) what we just found, we get
\begin{align}
   p(x,\varphi) &= \frac{1}{2\pi} \int_\mathbb{R} \chi_{\hat{\rho}}(iye^{-i\varphi}) e^{-iyx} dy.
   \label{prob_from_char}
\end{align}
Measurement or simulation of field quadrature values enables us to sample the characteristic function at specific phase-space points, thereby linking the measurement (or simulation) statistics to the state’s full phase-space description.

We now recognize how coherent states provide the closest quantum analogue to a classical EM field mode, as they combine minimum uncertainty with well-defined displacements in phase space. 
It is interesting that the structure of these states arises naturally from the action of the displacement operator on the vacuum, and that their statistical properties are elegantly captured by the characteristic function. 
These features make coherent states a fundamental reference point when classifying the classicality or nonclassicality of quantum states. 
In the following section, we delve further into exploring squeezed states, which preserve minimal uncertainty but redistribute it in a highly non-trivial manner.
\section{Squeezed states}
\label{sec:squeezed}

Before introducing squeezed states explicitly, we recall that the uncertainty principle does not constrain standard deviations of operators; it only constrains the product of them for conjugate variables. 
This asymmetry leaves room for states in which one quadrature is made more precise at the expense of increased fluctuations in the other.

We have seen that for coherent states (Sec. \ref{sec:coherent}), the $\hat X$ and $\hat P$ uncertainty product is minimized with values $\Delta \hat X  =  \Delta \hat P  = 1/2$, including vacuum.
Note that here we are considering the \textit{dimensionless} quadratures $\hat{X}=(\hat a+\hat a^\dagger)/2$ and $\hat{P}=(\hat a-\hat a^\dagger)/2i$ that satisfy the uncertainty relation [Eq.~\eqref{eq:RS_unce}] as follows
\begin{equation}
\label{eq:UncPrin}
\Delta \hat{X} \Delta \hat{P} \geq \frac{1}{4}.
\end{equation}

\begin{figure}[h] 
    \centering
    \begin{subfigure}[t]{0.48\linewidth}
        \tikzset{every picture/.style={line width=0.75pt}} %set default line width to 0.75pt        

\begin{tikzpicture}[x=0.75pt,y=0.75pt,yscale=-0.5,xscale=0.5]

%Shape: Axis 2D [id:dp7549838460702714] 
\draw  (65,225) -- (263,225)(84.8,36) -- (84.8,246) (256,220) -- (263,225) -- (256,230) (79.8,43) -- (84.8,36) -- (89.8,43)  ;
%Shape: Circle [id:dp8659159160396732] 
\draw  [draw opacity=0][fill={rgb, 255:red, 241; green, 162; blue, 38 }  ,fill opacity=0.5 ] (115,140.5) .. controls (115,114.27) and (136.27,93) .. (162.5,93) .. controls (188.73,93) and (210,114.27) .. (210,140.5) .. controls (210,166.73) and (188.73,188) .. (162.5,188) .. controls (136.27,188) and (115,166.73) .. (115,140.5) -- cycle ;
%Straight Lines [id:da8709442410029471] 
\draw [color={rgb, 255:red, 0; green, 0; blue, 0 }  ,draw opacity=1 ]   (112.28,198.6) -- (209.28,198.6)  ;
\draw [shift={(209.28,198.6)}, rotate = 180] [color={rgb, 255:red, 0; green, 0; blue, 0 }  ,draw opacity=1 ][line width=0.75]    (0,5.59) -- (0,-5.59)   ;
\draw [shift={(112.28,198.6)}, rotate = 180] [color={rgb, 255:red, 0; green, 0; blue, 0 }  ,draw opacity=1 ][line width=0.75]    (0,5.59) -- (0,-5.59)   ;
%Straight Lines [id:da7138122099912997] 
\draw [color={rgb, 255:red, 0; green, 0; blue, 0 }  ,draw opacity=1 ]   (222.28,187.6) -- (223.28,89.6)  ;
\draw [shift={(223.28,89.6)}, rotate = 90.58] [color={rgb, 255:red, 0; green, 0; blue, 0 }  ,draw opacity=1 ][line width=0.75]    (0,5.59) -- (0,-5.59)   ;
\draw [shift={(222.28,187.6)}, rotate = 90.58] [color={rgb, 255:red, 0; green, 0; blue, 0 }  ,draw opacity=1 ][line width=0.75]    (0,5.59) -- (0,-5.59)   ;

% Text Node
\draw (147,200) node [anchor=north west][inner sep=0.75pt]  [color={rgb, 255:red, 0; green, 0; blue, 0 }  ,opacity=1 ] [align=left] {$\displaystyle \Delta \hat{x}$};
% Text Node
\draw (227,130) node [anchor=north west][inner sep=0.75pt]  [color={rgb, 255:red, 0; green, 0; blue, 0 }  ,opacity=1 ] [align=left] {$\displaystyle \Delta \hat{p}$};
% Text Node
\draw (262,223) node [anchor=north west][inner sep=0.75pt]  [color={rgb, 255:red, 0; green, 0; blue, 0 }  ,opacity=1 ] [align=left] {$\displaystyle x$};
% Text Node
\draw (63,35) node [anchor=north west][inner sep=0.75pt]  [color={rgb, 255:red, 0; green, 0; blue, 0 }  ,opacity=1 ] [align=left] {$\displaystyle p$};

\end{tikzpicture}
        \caption{Coherent state: The uncertainties in both quadratures are equal.}
        \label{fig:coheren}
    \end{subfigure}
    \begin{subfigure}[t]{0.48\linewidth}
        \tikzset{every picture/.style={line width=0.75pt}} %set default line width to 0.75pt        

\begin{tikzpicture}[x=0.75pt,y=0.75pt,yscale=-0.5,xscale=0.5]

%Shape: Axis 2D [id:dp9317121486444707] 
\draw  (85,245) -- (283,245)(104.8,56) -- (104.8,266) (276,240) -- (283,245) -- (276,250) (99.8,63) -- (104.8,56) -- (109.8,63)  ;
%Straight Lines [id:da5148478534096741] 
\draw [color={rgb, 255:red, 0; green, 0; blue, 0 }  ,draw opacity=1 ]   (144.28,218.34) -- (194.28,217.34)  ;
\draw [shift={(194.28,217.34)}, rotate = 178.85] [color={rgb, 255:red, 0; green, 0; blue, 0 }  ,draw opacity=1 ][line width=0.75]    (0,5.59) -- (0,-5.59)   ;
\draw [shift={(144.28,218.34)}, rotate = 178.85] [color={rgb, 255:red, 0; green, 0; blue, 0 }  ,draw opacity=1 ][line width=0.75]    (0,5.59) -- (0,-5.59)   ;
%Straight Lines [id:da8240638887927583] 
\draw [color={rgb, 255:red, 0; green, 0; blue, 0 }  ,draw opacity=1 ]   (209.28,209.34) -- (209.28,69.34)  ;
\draw [shift={(209.28,69.34)}, rotate = 90] [color={rgb, 255:red, 0; green, 0; blue, 0 }  ,draw opacity=1 ][line width=0.75]    (0,5.59) -- (0,-5.59)   ;
\draw [shift={(209.28,209.34)}, rotate = 90] [color={rgb, 255:red, 0; green, 0; blue, 0 }  ,draw opacity=1 ][line width=0.75]    (0,5.59) -- (0,-5.59)   ;
%Shape: Ellipse [id:dp6680577518813838] 
\draw  [draw opacity=0][fill={rgb, 255:red, 41; green, 140; blue, 140 }  ,fill opacity=0.5 ] (144,139.67) .. controls (144,100.83) and (155.48,69.34) .. (169.64,69.34) .. controls (183.8,69.34) and (195.28,100.83) .. (195.28,139.67) .. controls (195.28,178.51) and (183.8,210) .. (169.64,210) .. controls (155.48,210) and (144,178.51) .. (144,139.67) -- cycle ;

% Text Node
\draw (150,221) node [anchor=north west][inner sep=0.75pt]  [color={rgb, 255:red, 0; green, 0; blue, 0 }  ,opacity=1 ] [align=left] {$\displaystyle \Delta \hat{x}$};
% Text Node
\draw (216,132) node [anchor=north west][inner sep=0.75pt]  [color={rgb, 255:red, 0; green, 0; blue, 0 }  ,opacity=1 ] [align=left] {$\displaystyle \Delta \hat{p}$};
% Text Node
\draw (282,243) node [anchor=north west][inner sep=0.75pt]  [color={rgb, 255:red, 0; green, 0; blue, 0 }  ,opacity=1 ] [align=left] {$\displaystyle x$};
% Text Node
\draw (83,55) node [anchor=north west][inner sep=0.75pt]  [color={rgb, 255:red, 0; green, 0; blue, 0 }  ,opacity=1 ] [align=left] {$\displaystyle p$};

\end{tikzpicture}
        \caption{Squeezed state: The uncertainty in one quadrature is smaller than the coherent state, but the other is greater. }
        \label{fig:squeezed}
    \end{subfigure}
    \caption{Illustrations of symmetric (coherent) and asymmetric (squeezed) quadrature uncertainties.}
    \label{fig:xp-space}
\end{figure}
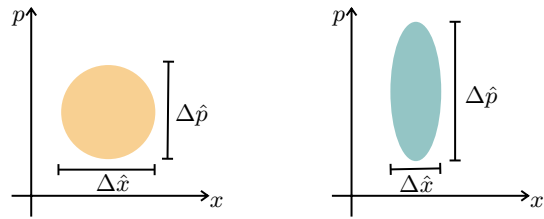

Notice that it is always possible for a state to have less uncertainty than $1/2$ in one of those measurements while still satisfying the uncertainty principle, Eq. \eqref{eq:UncPrin}, as long as there is more uncertainty in the complementary measurement to compensate.
States in which the uncertainty of one quadrature is reduced below the coherent state value, while still having minimal uncertainty overall in the product, are called squeezed-quadrature states or simply \textit{squeezed states}. 
It becomes clear where the name `squeezed' comes from when one examines Fig.~\ref{fig:xp-space}. 
Crucially, the overall volume in phase space is the same as for a coherent state.

The squeezed states satisfy the following properties:
(a) They can be expressed as a superposition of the energy eigenstates:
\begin{equation}\label{eq:squeezed}
        \ket{\xi} = \frac{1}{\sqrt{\cosh(r)}} \sum_{n=0}^{\infty} \frac{\sqrt{(2n)!}}{2^n n!} \left(-e^{i\phi} \tanh(r)\right)^n \ket{2n},
\end{equation}
with the squeezing parameter $\xi = r e^{i\theta}$.
(b) They can be generated by squeezing the quadratures of the ground state $\kvac$ by some finite amount in a certain direction given by the parameter $\xi$:
\begin{equation}
    \ket{\xi} = e^{\frac{1}{2} \left(\xi^* \hat{a}^2-\xi \hat{a}^{\dagger 2}\right)}\kvac,
    \label{eq:SqueezedDef}
\end{equation}
where $\hat{S}(\xi)=e^{\frac{1}{2}\left(\xi^* \hat{a}^2-\xi \hat{a}^{\dagger 2}\right)}$ is the so-called squeezing operator.%
\fnt{Prove that the action of the squeeze operator on the quadratures $\hat{X}$ and $\hat{P}$ is given by: $\hat{S}(\xi)^\dagger\hat{X} \hat{S}(\xi) = [\cosh(r) - \cos(\theta) \sinh(r)]\hat{X} - \sin(\theta) \sinh(r) \hat{P}$ and $\hat{S}(\xi)^\dagger \hat{P} \hat{S}(\xi) = [\cosh(r) + \cos(\theta) \sinh(r)]\hat{P} - \sin(\theta) \sinh(r) \hat{X}$. 
\underline{Hint:} Use the Baker-Campbell-Hausdorff (BCH) lemma $e^{\hat{B}} \hat{a} e^{-\hat{B}} = \hat{a} + [\hat{B}, \hat{a}] + \frac{1}{2!} [\hat{B}, [\hat{B}, \hat{a}]] + \frac{1}{3!} [\hat{B}, [\hat{B}, [\hat{B}, \hat{a}]]] + \mydots$}

Squeezed states exhibit what we will later define as non-classical behavior. 
These non-classical properties have played an important role in various quantum technology applications. 
For example, the use of squeezed states has demonstrated improvements in the enhancement of spatial resolution at the nanoscale in biological applications \cite{TaylorJanousekDariaEtAl_2014}. 
Squeezed states have also been employed in various quantum information applications, such as quantum dense coding \cite{PatraGuptRoyEtAl_2022} and quantum key distribution \cite{CerfLevyAssche_2001}.
Finally, the use of squeezed states has significantly improved the sensitivity of gravitational wave detectors—such as those run by the LIGO Scientific Collaboration—making their application in this field particularly noteworthy \cite{LIGO_2011}.

We do not need to dwell much more on squeezed states in this course. 
Here it suffices to say that squeezed-quadrature states are quantum states where the uncertainty in one of the two quadrature phases of an electromagnetic field is reduced below the standard limit imposed by the coherent states.
This squeezing in one quadrature necessarily enhances the uncertainty in the other conjugate quadrature due to the Heisenberg uncertainty principle, and this effect is a vital resource for various quantum technologies.
The coherent and squeezed states discussed so far have been conveniently represented as superpositions of vector states [Eq.~\eqref{eq:coherent} and Eq.~\eqref{eq:squeezed}]. 
However, this vector representation has its limitations. 
To effectively describe more general quantum states --- particularly those involving statistical uncertainty in the original information of the quantum system --- it becomes essential to use the concepts of statistical mixtures and the density operator introduced in Sec \ref{sec:Mixtures}.
\section{Convexity and classicality}
\label{sec:convexCL}
We have introduced the most classical pure states of the field --- coherent states (Sec. \ref{sec:coherent}) --- and now we would like to explore how to construct other states that still behave in a classical-like manner. 
An intuitive idea is to combine coherent states with different amplitudes $\alpha$, each weighted by some function $P(\alpha)$. 
However, as discussed in our introduction to mixed states, we must distinguish carefully between a \textit{superposition} and a \textit{statistical mixture}.
Simply adding kets with complex coefficients $a_i$, such that $\sum_i |a_i|^2=1$, we obtain a coherent superposition of states, and such superpositions typically give rise to highly nonclassical behavior due to interference terms. 
What we want instead is a classical mixture, where different coherent states contribute with certain classical probabilities rather than quantum amplitudes. 
To describe such mixtures correctly, we must use the density-operator formalism discussed before (Sec. \ref{sec:Mixtures}).
In this representation, a statistical ensemble of coherent states is expressed as
\begin{align}
    \hat{\rho}  
    &= \int_\mathbb{C} P(\alpha)\, |\alpha\rangle\langle\alpha|\, d^2 \alpha  
    \label{eq:rhoP}
\end{align}
where we introduce the notation $d^2\alpha$ that represents an integral over the real and imaginary parts of $\alpha$, i.e. $d^2\alpha = d{\rm{Re}}(\alpha) d{\rm{Im}}(\alpha)$.

Having $P(\alpha_i) = 0$ means the coherent state of amplitude $\alpha_i$, $|\alpha_i\rangle$, is not in the mixture. 
We could also have $P(\alpha) = \delta(\alpha - \alpha_j)$, which will correspond to having only $|\alpha_j\rangle$ in the mixture (a very boring mixture since we are not really mixing anything). 
We can see that in general we have $P(\alpha) \geq 0$ and $\int_{\mathbb{C}}P(\alpha)\, |\alpha\rangle\langle\alpha|\, d^2 \alpha = 1$. 
These are two of the defining properties of a probability distribution.
Any element that can be expressed as a probabilistic mixture of other elements necessarily belongs to the convex set generated by those elements.

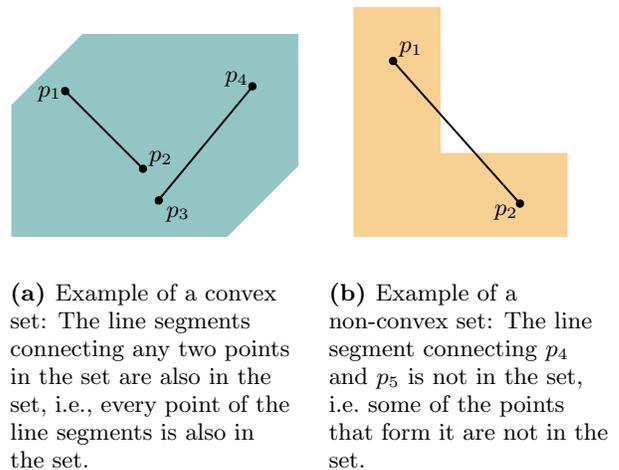
\begin{figure}[t]
    \centering
    \begin{subfigure}[t]{0.43\linewidth}
        \tikzset{every picture/.style={line width=0.75pt}} %set default line width to 0.75pt        

\begin{tikzpicture}[x=0.75pt,y=0.75pt,yscale=-0.8,xscale=0.8]

%Snip Diagonal Corner Rect [id:dp4736508578339359] 
\draw  [draw opacity=0][fill={rgb, 255:red, 41; green, 140; blue, 140 }  ,fill opacity=0.5 ] (76.8,93) -- (213,93) -- (213,93) -- (213,176.2) -- (168.2,221) -- (32,221) -- (32,221) -- (32,137.8) -- cycle ;
%Straight Lines [id:da2655845899173759] 
\draw    (64,127) -- (114,177) ;
%Straight Lines [id:da7693397218074285] 
\draw    (185,125) -- (123,200) ;
%Shape: Circle [id:dp02667113594502446] 
\draw  [fill={rgb, 255:red, 0; green, 0; blue, 0 }  ,fill opacity=1 ] (68,129) .. controls (68,127.9) and (67.1,127) .. (66,127) .. controls (64.9,127) and (64,127.9) .. (64,129) .. controls (64,130.1) and (64.9,131) .. (66,131) .. controls (67.1,131) and (68,130.1) .. (68,129) -- cycle ;
%Shape: Circle [id:dp9607115715574361] 
\draw  [fill={rgb, 255:red, 0; green, 0; blue, 0 }  ,fill opacity=1 ] (186,126) .. controls (186,124.9) and (185.1,124) .. (184,124) .. controls (182.9,124) and (182,124.9) .. (182,126) .. controls (182,127.1) and (182.9,128) .. (184,128) .. controls (185.1,128) and (186,127.1) .. (186,126) -- cycle ;
%Shape: Circle [id:dp7148559655069698] 
\draw  [fill={rgb, 255:red, 0; green, 0; blue, 0 }  ,fill opacity=1 ] (117,178) .. controls (117,176.9) and (116.1,176) .. (115,176) .. controls (113.9,176) and (113,176.9) .. (113,178) .. controls (113,179.1) and (113.9,180) .. (115,180) .. controls (116.1,180) and (117,179.1) .. (117,178) -- cycle ;
%Shape: Circle [id:dp37306149446806625] 
\draw  [fill={rgb, 255:red, 0; green, 0; blue, 0 }  ,fill opacity=1 ] (127,198) .. controls (127,196.9) and (126.1,196) .. (125,196) .. controls (123.9,196) and (123,196.9) .. (123,198) .. controls (123,199.1) and (123.9,200) .. (125,200) .. controls (126.1,200) and (127,199.1) .. (127,198) -- cycle ;

% Text Node
\draw (47,124) node [anchor=north west][inner sep=0.75pt]   [align=left] {$\displaystyle p_{1}$};
% Text Node
\draw (117,165) node [anchor=north west][inner sep=0.75pt]   [align=left] {$\displaystyle p_{2}$};
% Text Node
\draw (128,200) node [anchor=north west][inner sep=0.75pt]   [align=left] {$\displaystyle p_{3}$};
% Text Node
\draw (165,115) node [anchor=north west][inner sep=0.75pt]   [align=left] {$\displaystyle p_{4}$};

\end{tikzpicture}
        \caption{Example of a convex set: The line segments connecting any two points in the set are also in the set, i.e., every point of the line segments is also in the set.}
        \label{fig:convex}
    \end{subfigure}
    \hspace{3mm}
    \begin{subfigure}[t]{0.43\linewidth}
        \tikzset{every picture/.style={line width=0.75pt}} %set default line width to 0.75pt        

\begin{tikzpicture}[x=0.75pt,y=0.75pt,yscale=-0.8,xscale=0.8]

%Shape: L Shape [id:dp6425758145065906] 
\draw  [draw opacity=0][fill={rgb, 255:red, 241; green, 162; blue, 38 }  ,fill opacity=0.5 ] (23,67) -- (78,67) -- (78,159) -- (158,159) -- (158,212) -- (23,212) -- cycle ;
%Straight Lines [id:da282459090739581] 
\draw    (47,100) -- (128,191) ;
%Shape: Circle [id:dp0901110858701456] 
\draw  [fill={rgb, 255:red, 0; green, 0; blue, 0 }  ,fill opacity=1 ] (130,191) .. controls (130,189.9) and (129.1,189) .. (128,189) .. controls (126.9,189) and (126,189.9) .. (126,191) .. controls (126,192.1) and (126.9,193) .. (128,193) .. controls (129.1,193) and (130,192.1) .. (130,191) -- cycle ;
%Shape: Circle [id:dp9918057670385493] 
\draw  [fill={rgb, 255:red, 0; green, 0; blue, 0 }  ,fill opacity=1 ] (50,101) .. controls (50,99.9) and (49.1,99) .. (48,99) .. controls (46.9,99) and (46,99.9) .. (46,101) .. controls (46,102.1) and (46.9,103) .. (48,103) .. controls (49.1,103) and (50,102.1) .. (50,101) -- cycle ;

% Text Node
\draw (50,87) node [anchor=north west][inner sep=0.75pt]   [align=left] {$\displaystyle p_{1}$};
% Text Node
\draw (110,190) node [anchor=north west][inner sep=0.75pt]   [align=left] {$\displaystyle p_{2}$};

\end{tikzpicture} \vspace*{\baselineskip}
        \caption{Example of a non-convex set: The line segment connecting $p_4$ and $p_5$ is not in the set, i.e.\ some of the points that form it are not in the set.}
        \label{fig:nonconvex}
    \end{subfigure}
    \caption{Illustration of convex and non-convex sets using line-segment criteria.
    }
    \label{fig:convex_vs_nonconvex}
\end{figure}

Let us note that another concept is at play here: convexity. 
To understand what we mean by convexity in a physical sense, let us first discuss its abstract mathematical meaning from set theory. 
Loosely speaking, a set is convex if the straight line between any two points in the set always lies entirely within the set. 
And here you are free to imagine the set as simply a two-dimensional shape, as in Fig.~\ref{fig:convex_vs_nonconvex}. 
Making this idea mathematically precise is quite simple.
Let us consider the following definition:

\textit{Definition} \ul{Convex Set}: Let $S$ be a set. 
We say S is convex if $(1-p) s_1 + p\, s_2 \in S,$ $\forall s_1, s_2 \in S, p \in [0,1]$.

The way we are combining the points $s_1$ and $s_2$ is called a \textit{convex combination}. 
And really, this is effectively what we were writing in Eq.~\eqref{eq:rhoP}: a sum over different elements of a set (the collection of $\ket{\alpha}$), weighted by some factors (the $P\left(\alpha\right)$ weight function), where those factors are all normalized and constrained between 0 and 1.
If we form all possible statistical mixtures of coherent states, the resulting convex set of density operators still retains the classical characteristics of coherent states.
For this reason, the entire convex set is commonly referred to as the set of classical states of light.

\begin{figure}
    \centering
    \tikzset{every picture/.style={line width=0.75pt}} %set default line width to 0.75pt        

\begin{tikzpicture}[x=0.75pt,y=0.75pt,yscale=-1,xscale=1]
%uncomment if require: \path (0,182); %set diagram left start at 0, and has height of 182

%Shape: Ellipse [id:dp6045365792194101] 
\draw  [draw opacity=0][fill={rgb, 255:red, 241; green, 162; blue, 38 }  ,fill opacity=0.5 ] (16,90.5) .. controls (16,44.38) and (80.02,7) .. (159,7) .. controls (237.98,7) and (302,44.38) .. (302,90.5) .. controls (302,136.62) and (237.98,174) .. (159,174) .. controls (80.02,174) and (16,136.62) .. (16,90.5) -- cycle ;
%Shape: Ellipse [id:dp3796676749213428] 
\draw  [draw opacity=0][fill={rgb, 255:red, 41; green, 140; blue, 140 }  ,fill opacity=0.5 ] (16,90.5) .. controls (16,64.54) and (54.73,43.5) .. (102.5,43.5) .. controls (150.27,43.5) and (189,64.54) .. (189,90.5) .. controls (189,116.46) and (150.27,137.5) .. (102.5,137.5) .. controls (54.73,137.5) and (16,116.46) .. (16,90.5) -- cycle ;

% Text Node
\draw (51,71) node [anchor=north west][inner sep=0.75pt]  [color={rgb, 255:red, 0; green, 0; blue, 0 }  ,opacity=1 ] [align=left] {\begin{minipage}[lt]{70.2pt}\setlength\topsep{0pt}
\begin{center}
Classical\\(P-Nonnegative)
\end{center}

\end{minipage}};
% Text Node
\draw (210,79) node [anchor=north west][inner sep=0.75pt]  [color={rgb, 255:red, 0; green, 0; blue, 0 }  ,opacity=1 ] [align=left] {\begin{minipage}[lt]{44.68pt}\setlength\topsep{0pt}
\begin{center}
Quantum
\end{center}

\end{minipage}};

\end{tikzpicture}
    \caption{Quantum and classical (in the sense of P-non negativity) states form convex sets.}
    \label{fig:classical_quantum_convex}
\end{figure}
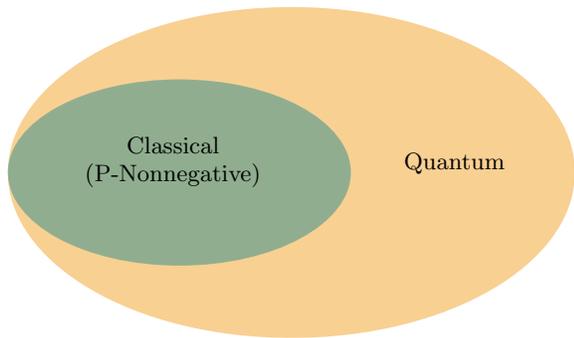

Our physical reality is, in fact, far more interesting and richer than what classical states alone can describe.
So far, we have demanded that $P\left(\alpha\right)$ satisfy the convexity conditions required to be a well-defined probability distribution.
However, many quantum states cannot be represented in this way unless these constraints are relaxed.
For example, if we allow $P(\alpha)$ to take negative values, then it is no longer a proper probability distribution; instead, it is referred to as a \textit{quasiprobability}.
States described by a negative $P$-function are known as \textit{nonclassical states}.
Although $P(\alpha)$ is not a true probability distribution in this case, it remains a useful phase-space representation of a quantum state, in analogy with classical statistical mechanics.
Such nonclassical states are also represented by positive semidefinite density operators \(\hat{\rho} \ge 0\), i.e.\ \(\langle \psi | \hat{\rho} | \psi \rangle \ge 0\) for all \(|\psi\rangle\), ensuring that all measurement outcome probabilities are non‑negative, and they are normalized \(\mathrm{Tr}(\hat{\rho}) = 1\), guaranteeing that the total probability over all possible outcomes is 1. 
The set of quantum states and the set nonclassical states are both convex and related as shown in Fig.~\ref{fig:classical_quantum_convex}.

In fact, some of the states we have studied already, such as the squeezed states (Sec. \ref{sec:squeezed}), can only be represented in the diagonal form [Eq. \eqref{eq:rhoP}] if we allow $P(\alpha)$ to have negative values. 
So the negativity in the quasiprobability representation is a sign that the state is truly, fundamentally quantum in some sense; that its behavior differs from classical theories. This phase-space picture of quantum theory offers an alternative perspective that complements the ones we have built so far and reveals hidden insights into quantum systems.
For this reason and many more, it is worth devoting some time to discussing quasiprobabilities in more detail.
\section{Quasiprobabilities and the notion of P-nonclassicality}
\label{sec:quasip}

Let us discuss more what we mean by a `phase space picture' of quantum mechanics, a key component of modern quantum optics.
Originating with Wigner’s work in the 1930s \cite{Wigner_1932}, the idea of assigning phase-space distributions to quantum states extends classical statistical concepts—such as those used in statistical mechanics, chaos theory, and thermodynamics—into the quantum realm \cite{Groenewold1946,Moyal1949,Glauber_1963,Sudarshan_1963,CahillGlauber_1969}.

In classical physics, we describe the state of a physical object and its dynamics using probability densities in the phase space.
In the classical domain, the probability density is a positive and normalized distribution, and we use these to calculate the state of the system and the value of the dynamical variables in the form of expectation values. 
Classically, the probability distribution can easily be associated with the probability of observing the system to be in certain well-defined states.
But as we discussed before, when we attempt to make an equivalent description for quantum states, we find that the distributions that play the same role behave quite differently, most notably, by possibly having negative values.

Every bosonic system can be spanned in terms of the generating set of coherent states $\{\ket{\alpha}\}$ in the diagonal form of Eq.~\eqref{eq:rhoP}. 
The $P(\alpha)$ is also called the Glauber-Sudarshan $P$ distribution \cite{Glauber_1963,Sudarshan_1963}.
Although it can potentially be negative, it is still well-normalized like a good probability distribution, and it
allows us to compute expectation values of observables as:
\begin{equation*}
\int d^{2}\alpha\,P(\alpha)=1,
\end{equation*}
and
\begin{equation*}
\langle:\hat{f}(\hat a,\hat a^\dagger):\rangle =\int d^{2}\alpha\,P(\alpha)\,f(\alpha,\alpha^*).
\end{equation*}

Here we have introduced some new notation. 
The function in the bra-ket being surrounded by dots, $:\hat{f}\left(\hat{a},\hat{a}^{\dagger}\right):$ means that it is \textit{normal-ordered}. This means that we have taken the operator $\hat{f}\left(\hat{a},\hat{a}^{\dagger}\right)$ under the requirement that in every term of the expansion of the operator, all creation operators are to the left of all annihilation operators. 
That may sound strange, but we have already seen a simple operator that naturally fulfills this condition: the number operator $\hat{N}=\hat{a}^{\dagger}\hat{a}$. 
Hence, the $P$ distribution is the one associated with normal ordering

It can be shown that there are in principle infinitely many ways to expand a given operator in terms of the field operators, 
and that in turn means that there are infinitely many possible orderings, depending on
the quantity of terms in the expansion with the creation (annihilation) operators to the left (right).
The reason we mention all of this is that every given order will have an associated quasiprobability distribution.

Research in quantum optics significantly benefited from the concept of phase-space quasiprobability distributions.
We previously introduced the $P$ distribution, but one can relate a large class of quasiprobability distributions to the $P$ distribution and to each other via a parameter $s$. 
Specifically, $s$ controls a convolution of $P(\alpha)$ with a Gaussian kernel:
\begin{equation}
 P(\alpha,s)=\frac{2}{\pi(1-s)}\int d^2\gamma\, P(\gamma,1)\, \exp\left(-\frac{2|\alpha-\gamma|^2}{(1-s)}\right),
 \end{equation}
where $ P(\gamma)=P(\gamma,1)$ ($s=1$) is the Glauber-Sudarshan $P$ distribution.

This is the so-called $s$-parametrization of quasiprobabilities \cite{CahillGlauber_1969}, and the different quasiprobabilities in this family have different properties. 
For instance, they become increasingly regular (meaning containing fewer singular values) as the parameter $s$ decreases from 1 towards $-1$. 
If you have heard of quasiprobabilities before, it is likely via the Wigner distribution we alluded to earlier. 
This turns out to be the case where $s=0$. 
And like the $P$ distribution, it is associated with a specific operator ordering, in this case a symmetric ordering where the creation and annihilation operators are distributed symmetrically in each term in the expansion. 
The reason the Wigner distribution in particular is so well-known is that it often retains classical-like features, while still capturing quantum interference effects. 
And again, although it is real and normalized, it can take on negative values—these negativities serve as a distinctive signature of nonclassicality.
Moreover, the Wigner function yields the correct marginal probability distributions for conjugate quadratures, reinforcing its interpretation as a phase-space representation.

There are infinitely many quasiprobabilities defined via $s$ that may be of interest, but one more that is useful to mention is the $s=-1$ case, called the Husimi $Q$ distribution \cite{Husimi_1940}. 
This is obtained using anti-normal ordering, where annihilation operators always precede creation ones in each term of the expansion. 
It even has a very simple definition: the $Q$ distribution of a state $\hat{\rho}$ is just $Q_{\hat{\rho}}\left(\alpha\right)=\braket{\alpha|\hat{\rho}|\alpha}/\pi$.
And surprisingly, $Q\left(\alpha\right)$ is actually always positive (and never singular).

The reason the $Q$ distribution is still considered a quasiprobability, even though it is everywhere nonnegative, is that it cannot satisfy the
classical additivity rules for mutually exclusive events.
For an ordinary probability distribution, if two events $a$ and $b$ are mutually exclusive, then $p(a \cup b) = p(a) + p(b)$.
In phase space, this corresponds to associating probabilities with disjoint
regions in a representation-independent way.
However, the $Q$ distribution does \textit{not} represent probabilities of
simultaneous values of noncommuting observables such as $x$ and $p$. 
Instead, $Q(\alpha) = \langle \alpha | \hat\rho | \alpha \rangle/\pi$ gives the outcome probabilities of measuring the coherent-state POVM
$\{|\alpha\rangle\langle\alpha|/\pi\}$.
Coherent states are overcomplete and non-orthogonal, so they do not correspond to mutually exclusive events. 
In particular, they satisfy $\langle \alpha | \beta \rangle \neq 0 \quad \text{for all } \alpha,\beta \in \mathbb{C}$, which implies that $|\alpha\rangle\langle\alpha| + |\beta\rangle\langle\beta|$ is not a projector onto a union of mutually exclusive outcomes. 
Consequently, one cannot interpret $Q(\alpha)$ as defining probabilities for disjoint regions of phase
space, and the classical additivity relation fails: the probability assigned to a region constructed from overlapping POVM elements is not the sum of the
individual $Q$-values.

Thus, positivity alone is insufficient. Because the coherent-state POVM fails to
provide mutually exclusive events, the $Q$ function does not satisfy the axioms of
classical probability and therefore remains a quasiprobability distribution.

But like with other quasiprobabilities, it can be quite useful. The connection to coherent state POVMs is important in its own right, its positivity is helpful for visualizing the semiclassical structure of states, and its smoothing properties suppress singularities present in other quasiprobabilities.
However, this same coarse-graining comes at the cost of losing access to the finer quantum features, such as the interference fringes and negativities that one sees with the Wigner representation.

Let us now explore in more detail how quasiprobabilities can be used for tests of nonclassicality.
From the density matrix representation of Eq.~(\ref{eq:rhoP}), we have defined classical fields as the ones that can be written as classical mixtures of coherent states. 
(Recall that coherent states are the `most classical' ones and hence serve as a classical reference.) 
For such fields, let us call the $P$ distribution $P(\alpha)=P_{\text{cl.}}$. 
And so naturally, in the case that $P$  does not resemble a classical probability distribution, we have a \textit{nonclassical field}.

Returning to the original formulation of nonclassicality, classical states are defined as those for which the function $P(\alpha)$ is positive definite \cite{TitulaerGlauber_1965}.
This means that, regardless of how the state is represented, if there exists any representation of the $P$-function that is manifestly nonnegative, then the state must be regarded as classical.
However, across many textbooks and research papers, one often encounters statements along the lines of: ``a state is nonclassical if its $P$-function becomes negative or takes a form more singular than a Dirac delta.''
This claim does not hold in general.
The classical thermal state, elegantly analyzed in Ref. \cite{Sperling_2016}, already demonstrates that this claim is not generally valid and the negative values are the signature of quantumness in this context.

If our task is to certify nonclassicality, then ultimately our goal is to show that there is a value of $s$ for which the given quasiprobability contains negative values.
In our search for negativities we can already conclude a few things. 
First, all $P(\alpha,s)$ for $s\le 0$ are always regular (lacking singularities), eliminating one major way negativities can sneak in to a function.
Next, the $Q$ function is proportional to the expectation value of the density operator on the coherence states, and therefore is by definition always nonnegative.
Additionally, Wigner functions are known to be nonnegative only for some nonclassical states.
One should keep in mind that the positivity of an $s$-parametrized quasiprobability with $s<1$ is not sufficient to identify a state as classical.

We have already established that the $P$ distribution's negativities are always true signs of quantumness.
But unfortunately, the $P$ distribution is in general strongly singular for many quantum states, even in the single mode scenario. 
As a consequence, this function is in most cases experimentally not accessible. 
With the s-parametrization one overcomes the singularities, but we lose the ability to test for quantumness.

Finally, let us consider the squeezed states (Sec. \ref{sec:squeezed}) of Eq.~\eqref{eq:SqueezedDef} again in this context.
They form a basis for continuous variable quantum information, science and technology and play a key role in quantum metrology, and they provide a good example of how tricky it can be to directly test nonclassicality from quasiprobabilities.
All possible quasiprobabilities for such a state are either positive or irregular.
For instance, it has a Gaussian Wigner function that is always positive, and its $P$ function
\begin{align*}
    P_{\text{SV}}(\alpha)=
    e^{-\frac{(\Delta \hat x)^2-(\Delta \hat p)^2}{8}(\partial_\alpha^2+\partial^2_{\alpha^*}) + \frac{(\Delta \hat x)^2+(\Delta \hat p)^2-2}{4} \partial_\alpha\partial_{\alpha^*}} \delta(\alpha),
\end{align*}
is an exponential of higher order derivatives of the delta distribution.
This function is so singular that it is not even possible to plot properly as a function of $\alpha$.
Highly singular distributions may involve derivatives of $\delta$-functions, infinite spikes, or objects that do not have well-defined values anywhere.
This makes it completely impossible to directly check for negativities.
As we can see, testing for quantum effects using quasiprobability can be a useful technique; however, it has its limitations.

% DAY 2
\section{Gaussianity}
\label{sec:gaussianiry}

The concept of Gaussianity in quantum optics is intrinsically linked to the Wigner function \cite{Wigner_1932}.  
States associated with Hamiltonians that are quadratic in the quadrature operators possess a Gaussian Wigner function (cf. Appendix \ref{apx:gaussian}) of the form 
\begin{equation*} 
	W(\mathbf{r}) = \frac{1}{ \sqrt{(2\pi)^{2N}\text{Det}(\sigma)}} e^{-\frac{1}{2}(\mathbf{r}-\mathbf{\bar{r}})^T \sigma^{-1}(\mathbf{r}-\mathbf{\bar{r}})},
\end{equation*}
where $\mathbf{\bar{r}}$ is the vector of expectation values $\braket{\mathbf{\hat r}}$ and $\mathbf{\hat r} = [\hat x_1, \hat p_1,\mydots,\hat x_n, \hat p_n]^T$, and $\sigma$ is the covariance matrix given by $\sigma_{kl}= (\braket{\{\mathbf{\hat r}_k,\mathbf{\hat r}_l\}}- \braket{\mathbf{\hat r}_k} \braket{\mathbf{\hat r}_l})/2$.
Here $\{\cdot,\cdot\}$ denotes the anticommutator and the expectation value is given by $\braket{\hat O} = \tr[\hat O\hat \rho]$.
For this reason, such states are commonly referred to as Gaussian states.
Gaussian states are fully specified by the first and second moments of the quadrature operators.
Thus, their description via the covariance matrix allows one to completely characterize an important class of infinite-dimensional quantum states using only a finite number of parameters.
In the same spirit, one can define Gaussian maps (processes or measurements) as operations that take a Gaussian state as input and produce another Gaussian state as output.

Why are Gaussian states so important in quantum information and optics?
They are efficiently generated in the lab, meaning they can be produced in a deterministic manner. 
And theoretically, every pure single-mode Gaussian state can be generated by the combined action of the squeezing, rotation and displacement operators on the vacuum state.
Many of the states we have encountered so far are Gaussian states, including the vacuum state, coherent states, thermal states, and squeezed states.
In Sec. \ref{apx:single_mode} and Sec. \ref{apx:two_mode}, we show how to use the Strawberry Fields (SF) Python library to simulate common Gaussian states and obtain their covariance matrix and mean vector. 
We also describe how to retrieve the Wigner function from these simulations for single-mode states.
\section{Combining states of light}
\label{sec:BS_transf}

When working with quantum optical systems, it quickly becomes essential to understand how different states of light combine and interact. 
The most fundamental --- and arguably most universal --- model for such combinations is the beam splitter. 
Although deceptively simple, the beam splitter provides the canonical description of how two optical modes interfere, mix, and redistribute their quantum properties. 
Its mathematical structure underlies not only passive optical components such as couplers, interferometers, and four-port devices, but also many of the cornerstone demonstrations of quantum mechanics, from Hong–Ou–Mandel interference to single-photon interference and entanglement generation. 
For this reason, we devote the following section to a detailed derivation of the beam-splitter transformation. 
This is the most technically involved derivation in these notes, but its generality and central role in quantum optics make it an essential foundation for understanding how quantum states of light are combined and manipulated in practice.

\begin{figure}
    \centering
    \tikzset{every picture/.style={line width=0.75pt}} %set default line width to 0.75pt        

\begin{tikzpicture}[x=0.75pt,y=0.75pt,yscale=-0.75,xscale=0.75]
%uncomment if require: \path (0,417); %set diagram left start at 0, and has height of 417

%Shape: Rectangle [id:dp28711848676862983] 
\draw  [color={rgb, 255:red, 41; green, 140; blue, 140 }  ,draw opacity=1 ][fill={rgb, 255:red, 41; green, 140; blue, 140 }  ,fill opacity=0.5 ] (246,131) -- (414.28,131) -- (414.28,291.71) -- (246,291.71) -- cycle ;
%Straight Lines [id:da3868616009122413] 
\draw [color={rgb, 255:red, 41; green, 140; blue, 140 }  ,draw opacity=1 ][fill={rgb, 255:red, 41; green, 140; blue, 140 }  ,fill opacity=0.5 ]   (246,131) -- (414.28,291.71) ;

%Straight Lines [id:da6401996546154392] 
\draw    (333.28,291.98) -- (333.28,351.98) ;
\draw [shift={(333.28,353.98)}, rotate = 270] [color={rgb, 255:red, 0; green, 0; blue, 0 }  ][line width=0.75]    (10.93,-3.29) .. controls (6.95,-1.4) and (3.31,-0.3) .. (0,0) .. controls (3.31,0.3) and (6.95,1.4) .. (10.93,3.29)   ;
%Straight Lines [id:da5978363608201134] 
\draw    (334.28,63.98) -- (334.28,124.98) ;
\draw [shift={(334.28,126.98)}, rotate = 270] [color={rgb, 255:red, 0; green, 0; blue, 0 }  ][line width=0.75]    (10.93,-3.29) .. controls (6.95,-1.4) and (3.31,-0.3) .. (0,0) .. controls (3.31,0.3) and (6.95,1.4) .. (10.93,3.29)   ;
%Straight Lines [id:da403259858852558] 
\draw    (187.28,209.71) -- (244.28,209.71) ;
\draw [shift={(246.28,209.71)}, rotate = 180] [color={rgb, 255:red, 0; green, 0; blue, 0 }  ][line width=0.75]    (10.93,-3.29) .. controls (6.95,-1.4) and (3.31,-0.3) .. (0,0) .. controls (3.31,0.3) and (6.95,1.4) .. (10.93,3.29)   ;
%Straight Lines [id:da7662083144710411] 
\draw    (413.28,210.71) -- (467.28,210.71) ;
\draw [shift={(469.28,210.71)}, rotate = 180] [color={rgb, 255:red, 0; green, 0; blue, 0 }  ][line width=0.75]    (10.93,-3.29) .. controls (6.95,-1.4) and (3.31,-0.3) .. (0,0) .. controls (3.31,0.3) and (6.95,1.4) .. (10.93,3.29)   ;

% Text Node
\draw (186,189) node [anchor=north west][inner sep=0.75pt]   [align=left] {$\displaystyle \mathbf{E}_{1}$};
% Text Node
\draw (453,185) node [anchor=north west][inner sep=0.75pt]   [align=left] {$\displaystyle \mathbf{E}_{3}$};
% Text Node
\draw (339,62) node [anchor=north west][inner sep=0.75pt]   [align=left] {$\displaystyle \mathbf{E}_{2}$};
% Text Node
\draw (343,331) node [anchor=north west][inner sep=0.75pt]   [align=left] {$\displaystyle \mathbf{E}_{4}$};

\end{tikzpicture}
    \caption{Beam splitter model showing classical input and output field amplitudes.}
    \label{fig:beam_splitter}
\end{figure}
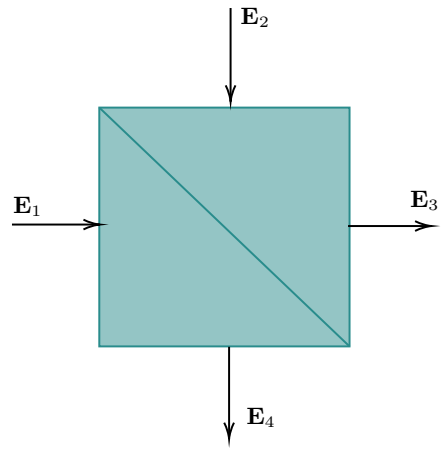

So far, we have been considering states that represent one system. 
But famously, many of the interesting features of quantum mechanics only manifest when we combine systems together.
An easy way to do that, since we are working with light, would be to use a \textit{beam splitter}.
Describing the use of this device classically is unambiguous.
Consider two fields $E_1$ and $E_2$ incident on a beam splitter as sketched in Fig.~\ref{fig:beam_splitter}. 
They will be transmitted with (complex) transmission coefficients $t_1$ and $t_2$ respectively and reflected with (complex) reflection coefficients $r_1$ and $r_2$ respectively.
Mathematically, we could write the transformation of the beam splitter as $E_3 = t_3 E_1 + r_3 E_2$ and $E_4 = r_4 E_1 + t_4 E_2$.

In terms of an input fields vector and a linear transformation (this is where the name \textit{linear optics} originates) we have
\begin{align*}
    \begin{pmatrix} 
        E_3 \\
        E_4
    \end{pmatrix} = 
    \begin{pmatrix} 
        t_3 & r_3 \\ 
        r_4 & t_4 
    \end{pmatrix}
    \begin{pmatrix}
        E_1 \\
        E_2
    \end{pmatrix}.
\end{align*}
We can describe the system in a quantum manner by associating an annihilation and a creator operator with each of the field modes. 
So, the transformation will look like
\begin{align}
    \begin{pmatrix} 
        \hat{a}_3\\
        \hat{a}_4 
    \end{pmatrix}
    = 
    \begin{pmatrix} 
        t_3 & r_3 \\
        r_4 & t_4 
    \end{pmatrix} 
    \begin{pmatrix}
        \hat{a}_1 \\
        \hat{a}_2 
    \end{pmatrix},
    \label{linearTransformationOfModes}
\end{align}
where $t_3,t_4,r_3,r_4 \in \mathbb{C}$.

Under the assumption of the quantum theory, the field modes must respect commutation relations in the input and in the output given by
\begin{align}
    &[\hat{a}_i, \hat{a}_j] = 0,\\
    &[\hat{a}_i, \hat{a}_j^\dagger] = \delta_{ij},
\end{align}
where $i$, $j$ can correspond either to the input modes (1 and 2) or the output modes (3 and 4). Using the transformation defined in Eq.~(\ref{linearTransformationOfModes}), we find:
\begin{align}
    &[\hat{a}_3, \hat{a}_3^\dagger] = |t_3|^2+|r_3|^2 = 1 \label{a3a3d} \\ 
    &[\hat{a}_4, \hat{a}_4^\dagger] = |t_4|^2+|r_4|^2 = 1 \label{a4a4d} \\ 
    &[\hat{a}_3, \hat{a}_4^\dagger] = t_3r_4^*+r_3t_4^* = 0. \label{a3a4d}
\end{align}
For complex numbers to be equal, their magnitudes and phases must both be equal. So considering $t_i = |t_i|e^{i\varphi_{t_i}}$ and $r_i = |r_i|e^{i\varphi_{r_i}}$, together with Eq.~\eqref{a3a4d} we have $|t_3||r_4|=|t_4||r_3|$, and $\varphi_{t_3}-\varphi_{t_4} = \varphi_{t_3}-\varphi_{t_4} \displaystyle \pm \pi$.
With these relations, Eqs.~\eqref{a3a3d} and \eqref{a4a4d}, imply $|t_3| = |t_4| = T = \cos(\theta)$ and $|r_3| = |r_4| = R = \sin(\theta)$ for $\theta \in [0, \frac{\pi}{2})$ (since the modulus of a complex number is nonnegative). 

Let us call $B$ the transformation matrix, defining $\varphi_0 = \frac{1}{2} (\varphi_{t_3}+\varphi_{t_4})$, $\varphi_T = \frac{1}{2}(\varphi_{t_3}-\varphi_{t_4})$ and $\varphi_R = \frac{1}{2}(\varphi_{r_3}-\varphi_{r_4} \displaystyle \mp \pi)$. We can now extract a global phase $\varphi_0$, obtaining
\begin{align}
    B = 
    \begin{pmatrix} 
        t_3 & r_3 \\
        r_4 & t_4 
    \end{pmatrix} 
    =
    e^{\varphi_0}
    \begin{pmatrix} 
        Te^{i\varphi_T} & Re^{i\varphi_R} \\
        -Re^{-i\varphi_R} & Te^{-i\varphi_T} 
    \end{pmatrix}.
    \label{generalU2}
\end{align}
The output modes will therefore be given by
\begin{align*}
    \hat{a}_3 
    &= Te^{i\varphi_T}\hat{a}_1 + Re^{i\varphi_R}\hat{a}_2 \\
    &= e^{i\varphi_T}(T\hat{a}_1 + Re^{i(\varphi)}\hat{a}_2)\\
    \hat{a}_4 
    &= -Re^{i\varphi_R}\hat{a}_1 + Te^{i\varphi_T}\hat{a}_2 \\
    &= e^{-i\varphi_T}(-Re^{i(\varphi)}\hat{a}_1+T\hat{a}_2),
\end{align*}
where $\varphi = \varphi_R-\varphi_T$. 
The resulting form for $B$ is then
\begin{align*}
    B = 
    \begin{pmatrix} 
        T & Re^{i\varphi} \\
        -Re^{-i\varphi} & T 
    \end{pmatrix}.
\end{align*}
Note that $B$ is unitary, i.e.\ $B^{-1} = B^\dagger$, and that its determinant is constrained to be $T^2 + R^2 = 1$. 
As the name suggests, we can use a beam splitter to split a beam. This means we can apply the setup from Fig.~\ref{fig:beam_splitter} with only one of the two possible inputs, and still get two outputs.
But we can easily check that it's not possible to satisfy the commutation relations if we have $\hat{a}_3 = t\hat{a}_1$ and $\hat{a}_4 = r\hat{a}_1$, since $[\hat{a}_3, \hat{a}_4^\dagger] = t r^* \neq 0$. 

A natural question now arises: What is the mathematical representation of the physical \textit{process} of splitting a beam?
We know that one of the input ports of a beam splitter must be in the vacuum state. 
However, our description so far has been formulated entirely in terms of operators—which is desirable, because the behavior of the optical device should not depend on the particular input states. 
This leads to an apparent difficulty: if we only specify how operators transform, how can we actually calculate what happens to arbitrary input states?

What we ultimately want is the following: given two input states, determine the output states after the beam splitter. 
From standard quantum mechanics, we know that physical transformations of closed systems --— such as our idealized beam splitter --- are represented by unitary evolution operators. 
Thus, we need to identify the unitary $\hat U_B$  that takes a two-mode input state
$\ket{\psi_{in}} \rightarrow \ket{\psi_{out}} = \hat{U}_B \ket{\psi_{in}}$. 
But we do not yet know how the states themselves should transform, and we want a description that does not depend on any specific choice of input state. 
Therefore, we move to the Heisenberg picture, where the states remain fixed and the operators evolve. 
Since we already know how the field operators must transform at a beam splitter, the Heisenberg picture allows us to infer the form of $\hat U_B$.

Where should we begin? Earlier, we observed that the beam splitter transformation belongs to the group 
$SU(2)$. 
Angular momentum operators also form a representation of $SU(2)$, so this provides a natural starting point: the generators of $SU(2)$ are the Pauli matrices,
\begin{align*}
    &\sigma_x = 
    \begin{pmatrix} 
        0 & 1 \\
        1 & 0 
    \end{pmatrix},
    &\sigma_y = 
    \begin{pmatrix} 
        0 & -i \\
        i & 0 
    \end{pmatrix},
    &&\sigma_z = 
    \begin{pmatrix} 
        1 & 0 \\
        0 & -1 
    \end{pmatrix}.
\end{align*}

Motivated by this structure, we can try to construct analogous generators using the creation and annihilation operators of our two optical modes. With a bit of insight --- and some luck --- we arrive at the Jordan–Schwinger representation, which expresses the 
$SU(2)$ generators directly in terms of pairs of bosonic operators
\begin{align*}
\hat{L}_i =
    \begin{pmatrix} 
        \hat{a}_1 \\
        \hat{a}_2
    \end{pmatrix}^\dagger
    \sigma_i
    \begin{pmatrix} 
        \hat{a}_1 \\
        \hat{a}_1
    \end{pmatrix},
\end{align*}
where $i \in \{x,y,z\}$.%
\fnt{Show that the operators $\hat{L}_i$, with $i \in \{x, y, z\}$ satisfy the $\mathfrak{su}(2)$ commutation relations, i.e.\ $[\hat{L}_i, \hat{L}_j] = i \epsilon_{ijk}\hat{L}_k$.}
Explicitly, we have: $\hat{L}_x = \frac{1}{2}(\hat{a}_1^\dagger\hat{a}_2+\hat{a}_2^\dagger\hat{a}_1)$, $\hat{L}_y = \frac{i}{2}(\hat{a}_2^\dagger \hat{a}_1-\hat{a}_1^\dagger \hat{a}_2)$, and $\hat{L}_z = \frac{1}{2}(\hat{a}_1^\dagger \hat{a}_1-\hat{a}_2^\dagger \hat{a}_2)$.
Let's explore how these generators may transform our field modes
\fnt{Verify these relations. Take into account that in this context, the left-hand side of each of the equations means that the operators are acting in each one of the components of this `vector' of operators.}
\begin{align*}
    &e^{-i\Phi\hat{L}_z} \begin{pmatrix} \hat{a}_1 \\ \hat{a}_2 \end{pmatrix} e^{i\Phi\hat{L}_z} = 
    \begin{pmatrix} 
        e^{i\frac{\Phi}{2}} & 0 \\
        0 & e^{-i\frac{\Phi}{2}} 
    \end{pmatrix}
    \begin{pmatrix} \hat{a}_1 \\ \hat{a}_2 \end{pmatrix}, \\
    &e^{-i\Theta\hat{L}_y} \begin{pmatrix} \hat{a}_1 \\ \hat{a}_2 \end{pmatrix} e^{i\Theta\hat{L}_y} = 
    \begin{pmatrix} 
        \cos(\frac{\Theta}{2}) & \sin(\frac{\Theta}{2}) \\
         -\sin(\frac{\Theta}{2}) & \cos(\frac{\Theta}{2})
    \end{pmatrix}
    \begin{pmatrix} \hat{a}_1 \\ \hat{a}_2 \end{pmatrix}, \\
    &e^{-i\Omega\hat{L}_x} \begin{pmatrix} \hat{a}_1 \\ \hat{a}_2 \end{pmatrix} e^{i\Omega\hat{L}_x} = 
    \begin{pmatrix} 
        \cos(\frac{\Omega}{2}) & -i\sin(\frac{\Omega}{2}) \\
         -i\sin(\frac{\Omega}{2}) & \cos(\frac{\Omega}{2})
    \end{pmatrix}
    \begin{pmatrix} \hat{a}_1 \\ \hat{a}_2 \end{pmatrix}.
\end{align*}

Notice that $\hat U_B = e^{i\Phi\hat{L}_z} e^{i\Theta\hat{L}_y} e^{i\Psi\hat{L}_z}$, so transforming our operators in the Heisenberg picture, i.e.\ making $U_B^\dagger \begin{pmatrix} \hat{a}_1 \\ \hat{a}_2 \end{pmatrix} U_B  = \begin{pmatrix} \hat{a}_3 \\ \hat{a}_4 \end{pmatrix}$, is equivalent to having the matrix-`vector' product:
\begin{align*}
     &\begin{pmatrix} \hat{a}_3 \\ \hat{a}_4 \end{pmatrix} = 
    \begin{pmatrix} 
        e^{\frac{i}{2}(\Phi+\Psi)} \cos(\frac{\Theta}{2}) & e^{\frac{i}{2}(\Phi-\Psi)} \sin(\frac{\Theta}{2})\\
        -e^{-\frac{i}{2}(\Phi-\Psi)} \sin(\frac{\Theta}{2})& e^{-\frac{i}{2}(\Phi+\Psi)} \cos(\frac{\Theta}{2}) 
    \end{pmatrix}
    \begin{pmatrix} \hat{a}_1 \\ \hat{a}_2 \end{pmatrix}.
\end{align*}

Comparing this to Eq.~(\ref{generalU2}), we see that this is exactly what we were looking for (ignoring the global phase $\varphi_0$), and defining $\Theta = 2 \theta$, $\Phi+\Psi = 2 \varphi_T$ and $\Phi-\Psi = 2\varphi_R$.

We have found the $\hat U_B$ that allows us to evolve states and operators. 
Let us consider Fock states first.
As discussed before, they form a basis on which we can expand any other state. 
We can start by checking what happens to the input state $\ket{n_1, n_2}$ 
\begin{widetext}
    \begin{align*}
    \ket{n_1, n_2} \xrightarrow[]{\text{BS}} \ket{n_1, n_2}' &= \hat U_B \ket{n_1, n_2} 
    = \frac{1}{\sqrt{n_1!}} \frac{1}{\sqrt{n_2!}} \hat U_B \hat{a}_1^{\dagger n_1}\hat{a}_2^{\dagger n_2} \ket{0,0}
    = \frac{1}{\sqrt{n_1! n_2!}} \hat U_B \hat{a}_1^{\dagger n_1} \hat{a}_2^{\dagger n_2} \hat U_B^\dagger \hat U_B\ket{0,0}
    \\ &
    = \frac{1}{\sqrt{n_1! n_2!}} \left(\hat U_B \hat{a}_1^{\dagger n_1} \hat U_B^\dagger\right) \left( \hat U_B\hat{a}_2^{\dagger n_2} \hat U_B^\dagger\right) \ket{0,0} 
    = \frac{1}{\sqrt{n_1! n_2!}} \left(\hat U_B \hat{a}_1^\dagger \hat U_B^\dagger\right)^{n_1} \left( \hat U_B\hat{a}_2^\dagger \hat U_B^\dagger\right)^{n_2} \ket{0,0} 
    \\ & 
    = \frac{1}{\sqrt{n_1! n_2!}} \left(\hat U_B^\dagger \hat{a}_1 \hat U_B\right)^{\dagger n_1} \left( \hat U_B^\dagger \hat{a}_2 \hat U_B\right)^{\dagger n_2}\ket{0,0}.
\end{align*}
\end{widetext}
This is precisely where the unitary operator $\hat{U}_B$ we derived becomes especially useful. 
We already know how $\hat{a}_1$ and $\hat{a}_2$ transform under $\hat{U}_B$, and the transformation has exactly the same form as applying the matrix $B$ to the vector of mode operators. 
This makes the calculation remarkably straightforward.

With this in mind, we can manipulate the expression for the output state. 
The only subtlety is that we need the transformation $\hat{U}_B\, \hat{a}_i\, \hat{U}_B^\dagger$, rather than $\hat{U}_B^\dagger\, \hat{a}_i\, \hat{U}_B$.
Therefore, we must use the inverse transformation, which corresponds to $B^{-1} = B^\dagger$, and we must also account for the commutation relations between the two modes.

Carrying this out, we can finally express the output state in the Fock basis, 
\begin{widetext}
\begin{align}
   \ket{n_1, n_2}' &
    = \frac{1}{\sqrt{n_1! n_2!}} \left(T \hat{a}_1^\dagger - Re^{-i\varphi}\hat{a}_2^\dagger \right)^{n_1} \left(Re^{i\varphi}\hat{a}_1^\dagger + T \hat{a}_2^\dagger \right)^{n_2} \ket{0,0} \\&
    = \frac{1}{\sqrt{n_1! n_2!}}\sum_{k_1=0}^{n_1}\sum_{k_2=0}^{n_2} \binom{n_1}{k_1}\binom{n_2}{k_2}(-1)^{(n_1-k_1)}T^{(k_1+n_2-k_2)} 
    R^{(n_1-k_1+k_2)}e^{i\varphi (k_2+k_1-n_1)}  \\ & 
    \hspace{50mm}\times\sqrt{(k_1+k_2)!(n_1+n_2-k_1-k_2)!} \ket{k_1+k_2,n_1+n_2-k_1-k_2} \label{beamSpliterFock}
\end{align}
\end{widetext}
This formula is quite involved, so let us work through a simple example. 
Consider sending a single photon into a $50{:}50$ beam splitter. 
In this case, the input state is $\ket{1,0}$ and the beam splitter parameters are $T = R = 1/\sqrt{2}$. 
For convenience, we also choose the phase $\varphi = \pi$.
Replacing these values in Eq. \eqref{beamSpliterFock}, we get:
\fnt{Calculate the output state of a 50:50 beam splitter, when the input state is $\ket{1,1}.$ Comment on the result, and then Google about the Hong–Ou–Mandel effect.}
\begin{align*}
    \ket{1, 0}' &= \frac{1}{\sqrt{2}} (\ket{0,1}+\ket{1,0}).
\end{align*}
In general, if we try to split $n$ photons, we get:
\begin{align*}
    \ket{n, 0}' &= \frac{1}{\sqrt{n!}}\sum_{k=0}^n \binom{n}{k}T^kR^{n-k}e^{-i\varphi (n-k)}\\
     & \quad \quad(-1)^{(n-k)}\sqrt{k!(n-k)!} \ket{k,n-k}\\
    &= \sum_{k=0}^n \sqrt{\binom{n}{k}} T^k (-R e^{-i\varphi})^{(n-k)}\ket{k,n-k}.
\end{align*}
Beam splitters thus reveal an important --- and remarkable --- feature of quantum optics: even when we inject something as simple as $n$ photons in one input port and vacuum in the other, the output cannot, in general, be written as a product of two independent states. 
Instead, the beam splitter produces superpositions that correlate the two output modes in a fundamentally quantum way. 
This inseparability is not merely a mathematical curiosity; it is a direct manifestation of how interference and quantum statistics intertwine. 
In fact, beam splitters provide one of the simplest and most versatile tools for generating correlated and even entangled states of light. This observation naturally leads us to the next topic: understanding and characterizing entanglement.
In Sec. \ref{apx:BS} we show how to implement beam splitter transformations using Strawberry Fields to mix different Gaussian states.
\section{Entanglement}
\label{sec:entanglement}

Entanglement is a fundamental phenomenon in quantum mechanics, lying at the heart of both foundational issues and many emerging technologies, and featured heavily in modern quantum optics research \cite{Bruss2002,PlenioVirmani2007,HorodeckiHorodeckiHorodeckiHorodecki_2009,GuhneToth2009,EltschkaSiewert2014,FriisVitaglianoMalikHuber2018}.
It is a nonclassical feature that exists only in composite systems.

Let us consider two systems, $A$ and $B$, which are conventionally assigned to Alice and Bob.
These systems might be polarized photons, electrons with spin, atomic excitations, or any system with physical degrees of freedom that can occupy two, three, or even infinitely many distinguishable states.
The composite system \(AB\) is described by a state vector \(\ket{\psi_{AB}}\) in the Hilbert space \(\mathcal{H}_A \otimes \mathcal{H}_B\).
Suppose the two parts can be individually addressed and distinguished by their physical properties.
If the global state is pure and can be written as a product of states associated with each subsystem, i.e.\
\begin{equation*}
    \rho_{AB} 
    = \ket{\psi_{AB}}\bra{\psi_{AB}} 
    = \ket{\psi_A}\!\bra{\psi_A} \otimes \ket{\psi_B}\!\bra{\psi_B}
    = \rho_A \otimes \rho_B,
\end{equation*}
then we say that the state is \textit{separable}.
Otherwise, if no such factorization exists, the state is said to be \textit{entangled}.
The tensor product structure expresses the idea that two physically independent systems can be described jointly by 
a \textit{product state} of their local (reduced) density operators
\(\rho_A\) and \(\rho_B\). 
That is, all the information contained in the joint state is already encoded in its marginals.
Notice that, in what follows, we will move between the state vector and density matrix formalisms whenever convenient.

At first sight, it may seem straightforward to determine whether a bipartite quantum system is entangled. 
One might expect to simply check whether the state is a product state by 1.\ computing the reduced density operators via partial traces, then 2.\ calculating their tensor product, and 3.\ comparing the result with the original state. 
If they coincide, the state is clearly separable and therefore not entangled. 
From this perspective, the task appears trivial.
However, if entanglement were truly so easy to identify, it would be difficult to justify the vast amount of research devoted to it. 

Why has this problem remained a central topic in quantum information theory for decades? 
The crucial issue lies in the implicit assumption that the systems under consideration are in pure states. 
In practice, this assumption is, at best, a convenient idealization. 
Pure states do not occur in nature in any exact or experimentally accessible sense; every preparation procedure in the laboratory inevitably yields a \textit{mixed state}. 
This seemingly simple fact complicates the characterization of entanglement in a profound way.
Consequently, one must generalize the notion of separability to mixed states. 

A mixed bipartite state $\rho_{AB}$ is called \textit{separable} if it can be written as a convex combination of product states:
\begin{equation}
    \rho_{AB} = \sum_{i} p_{i}\,
    \ket{\psi_{A}^{(i)}}\bra{\psi_{A}^{(i)}} \otimes
    \ket{\psi_{B}^{(i)}}\bra{\psi_{B}^{(i)}},
\end{equation}
with $p_{i} \geq 0$ and $\sum_{i} p_{i} = 1$. 
This definition means that the state can, in principle, be prepared by \textit{local operations and classical communication} (LOCC), i.e. Alice prepares her system in the state $\ket{\psi_{A}^{(i)}}$, Bob independently prepares his on $\ket{\psi_{B}^{(i)}}$, and they coordinate their choices using shared classical information (for instance, by agreeing on the probabilities $p_{i}$ in advance).

It is important to note that such states may exhibit \textit{classical correlations}, in contrast to the case of pure product states, where no correlations are present at all. 
It is worth emphasizing this distinction. By classical correlations, we mean the kinds of everyday correlations produced by conventional statistics. 
For instance, if I draw four aces from a standard playing card deck, I know for a fact that my probability of drawing another ace is now zero. 
This is because there is a classical correlation between the cards in my hand and the cards in the deck. 
By obtaining information about the cards in my hand, I can learn information about the contents of the deck. 
Crucially, genuinely quantum correlations do \textit{not} work this way. 
Entanglement means there is \textit{more} information I can infer from the deck-hand system than I could from classical correlations alone. 
Discerning classical from quantum correlations is the difficult task at the heart of entanglement detection, and explains why so much time and effort is devoted to the study of this subject.
When mixed states are involved, determining whether correlations are classical or genuinely quantum becomes significantly more subtle.
From a computational point of view, determining whether a general mixed state is separable is an NP-hard problem \cite{Gurvits2003}.
This implies that, in practice, no efficient algorithm is expected to exist for deciding separability in full generality.

Why is the problem so difficult? 
A key reason is that the decomposition of a mixed state into pure states is never unique. 
If every density matrix admitted only a single pure-state decomposition, one could simply find that decomposition and test whether it consists solely of product states. 
However, a mixed state can be written as a convex combination of pure states in infinitely many different ways.
Formally, one can define the set of all pure-state ensembles corresponding to a density matrix $\hat \rho$ as
\begin{align*}
    \mathcal{D}[\rho] := 
    \Big\{\,\{(p_i,\psi^i)\}_i \ \Big|\ 
    \hat \rho = \sum_i p_i \, &\ket{\psi^i}\bra{\psi^i},
    \\
    &p_i \ge 0,\ \sum_i p_i = 1 \Big\}.
\end{align*}
Each such ensemble represents a distinct physical preparation procedure that leads to the same mixed state $\rho$, and there are infinitely many of them in general.
This non-uniqueness has a direct physical interpretation: a given density matrix can arise from infinitely many distinct probabilistic mixtures of pure states, and these preparations are indistinguishable by any measurement. 
Thus, different ensembles may generate the same physical state, yet no observation can reveal which specific mixture was used.

Another way of thinking about this is that a given quantum state can be prepared in infinitely many ways. 
And among all these possible decompositions, the central question is whether at least one of them involves only product states, i.e.\ whether entanglement was not required in its preparation. 
Determining this is extraordinarily challenging, and it lies at the core of the entire field of entanglement theory.

Even though a complete solution to this problem is out of reach in general, a great deal of meaningful progress is still possible. 
For instance, one can develop methods to detect entanglement, characterize its structure, and even quantify it in many relevant cases \cite{Bruss2002,PlenioVirmani2007,HorodeckiHorodeckiHorodeckiHorodecki_2009,GuhneToth2009,EltschkaSiewert2014,FriisVitaglianoMalikHuber2018}.
What remains unattainable, however, is a necessary and sufficient criterion that is both universally valid and efficiently computable for deciding whether an arbitrary state is entangled or separable.
This limitation persists even under the most favorable assumptions: 
even if one had complete experimental access to the system (that is, full knowledge of the density matrix obtained through all possible measurements), there is, in general, no simple and definitive test that decides separability for mixed states of arbitrary dimension.
Nevertheless, several powerful tools have been developed to detect entanglement in practice, and in the following we introduce some of the most relevant criteria and techniques, first introducing the Peres-Horodecki criterion and then discussing some methods that rigorously verify entanglement in the case of bipartite infinite dimensional systems.
They provide necessary and sufficient conditions for entanglement in Gaussian states by placing bounds on the combined variances of appropriately chosen quadrature operators. 
Violation of these bounds constitutes an operational and experimentally accessible demonstration of bipartite CV entanglement.

\paragraph*{Peres-Horodecki's criterion:}
As we have seen, characterizing entanglement for mixed quantum states is a highly nontrivial task. 
While separability is straightforward to define in principle, deciding whether a given density operator can be written as a convex mixture of product states is, in general, computationally hard. 
This motivates the search for operational criteria that allow one to efficiently detect entanglement, even if they do not provide a complete solution in all cases. 
One of the most important and widely used such criteria is based on the behavior of a quantum state under partial transposition.

The Peres-Horodecki criterion states that if a bipartite state is separable, then its partial transpose must remain a positive semidefinite operator \cite{Peres1996, HorodeckiHorodeckiHorodecki1996}. 
This condition is necessary for separability in all dimensions and, remarkably, it is also sufficient for low-dimensional systems, specifically for $2 \times 2$ and $2 \times 3$ Hilbert spaces. 
As a result, the criterion provides a simple and powerful test for entanglement in many physically relevant situations: the appearance of negative eigenvalues in the partially transposed density matrix immediately signals entanglement. 
Beyond its practical usefulness, the Peres-Horodecki criterion also plays a central conceptual role, revealing deep connections between entanglement, positivity, and the mathematical structure of quantum states.

\paragraph*{Simon's criterion:}
This criterion was first published by R. Simon in 2000 \cite{Simon_2000}. 
Let us briefly summarize the derivation of this test (the discussion below is inspired by, and closely follows, the treatment of this topic given in Ch.~7 of Ref.~\cite{Serafini_2023}). 
First, we have seen that every quantum separable state $\hat{\rho}$ should have an associated Wigner function $W_{\hat{\rho}}(\mathbf{r})$. 
Here, for simplicity, we restrict our discussion to two-mode systems, where $\mathbf{r}=(x_1,p_1,x_2,p_2)^T$. 
To implement a partial transposition in the quantum phase space, we use the fact that this operation on a single-mode canonical operators act as:
\begin{equation}\label{PTonqp}
PT: \quad \hat{x} \rightarrow \hat{x} \quad , \quad \hat{p} \rightarrow -\hat{p},
\end{equation}
therefore, if we apply a partial transposition on the second subsystem (mode), the Wigner function transforms as follows:
\begin{equation}\label{PTonWigner}
P T: \quad W_{\hat{\rho}}\left( (x_1, p_1, x_2, p_2)^T\right) \rightarrow W_{\hat{\rho}}\left((x_1, p_1, x_2,-p_2)^T\right),
\end{equation}
and if we define the $4\times4$ phase-space mirror reflection matrix as:
\begin{equation*}\label{mirrorreflc}
T_{4\times4}=
\begin{pmatrix}
1 & 0 & 0 & 0 \\
0 & 1 & 0 & 0 \\
0 & 0 & 1 & 0 \\
0 & 0 & 0 & -1\\
\end{pmatrix},
\end{equation*}
then the partial transposition of the Wigner function can be simply expressed as:
\begin{equation}\label{PTonWigner2}
P T: \quad W_{\hat{\rho}}(\mathbf{r}) \rightarrow W_{\hat{\rho}}(T\cdot\mathbf{r}).
\end{equation}\\
By virtue of the Peres-Horodecki criterion, if the state $\hat{\rho}$ associated with $W_{\hat{\rho}}(\mathbf{r})$  is separable, then its partially transposed state $\hat{\rho}^{T_2}$ will be associated with a valid Wigner function $W_{\hat{\rho}^{T_2}}(\mathbf{r})=W_{\hat{\rho}}(T\cdot\mathbf{r})$. 
That is, it will be associated with a physically valid state.\\

Every physically valid quantum system with covariance matrix $\sigma$ should satisfy the Robertson-Schrödinger uncertainty relation:
\begin{equation}\label{RSuncertainty}
    \sigma + i\frac{ \hbar}{2} \Omega \geq 0,
\end{equation}
where the symplectic (real, canonical, anti-symmetric) form $\Omega =\bigoplus^n_{j=1} \begin{pmatrix}
0 & 1\\
-1 & 0
\end{pmatrix}$.
And so the covariance matrix $\tilde{\sigma}$ associated with $W_{\hat{\rho}^{T_2}}(\mathbf{r})$ should still satisfy the Robertson-Schrödinger relation if the state is separable:
\begin{equation}\label{RSuncertainty2}
    \tilde{\sigma} + i\frac{ \hbar}{2} \Omega \geq 0.
\end{equation}
This requirement is only preserved under independent local linear canonical transformations on the covariance matrix \cite{Simon_2000}, i.e., transformations of the form
\begin{equation}\label{Sloc}
S_{loc}=
\begin{pmatrix}
S_1 & 0\\
0 & S_2\\
\end{pmatrix},    
\end{equation}
where $S_1$ and $S_2 $ are $2\times2$ matrices that satisfy
\begin{equation*}
    S_1 \left(\begin{array}{cc}
0 & 1 \\
-1 & 0\\
\end{array}\right) S_1^T  = S_2 \left(\begin{array}{cc}
0 & 1 \\
-1 & 0\\
\end{array}\right) S_2^T = \left(\begin{array}{cc}
0 & 1 \\
-1 & 0\\
\end{array}\right).
\end{equation*}
These transformations are part of the set of symplectic transformations, which are particularly useful when working with Gaussian states.
Since Gaussian states are completely characterized by their first and second moments, and symplectic transformations act linearly on the quadrature operators, they induce affine transformations of the covariance matrix (that preserve the canonical commutation relations). This makes them the natural framework for describing the evolution and manipulation of such states (for a quick reference cf.\ \cite{Bohr-Brask2022}).

Our task now is to examine these transformations on the covariance matrices $\sigma$ and $\tilde{\sigma}$ to obtain conditions that are equivalent to Eqs.~\eqref{RSuncertainty} and \eqref{RSuncertainty2}, but easier to apply.
Let us express the covariance matrix in block form:
\begin{equation}
\sigma=\left(\begin{array}{cc}\sigma_A & \sigma_{AB} \\ \sigma_{AB}^T & \sigma_B\end{array}\right),    
\end{equation}
with $\sigma_{A}$, $\sigma_{B}$ and $\sigma_{AB}$ $2\times2$ matrices. The transformation $S_{loc} \sigma  S_{loc}^T$ leads to the block transformations:
\begin{equation}\label{transsigma}
\sigma_A \rightarrow S_1 \sigma_A S_1^T, \quad \sigma_B \rightarrow S_2 \sigma_B S_2^T, \quad \sigma_{AB} \rightarrow S_1 \sigma_{AB} S_2^T.
\end{equation}
The covariance matrix of the partial transposed state is:
\begin{equation}
    \tilde{\sigma} = T_{4\times4}\sigma T_{4\times4}= \left(\begin{array}{cc}\sigma_A & \sigma_z\sigma_{AB} \\ (\sigma_z\sigma_{AB})^T & \sigma_z \sigma_B \sigma_z\end{array}\right),
\end{equation}
with $\sigma_z$ the Pauli matrix $\begin{pmatrix}
1 & 0\\
0 & -1\\
\end{pmatrix}$. In this case, the  $S_{loc} \tilde{\sigma}  S_{loc}^T$ transformation leads to the block transformations: 
\begin{align}\label{transsigmapt}
    \sigma_A &\rightarrow S_1 \sigma_A S_1^T, \nonumber\\
    \sigma_B &\rightarrow S_2 \sigma_z\sigma_B \sigma_z S_2^T,\\
    \sigma_{AB} &\rightarrow S_1\sigma_{AB}  \sigma_z S_2^T.\nonumber
\end{align}
From Eqs.~\eqref{transsigma} and \eqref{transsigmapt}, we have the  following invariants under the independent local linear canonical transformations:  
\begin{gather}
    \textrm{Det}\ \sigma_A, \quad
    \textrm{Det}\ \sigma_B = \textrm{Det}\ \sigma_z \sigma_B \sigma_z,\nonumber\\
    \quad \textrm{Det}\ \sigma_{AB}  = -\textrm{Det}\ \sigma_{AB} \sigma_z, \quad\textrm{and}\nonumber\\
    \textrm{Tr}(\sigma_A J \sigma_{AB} J \sigma_B J \sigma_{AB}^T J). \label{invariants}
\end{gather}
Here we have used the fact that any covariance matrix can be transformed by a convenient local canonical transformation $S_{loc}$ into the Simon normal form \cite{Simon_2000}:
\begin{equation}
\sigma_{sf}=
\begin{pmatrix}
a & 0 & c_1 & 0 \\
0 & a & 0 & c_2\\
c_1 & 0 & b & 0 \\
0 & c_2 & 0 & b\\
\end{pmatrix}.
\end{equation}
\begin{widetext}
Using this normal form it one can see that Eq.~\eqref{RSuncertainty} is equivalent in terms of the invariants in Eqs.~\eqref{invariants} to:
\begin{equation}\label{uncertainity}
    \textrm{Det}\ \sigma_A \textrm{Det}\ \sigma_B + \left ( \frac{\hbar^2}{4}-\textrm{Det}\  \sigma_{AB} \right )^2 - 
    \textrm{Tr}(\sigma_A J \sigma_{AB} J \sigma_B J \sigma_{AB}^T J) \geq \frac{\hbar^2}{4}(\textrm{Det}\  \sigma_A + \textrm{Det}\ \sigma_B),
\end{equation}
and Eq.~\eqref{RSuncertainty2} is equivalent to: 
\begin{equation}\label{uncertainity2}
    \textrm{Det}\ \sigma_A \textrm{Det}\ \sigma_B + \left ( \frac{\hbar^2}{4}+\textrm{Det}\  \sigma_{AB} \right )^2 - \textrm{Tr}(\sigma_A J \sigma_{AB} J \sigma_B J \sigma_{AB}^T J) \geq \frac{\hbar^2}{4}(\textrm{Det}\  \sigma_A + \textrm{Det}\ \sigma_B).
\end{equation}
Then, due to the equivalence between Eq.~\eqref{RSuncertainty2} and Eq.~\eqref{uncertainity2}, we can conclude that if a state is separable, it satisfies Eq. \eqref{uncertainity2}.
Both relations in Eq.~\eqref{uncertainity} and Eq.~\eqref{uncertainity2} can be combined into a more general relation: 
\begin{equation}\label{simoncriterion}
    \textrm{Det}\ \sigma_A \textrm{Det}\ \sigma_B + \left ( \frac{\hbar^2}{4}-|\textrm{Det}\  \sigma_{AB}| \right )^2 - 
    \textrm{Tr}(\sigma_A J \sigma_{AB} J \sigma_B J \sigma_{AB}^T J) \geq \frac{\hbar^2}{4}(\textrm{Det}\  \sigma_A + \textrm{Det}\ \sigma_B).
\end{equation}
\end{widetext}
This is what we usually refer to as the Simon Criterion, and it is a necessary condition on the elements of the covariance matrix for a separable state. 
It is a direct implication of the Peres-Horodecki criterion.

Simon proved that such criterion is a necessary and sufficient separability condition for bipartite Gaussian states \cite{Simon_2000}. 
In Appendix \ref{apx:simon}, we show how this criterion works for two-mode Gaussian states.

\paragraph*{Logarithmic Negativity:}
Another criterion that we will use in the simulation labs (Sec.~\ref{sec:labs}) to characterize entanglement (and that has been related to the partial transposition) is called the logarithmic negativity \cite{VidalWerner2002}. 
Let us defineit as follows:
\begin{equation}
\label{eq:log2neg}
    E_{\mathcal{N}} = \log_2 \left\| \hat{\rho}^{\mathrm{T}_p} \right\|_1,
\end{equation}
where $\left\| \hat{\rho}^{\mathrm{T}_p} \right\|_1$ denotes the trace norm of the partially transposed density matrix $\hat{\rho}^{\mathrm{T}_p}$ \cite{Plenio_2005}.

The trace norm of $\hat{\rho}^{\mathrm{T}_p}$ corresponds to the sum of the absolute values of its eigenvalues:  $\left\| \hat{\rho}^{\mathrm{T}_p} \right\|_1 = \sum_{j} |\lambda_j|$,
with $\{\lambda_j\}$ being the set of eigenvalues of $\hat{\rho}^{\mathrm{T}_p}$.
If a state is separable, then by virtue of the Peres-Horodecki criterion, the partial transpose $\hat{\rho}^{\mathrm{T}_p}$ must be a valid density operator.
Consequently, its eigenvalues must be non-negative, rendering the absolute value redundant, such that
\begin{equation}
\label{eq:trace_norm_sep}
    \left\| \hat{\rho}^{\mathrm{T}_p} \right\|_1 = \sum_{j} \lambda_j.
\end{equation}
Additionally, the trace of $\hat{\rho}^{\mathrm{T}_p}$ must remain unity to be a legitimate density operator
\begin{equation}
\label{eq:trace_pt}
    \text{Tr}(\hat{\rho}^{\mathrm{T}_p}) = \sum_{j} \lambda_j = 1.
\end{equation}
Substituting Eq. \eqref{eq:trace_pt} into Eq. \eqref{eq:trace_norm_sep} yields: $\left\| \hat{\rho}^{\mathrm{T}_p} \right\|_1 = 1$.
Finally, using this result in Eq. \eqref{eq:log2neg} gives us $E_{\mathcal{N}} = 0$.
Thus, the logarithmic negativity provides an alternative formulation of the Peres-Horodecki criterion: $E_{\mathcal{N}} = 0$ for all separable states.
Conversely, whenever $E_{\mathcal{N}} > 0$, the state is necessarily entangled.
However, there also exist entangled states for which $E_{\mathcal{N}} = 0$.
Consequently, much like the Peres–Horodecki criterion itself, logarithmic negativity is not sufficient to detect all forms of entanglement. 
Nevertheless, it remains a valuable tool for characterizing entanglement, as it is an entanglement monotone that does not increase under local operations and classical communication (LOCC) \cite{Plenio_2005}.

Gaussian states have a convenient representation of logarithmic negativity \cite{AdessoSerafiniIlluminati2004}, given by:
\begin{equation}\label{logneg}
    E_{\mathcal{N}}=\sum_{j=1}^{m+n} \textrm{max} \{0,-\textrm{log}_2 ({\tilde{\nu_j})}\},
\end{equation}
where $m$ and $n$ represent the numbers of modes in each partition, and ${\tilde{\nu_j}}$ denote the symplectic eigenvalues associated with the covariance matrix of the partially transposed state $\tilde{\sigma}$. 
The absolute value of the different eigenvalues of the matrix $i\Omega \tilde{\sigma}$ corresponds to $\tilde{\nu_j}$.
Appendix \ref{apx:log_neg} details the implementation of this criterion using the Python library The Walrus.

Entanglement manifests itself differently depending on whether quantum information is encoded in discrete or continuous degrees of freedom. 
In DV systems, such as polarization or spin qubits, entanglement is typically characterized using finite-dimensional state spaces and measures that rely on explicit state reconstruction or combinatorial criteria. 
In contrast, CV systems --- such as optical field quadratures --- are naturally described in infinite-dimensional Hilbert spaces, where entanglement is often captured, as we described earlier, through second moments and phase-space representations. 
This structural difference makes CV entanglement particularly amenable to experimental characterization using tools such as homodyne detection, covariance matrices, and symplectic methods. 
While DV and CV approaches are conceptually distinct, they are complementary: DV systems excel in digital quantum protocols, whereas CV systems offer deterministic state preparation, high detection efficiencies, and direct access to collective observables. 
For these reasons, CV entanglement provides a powerful and experimentally accessible framework for exploring nonclassical correlations.
\section{Nonclassicality as a resource for entanglement}
\label{sec:NCLvsENT}

The connection between nonclassicality (Sec. \ref{sec:quasip}) and entanglement (Sec. \ref{sec:entanglement}) has often been motivated by the superposition principle, which underlies both phenomena \cite{RaimondBruneHaroche_2001}. 
While nonclassicality can arise already in single-mode systems and entanglement, as a form of correlation, necessarily requires composite ones, both properties ultimately stem from the impossibility of expressing the state as a classical mixture of coherent states \cite{Glauber_1963, Sudarshan_1963} or, respectively, as a separable mixture of product states \cite{SanperaTarrachVidal1998, SperlingWalmsley_2018}. 
There exists a precise operational relation between these two concepts by demonstrating that single-mode nonclassicality can be quantitatively converted into entanglement.
A key result is that when vacuum is one of the input fields of a beam splitter and a nonclassical single-mode state is sent through the remaining beam splitter input port, the amount of nonclassicality --- defined via the minimal number of superpositions of classical states required to construct it --- is mapped onto exactly the same amount of bipartite entanglement at the output \cite{VogelSperling_2014}.
This extends beyond the previously known fact that nonclassical inputs can generate entanglement under linear optical transformations \cite{KimSonBuzekKnight_2002,Wang2002,TahiraIkramNhaZubairy2009,JiangLangCaves2013}: the correspondence is demonstrated to be exact and quantitative for such a case.

Let's see a couple of examples of this: First, consider the action of a symmetric $50:50$ beam splitter when the inputs are a coherent state $\ket{\alpha}$ and vacuum $\ket{0}$. 
In the output ports we get a product of coherent states,
\begin{align*}
    \Ket{\alpha, 0} \xrightarrow{BS} \Ket{\frac{\alpha}{\sqrt{2}}, \frac{\alpha}{\sqrt{2}}}.
\end{align*}
We see that our input was one coherent state, and since the output is separable, the Schmidt rank is also 1.
Let us recall that the Schmidt rank of a bipartite pure state is the number of nonzero coefficients in its Schmidt decomposition. 
It quantifies the minimum number of product states needed to represent the state and provides a measure of its entanglement.
A value of one implies that the state is separable, whereas any value greater than one implies that the state is entangled.

Let's see another example. 
Consider a superposition of two coherent states of opposite phase and vacuum:
\begin{align*}
    \mathcal{N}_\alpha&(\Ket{\alpha} - \Ket{-\alpha}) \otimes \Ket{0} \\
    &\xrightarrow{BS}\mathcal{N}_\alpha\left(\Ket{\frac{\alpha}{\sqrt{2}}, \frac{\alpha}{\sqrt{2}}}-\Ket{-\frac{\alpha}{\sqrt{2}}, -\frac{\alpha}{\sqrt{2}}}\right),
\end{align*}
with $\mathcal{N}_\alpha = [2(1-\exp[{-2|\alpha|^2}])]^{-1/2}$. 
Here again, we can see that the superposition is of two coherent states, leading to a Schmidt rank of 2.

As a final example, let us consider a squeezed-state mode as input. 
In the output channels of the beam splitter we obtain the entangled state:
\begin{equation}
 |\xi,0\rangle \,\stackrel{\rm BS}{\longmapsto}\, \int \frac{{\rm d}^2 \alpha}{\pi\sqrt{\mu}} e^{-\frac{\nu}{2\mu} \alpha^{\ast\,2}-\frac{|\alpha|^2}{2}}\Ket{\frac{\alpha}{\sqrt{2}},\frac{\alpha}{\sqrt{2}}}.
\end{equation}
Expanding the state in terms of coherent states as
\begin{equation}
 \ket{\xi} = \int \frac{{\rm d}^2 \alpha}{\pi\sqrt{\mu}} e^{-\frac{\nu}{2\mu} \alpha^{\ast\,2}-\frac{|\alpha|^2}{2}}|\alpha\rangle.
\end{equation}
Describing the resulting entangled output state requires an infinite sum of tensor products of coherent states, since the input state $\ket{\xi}$ can only be written as a superposition involving infinitely many coherent states~\cite{GehrkeSperlingVogel2012}.
Consequently, an input state that is highly nonclassical, characterized by $r=\infty$, generates an equally strong degree of entanglement between the output modes.

This analysis further extends to multiport interferometers, where the same measure of nonclassicality at the input can be distributed as genuine multipartite entanglement among multiple output modes. 
Conversely, the degree of multimode nonclassicality acts as an upper bound on the entanglement that may be produced by linear mixing \cite{VogelSperling_2014}.

An important implication is that nonclassicality and entanglement become operationally equivalent resources: producing a given amount of entanglement is no more demanding than preparing a single-mode state with the corresponding degree of nonclassicality. 
This is particularly relevant from a practical standpoint, as engineering single-mode states is often experimentally less demanding than directly creating multipartite entanglement. 
Altogether, the results indicate that suitably prepared nonclassical states can serve as a versatile and accessible resource for generating entanglement of a desired strength and structure in quantum technologies.

To illustrate the role of nonclassicality as a resource, we present a simple example showing how entanglement can be activated using nonclassical correlations and beam splitters.
When two single-mode squeezed vacuum states, squeezed in orthogonal quadratures, are the input to a $50:50$ beam splitter, it yields entangled output modes; here, the so-called two-mode squeezed vacuum state is produced.
The entanglement is very sensitive to randomization of the relative phase, such that if, after the beam splitter, a full phase randomization of the state is performed, the new resulting state is then separable \cite{AgudeloSperlingVogel_2013}.
The Wigner function of the output is no longer Gaussian, but it is nonnegative.
Additionally, the reduced states of the output are thermal states.
Consequently, the local statistics of each mode exhibit the characteristics of a classical Gaussian state.
Nevertheless, such states still exhibit intensity–intensity (or equivalently, photon–photon) correlations.
Its quantum correlations can be verified through the lens of $P$-nonclassicality \cite{AgudeloSperlingVogel_2013}.
The nonclassicality of the fully randomized two-mode squeezed vacuum state has been experimentally certified \cite{KohnkeAgudeloSchunemannSchlettweinVogelSperlingHage_2021}.
Interfering each phase-randomized output mode separately with a vacuum state at a symmetric beam splitter enables the activation of entanglement across selected bipartitions of the final output modes.
This demonstrates that, while phase randomization renders the reduced states thermal and suppresses entanglement, the fundamental $P$-nonclassicality remains and continues to serve as a resource for producing quantum correlations that can subsequently be converted into entanglement.

Building on this scenario, we verify that the interference of two squeezed states with these features at a symmetric beam splitter leads to the generation of bipartite entanglement; see Section~\ref{apx:simon}.
In Section \ref{apx:reducedStates}, we confirm that the output modes are indeed thermal.
Later, in Section~\ref{apx:network}, we extend the configuration introduced above to model the interference of multiple modes within a beam splitter network.
This arrangement allows us to analyze entanglement across various bipartitions of the output modes using Simon’s criterion and logarithmic negativity.
At present, the phase-randomization procedure is not included in the simulation.%
\fnt{Implement a full phase randomization as indicated in Ref. \cite{AgudeloSperlingVogel_2013}. 
Mixed the output modes with vacuum in two new beam splitters and certify entanglement between the outputs of beam splitters two and three.}  

\section{Simulation labs}
\label{sec:labs}
Having established the theoretical framework for continuous-variable quantum systems—ranging from the quantization of the electromagnetic field to the formalism of Gaussian states and entanglement criteria—we now turn our attention to the numerical application of these concepts.
While analytical solutions offer deep insight into the fundamental properties of quantum states, computational simulations are an indispensable tool for analyzing complex optical networks and modeling experimental outcomes.
In this section, we will translate the mathematical descriptions of quantum states into executable code.
This approach serves two distinct purposes:
\begin{enumerate}
    \item \textbf{Simulating Measurement Statistics:} In an actual laboratory setting, quantum states are not accessed directly; they are reconstructed via statistical analysis of measurement data. By generating random numbers from specific probability distributions, we can simulate these results.
    \item \textbf{Modeling Optical Networks:} As the number of modes and optical elements increases, calculating covariance matrices and Wigner functions by hand becomes intractable. We will utilize specialized software libraries to model Gaussian states, visualize their phase-space representations, and quantify correlations in multi-mode systems.
\end{enumerate}
For this part, we recommend typing the code yourself as you read using \href{https://colab.google/}{Google Colab}, so you don't need a complicated setup.

\begin{widetext}
    \subsection{Data Simulation}

To generate random numbers representing a state, we need its cumulative distribution function, $F(x,\varphi)$. For the specific state to be generated, the characteristic function, $\chi(\beta)$, must be known. The cumulative distribution function, $F(x,\varphi)$, can be calculated from it in two steps.

\begin{itemize}
    \item Calculate the quadrature probability density function $p(x,\varphi)$ for the desired state using Eq.~\eqref{prob_from_char}:  
    $$
    p(x,\varphi) = \frac{1}{2\pi} \int_{-\infty}^{\infty} \mathrm{d}y\, e^{-i y x} \chi\left(i y e^{-i\varphi}\right)
    $$

    \item Integrate the density $p(x,\varphi)$ up to $x$:  
    $$
    F(x,\varphi) = \int_{-\infty}^x \mathrm{d}\tilde{x}\, p(\tilde{x},\varphi)
    $$
\end{itemize}

\textbf{Example}:  
A thermal state is parametrized with $\bar n = \left[\exp\left(\frac{\hbar \omega}{k_{\rm B}T}\right)-1\right]^{-1}$, the mean number of thermal photons.  
We get:
\begin{align*}
    \Phi_{\rm th}(\beta) &= \exp(-(\bar{n}+1/2)|\beta|^2) \\
    p(x,\varphi) &= \frac{1}{2\pi} \int_{-\infty}^{\infty} \mathrm{d}y\, \exp\left(-i y x \right)\exp(-\bar{n}y^2-\frac{y^2}{2}) = \frac{1}{\sqrt{\pi(4\bar{n}+2)}} \exp\left(-\frac{x^2}{4\bar{n}+2}\right) \\
    F(x,\varphi) &= \int_{-\infty}^x \mathrm{d}\tilde{x}\, \frac{1}{\sqrt{\pi(4\bar{n}+2)}} \exp\left(-\frac{\tilde{x}^2}{4\bar{n}+2}\right) = \frac{1}{2} + \frac{1}{2} \mathrm{erf}\left(\frac{x}{4\bar{n}+2}\right)
\end{align*}

Now we have to generate two sets (of the same size) of uniformly distributed random numbers.  
One set over the interval $[0, \pi]$ represents the phases $\varphi_j$ and the other set over $[0, 1]$ gives the values $F(x_j, \varphi_j)$.  
By numerically resolving equations of the type $F(x_j, \varphi_j) = p_0 \in [0, 1]$ for a fixed $\varphi = \varphi_j \in [0, \pi]$ with respect to $x_j$ for the desired state, you obtain the corresponding quadrature values $x_j$.

\textbf{Task 1.} Perform the simulation of the data for \textbf{one} of the states listed below.

\begin{itemize}
    \item The Fock states $|n\rangle$ (for one $n$ from the interval $2\leqslant n\leqslant10$):  
    $$\chi_n(\beta) = \exp[-\frac{|\beta|^2}{2}]{\rm L}_n(|\beta|^2),$$  
    where ${\rm L}_n(z)$ denotes the $n$-th Laguerre polynomial.

    \item The single photon added thermal state (SPATS):  
    $$\chi_{\rm SPATS}(\beta) = \exp[-\frac{|\beta|^2}{2}][1-(1+\bar{n})|\beta|^2] \exp[-\bar{n}|\beta|^2]$$

    \item The squeezed vacuum state with real parameter $\ket{r,0}$:  
    $$\chi_{\rm sv}(\beta) =\exp[-\frac{|\beta|^2}{2}] \exp\left[-\frac{(\beta+\beta^*)^2 e^{-2r}}{8} + \frac{(\beta-\beta^*)^2 e^{2r}}{8} + \frac{|\beta|^2}{2}\right]$$

    \item The superposition of coherent states, $|\psi\rangle = \frac{1}{\sqrt{\mathcal{N}}}\left(|\alpha\rangle + e^{i\theta}|-\alpha\rangle\right)$, where the normalization constant is given by $\mathcal{N} = 2 + 2\cos(\theta) e^{-2|\alpha|^2}$:  
    $$\chi_{\psi}(\beta) = \frac{1}{\mathcal{N}} \exp[-\frac{|\beta|^2}{2}] \left[e^{\beta\alpha^*-\beta^*\alpha} + e^{i\theta} e^{\beta\alpha^*+\beta^*\alpha} e^{-2|\alpha|^2} + e^{-i\theta} e^{-\beta\alpha^*-\beta^*\alpha} e^{-2|\alpha|^2} + e^{-\beta\alpha^*+\beta^*\alpha}\right]$$
\end{itemize}

\subsubsection{Import Libraries}
\begin{minted}{python}
import numpy as np                                          # Numerical operations
from scipy.special import erf, hermite, factorial           # Special functions
from scipy.optimize import bisect                           # Root finding (for inverse CDF)
import pandas as pd                                         # Data manipulation and storage
import matplotlib.pyplot as plt                             # Plotting
import time                                                 # Timing code execution
\end{minted}

\subsubsection{Find and code the CDF}

\begin{itemize}
    \item Fock State
    \begin{itemize}
        \item Characteristic function:  
        $$\chi_n(\beta) = \exp\left[-\frac{|\beta|^2}{2}\right]{\rm L}_n(|\beta|^2)$$
    
        \item Probability density:  
        $$p_n(x,\varphi) = \sum_{k=0}^{n} \dfrac{1}{\sqrt{2\pi}} \dfrac{n!}{2^k k!^2 (n-k)!} \exp\left(-\dfrac{x^2}{2}\right) {\rm H}_{2k}\left(\dfrac{x}{\sqrt{2}}\right)$$
    
        \item Cumulative distribution function:  
        $$F_n(x,\varphi) = \dfrac{1}{2} + \dfrac{1}{2} \mathrm{erf}\left(\dfrac{x}{\sqrt{2}}\right) - \dfrac{1}{\sqrt{\pi}} \exp\left(-\dfrac{x^2}{2}\right) \sum_{k=1}^{n} \dfrac{n!}{2^k k!^2 (n-k)!} {\rm H}_{2k-1}\left(\dfrac{x}{\sqrt{2}}\right)$$
    \end{itemize}

    \begin{minted}{python}
def fock_state_cdf(x, fock_number):
    result = 0.0
    for k in range(1, fock_number+1):
        H_poly = hermite(2*k-1)
        result -= (H_poly(x/np.sqrt(2)) * factorial(fock_number) 
                 / (2**k * factorial(k)**2 * factorial(fock_number-k)))
    result *= np.exp(-x**2/2) / np.sqrt(np.pi)
    result += 0.5 + 0.5 * erf(x/np.sqrt(2))
    return result
    \end{minted}

    \item SPATS
    \begin{itemize}
        
        \item Characteristic function:  
            $$\chi_{\rm SPATS}(\beta) = \exp[-\frac{|\beta|^2}{2}][1-(1+\bar{n})|\beta|^2]\, \exp[-\bar{n}|\beta|^2]$$

        \item Probability density:  
            $$p_{\rm SPATS}(x,\varphi) = \dfrac{1}{\sqrt{\pi(4\bar{n}+2)}}\exp\left(-\dfrac{x^2}{4\bar{n}+2}\right)\cdot\left[1-\dfrac{1+\bar{n}}{1+2\bar{n}}\left(1-\dfrac{x^2}{1+2\bar{n}}\right)\right]$$
        
        \item Cumulative distribution function:  
            $$F_{\rm SPATS}(x,\varphi) = \dfrac{1}{2}+\dfrac{1}{2}\mathrm{erf}\left(\dfrac{x}{\sqrt{4\bar{n}+2}}\right)-\dfrac{x}{\sqrt{\pi(4\bar{n}+2)}}\dfrac{1+\bar{n}}{1+2\bar{n}}\exp\left(-\dfrac{x^2}{4\bar{n}+2}\right)$$

    \end{itemize}
        \begin{minted}{python}
def spats_cdf(x, avg_pn):
    sigma = 4 * avg_pn + 2
    scale = (1 + avg_pn) / (1 + 2*avg_pn)
    return (0.5 + 0.5 * erf(x / np.sqrt(sigma))
            - x * scale / np.sqrt(np.pi * sigma) * np.exp(-x**2 / sigma))
    \end{minted}

    \item Squeezed Vacuum with real squeezing parameter

    \begin{itemize}
        \item Characteristic function:  
            $$\chi_{\rm sv}(\beta) = \exp[-\frac{|\beta|^2}{2}] \exp\left[-\frac{(\beta+\beta^\ast)^2 e^{-2r}}{8} + \frac{(\beta-\beta^\ast)^2 e^{2r}}{8} + \frac{|\beta|^2}{2}\right]$$

        \item  Probability density:  
            $$p_{\rm sv}(x,\varphi) = \dfrac{1}{\sqrt{2\pi}}\dfrac{1}{\sqrt{e^{2r}\cos^2(\varphi)+e^{-2r}\sin^2(\varphi)}}\exp\left[-\dfrac{x^2}{2e^{2r}\cos^2(\varphi)+2e^{-2r}\sin^2(\varphi)}\right]$$

        \item  Cumulative distribution function:  
            $$F_{\rm sv}(x,\varphi) = \dfrac{1}{2}+\dfrac{1}{2}\mathrm{erf}\left[\dfrac{x}{\sqrt{2e^{-2r}\cos^2(\varphi)+2e^{2r}\sin^2(\varphi)}}\right]$$
        \end{itemize}

        \begin{minted}{python}
def squeezed_state_cdf(x, phi, r):
    V_s = np.exp(-2 * r)  # Variance of the squeezed quadrature
    V_a = np.exp(2 * r)   # Variance of the anti-squeezed quadrature
    variance = 2 * V_s * np.cos(phi)**2 + 2 * V_a * np.sin(phi)**2 
    return 0.5 + 0.5 * erf(x / np.sqrt(variance))
        \end{minted}

        \item Superposition of Coherent States

        \begin{itemize}
            \item Characteristic function:  
                $$\chi_{\psi}(\beta) = \dfrac{1}{\mathcal{N}} \exp\left[-\frac{|\beta|^2}{2}\right] \left[e^{\beta\alpha^\ast-\beta^\ast\alpha} + e^{i\theta} e^{\beta\alpha^\ast+\beta^\ast\alpha} e^{-2|\alpha|^2} + e^{-i\theta} e^{-\beta\alpha^\ast-\beta^\ast\alpha} e^{-2|\alpha|^2} + e^{-\beta\alpha^\ast+\beta^\ast\alpha}\right]$$
            
            \item Probability density:  
                $$p_\psi(x,\varphi) = \dfrac{1}{\sqrt{2\pi}}\dfrac{1}{\mathcal{N}}\Big[ \exp\left[-\dfrac{(\alpha e^{i\varphi}+\alpha^\ast e^{-i\varphi}-x)^2}{2}\right] + e^{i\theta}e^{-2|\alpha|^2}\exp\left[-\dfrac{(-\alpha e^{i\varphi}+\alpha^\ast e^{-i\varphi}-x)^2}{2}\right]$$
            
                $$+ e^{-i\theta}e^{-2|\alpha|^2}\exp\left[-\dfrac{(\alpha e^{i\varphi}-\alpha^\ast e^{-i\varphi}-x)^2}{2}\right] + \exp\left[-\dfrac{(-\alpha e^{i\varphi}-\alpha^\ast e^{-i\varphi}-x)^2}{2}\right] \Big]$$
            
            \item Cumulative distribution function:  
                $$F_\psi(x,\varphi) = \dfrac{1}{2\mathcal{N}}\Big[2 + \mathrm{erf}\left(\dfrac{x-\alpha e^{i\varphi}-\alpha^\ast e^{-i\varphi}}{\sqrt{2}}\right) + e^{i\theta}e^{-2|\alpha|^2}\mathrm{erf}\left(\dfrac{x+\alpha e^{i\varphi}-\alpha^\ast e^{-i\varphi}}{\sqrt{2}}\right)$$
                $$ + e^{-i\theta}e^{-2|\alpha|^2}\mathrm{erf}\left(\dfrac{x-\alpha e^{i\varphi}+\alpha^\ast e^{-i\varphi}}{\sqrt{2}}\right) + \mathrm{erf}\left(\dfrac{x+\alpha e^{i\varphi}+\alpha^\ast e^{-i\varphi}}{\sqrt{2}}\right) \Big]$$
            \end{itemize}

            \begin{minted}{python}
def supcoh_state_cdf(x, phi, alpha, theta):
    N = 2 + 2 * np.cos(theta) * np.exp(-2 * np.abs(alpha)**2)
    c1 = np.exp(1j * theta) * np.exp(-2 * np.abs(alpha)**2)
    c2 = np.exp(-1j * theta) * np.exp(-2 * np.abs(alpha)**2)
    gamma = alpha * np.exp(1j * phi)

    result = (
         2 +
         erf((x - 2 * np.real(gamma)) / np.sqrt(2)) +
         c1 * erf((x + 2j * np.imag(gamma)) / np.sqrt(2)) +
         c2 * erf((x - 2j * np.imag(gamma)) / np.sqrt(2)) +
         erf((x + 2 * np.real(gamma)) / np.sqrt(2))
        )
    
    return np.real(result) / (2 * N) # The imaginary part is zero always
\end{minted}
\end{itemize}

\subsubsection{Invert the CDF to find the quadrature value}

Now, the idea is the following: we can generate values of the quadrature $\hat{X}_{\varphi}$ with the desired distribution by using the inversion method \cite{Devroye1986}, in which for any random variable $X \in \mathbb{R}$, the random variable $F_x^{-1}(U)$ has the same distribution as $X$, where $F_x^{-1}$ is the generalized inverse of the cumulative distribution function $F_X$ of $X$ and $U$ is uniform on $[0,1]$.
We also need to set up some parameters with the desired values. In code, this looks like:

\begin{minted}{python}
np.random.seed(42)                                    # For reproducibility
num_samples = 1000000                                 # Number of samples to generate
tolerance = 1e-12                                     # Tolerance for numerical methods
phases = 2 * np.pi * np.random.rand(num_samples) - np.pi # Random phases in [0, pi)
target_cdf_values = np.sort(np.random.rand(num_samples)) # Sorted random numbers in [0, 1)
search_width = 50                                     # Width for root finding search interval
sample_values = np.zeros(num_samples)                 # Array to hold sampled quadrature values


def cdf_inversion(sample_values, target_cdf_values, phases, cdf_func, arguments, tolerance):
    num_samples = len(sample_values)
    start_time = time.time()
    last_report = start_time
  
    for i in range(num_samples):
        sample_values[i] = bisect(
            lambda x: cdf_func(x, phases[i], **arguments) - target_cdf_values[i],
            -search_width, search_width, xtol=tolerance
        )
    
        now = time.time()
        if now - last_report > 10:
            print(f'Progress: {i+1}/{num_samples} ({100*(i+1)/num_samples:.1f}%)
                    time elapsed: {now - start_time:.2f} seconds.')
            last_report = now
    
    print(f'Progress: {i+1}/{num_samples} ({100*(i+1)/num_samples:.1f}%)')
    print(f'CDF inversion completed in {time.time() - start_time:.2f} seconds.')
    return sample_values
\end{minted}

\subsubsection{Use the function!}

Now we can just execute the function with the corresponding parameters to simulate the desired state.

\begin{minted}{python}
fock_arguments = {
    'fock_number': 3
}

spats_arguments = {
    'avg_pn': 3
}

sv_arguments = {
    'r': 1
}

supcoh_arguments = {
    'alpha': 16+2j,
    'theta': 0
}

cdf_func = squeezed_state_cdf

arguments = sv_arguments

sample_values = cdf_inversion(sample_values, target_cdf_values,
                                phases, cdf_func, arguments, tolerance) 
df = pd.DataFrame({'sample_value': sample_values, 'phase': phases})
\end{minted}

\subsubsection{Evaluation and Data Analysis}

\textbf{Task 2.} Check if the data sets represent physically correct states according to Heisenberg's uncertainty relation.

Considering the commutator relation $[\hat{X}_{\varphi},\hat{X}_{\varphi+\pi/2}] = -2i$, it results:
\begin{equation}
\mathrm{Var}[\hat{X}_{\varphi}] \cdot \mathrm{Var}[\hat{X}_{\varphi+\pi/2}] \geq 1
\end{equation}

Discuss the variances of the quadrature operator.

The analytical expressions are:

\begin{table}[h]
\centering
\begin{tabular}{|l|c|}
\hline
\textbf{State} & $\langle (\Delta\hat{x})^2\rangle$ \\
\hline
Fock & $2n + 1$ \\
\hline
SPATS & $4\bar{n} + 3$ \\
\hline
Squeezed vacuum & $| e^{i\varphi} \cosh r - e^{-i\varphi} \sinh r|^2$ \\
\hline
Superposition of coherent states & $(\alpha\,e^{i\varphi} - \alpha^* e^{-i\varphi})^2 \left[1 + \frac{\sin^2\theta}{\mathcal{N}^2}\,e^{-4|\alpha|^2}\right] + \frac{4|\alpha|^2}{1 + e^{-2|\alpha|^2} \cos\theta} + 1$ \\
\hline
\end{tabular}
\end{table}

\begin{minted}{python}
# Theoretical variance
def fock_state_var_ht(fock_number):
    return 2 * fock_number + 1
    
def spats_var_th(avg_pn):
    return  4 * avg_pn + 3

def squeezed_state_var_th(phi, r):
    return np.abs(np.exp(1j*phi)*np.cosh(r) - np.exp(-1j*phi)*np.sinh(r))**2

def supcoh_state_var_th(phi, alpha, theta):
    N = 2 + 2 * np.exp(-2 * np.abs(alpha)**2) * np.cos(theta)
    k0 = alpha * np.exp(1j * phi)
    k1 = 1+(np.exp(-2 * np.abs(alpha)**2) * np.sin(theta)/N)**2
    k2 = (4 * np.abs(alpha)**2)/(N/2)
    return np.real((2j*np.imag(k0))**2 * k1 + k2 + 1)

def compute_variance_by_state(cdf_func, sample_values, phases, arguments, num_bins=100):
    """
    Compute variance for fock/spats (single value) or sv/supcoh (per phase bin).
    Returns DataFrame with columns: phase, theoretical_variance, calculated_variance
    """
    if cdf_func in [fock_state_cdf, spats_cdf]:
        # Single variance, phase not relevant
        if cdf_func == fock_state_cdf:
            theoretical = fock_state_var_ht(arguments['fock_number'])
        else:
            theoretical = spats_var_th(arguments['avg_pn'])
        calculated = np.var(sample_values)
        return pd.DataFrame({'phase': [None], 'theoretical_variance': [theoretical],
                            'calculated_variance': [calculated]})
    else:
        # Per phase bin
        bins = np.linspace(np.min(phases), np.max(phases), num_bins + 1)
        bin_centers = 0.5 * (bins[:-1] + bins[1:])
        theoretical = np.zeros(num_bins)
        calculated = np.zeros(num_bins)
        for i in range(num_bins):
            mask = (phases >= bins[i]) & (phases < bins[i+1])
            if np.sum(mask) > 1:
                calculated[i] = np.var(sample_values[mask])
                if cdf_func == squeezed_state_cdf:
                    theoretical[i] = squeezed_state_var_th(bin_centers[i], arguments['r'])
                else:
                    theoretical[i] = supcoh_state_var_th(bin_centers[i], arguments['alpha'],
                                                            arguments['theta'])
            else:
                calculated[i] = np.nan
                theoretical[i] = np.nan
        return pd.DataFrame({'phase': bin_centers, 'theoretical_variance': theoretical, 
                                'calculated_variance': calculated})

variance_df = compute_variance_by_state(cdf_func, sample_values, phases, arguments, num_bins=100)
print('Computed Variance:\n', variance_df)

# Check Heisenberg uncertainty relation by multiplying variances at phi=x and phi=x+pi/2 for
# squeezed and supcoh states
if len(variance_df) > 1 and 'phase' in variance_df.columns:
    n_bins = len(variance_df)
    shift = n_bins // 4
    shifted_variance = np.full(n_bins, np.nan)
    variance_product = np.full(n_bins, np.nan)
    for i in range(n_bins):
        j = (i + shift) % n_bins
        shifted_variance[i] = variance_df['calculated_variance'][j]
        if (not np.isnan(variance_df['calculated_variance'][i]) and not
            np.isnan(variance_df['calculated_variance'][j])):
            variance_product[i] = variance_df['calculated_variance'][i]
                                    * variance_df['calculated_variance'][j]
    variance_df['shifted_variance'] = shifted_variance
    variance_df['variance_product'] = variance_product
    print('Variance product for Heisenberg relation:\n', variance_df[['phase',
                        'calculated_variance', 'shifted_variance', 'variance_product']])
\end{minted}

\textbf{Task 3.} Visualize the data sets in $x$-$\varphi$-diagrams (linear and/or polar). 
	Discuss the correlations between the parameters randomly generated $F$ and $\varphi$ in the cases of physical and non-physical states, compare with the vacuum state.
	Check the phase and the quadrature distributions through their histograms.

\begin{minted}{python}
# Visualize the data sets in x-phi diagrams (linear and polar)
import matplotlib.pyplot as plt

# Linear scatter plot of quadrature vs phase
plt.figure(figsize=(8, 5))
plt.scatter(phases, sample_values, s=1, alpha=0.5)
plt.xlabel('Phase $\\phi$')
plt.ylabel('Quadrature $x$')
plt.title('Scatter plot: Quadrature vs Phase')
plt.grid(True)
plt.show()

# Polar plot of quadrature vs phase
plt.figure(figsize=(7, 7))
plt.subplot(111, polar=True)
plt.scatter(phases, sample_values, s=1, alpha=0.5)
plt.title('Polar plot: Quadrature vs Phase')
plt.show()

# Histogram of quadrature values
plt.figure(figsize=(8, 5))
plt.hist(sample_values, bins=100, alpha=0.7, color='tab:blue')
plt.xlabel('Quadrature $x$')
plt.ylabel('Counts')
plt.title('Histogram of Quadrature Values')
plt.grid(True)
plt.show()

# Histogram of phase values
plt.figure(figsize=(8, 5))
plt.hist(phases, bins=100, alpha=0.7, color='tab:orange')
plt.xlabel('Phase $\\phi$')
plt.ylabel('Counts')
plt.title('Histogram of Phase Values')
plt.grid(True)
plt.show()

# Plot of variance vs phase
if len(variance_df) > 1 and 'phase' in variance_df.columns:
    plt.figure(figsize=(8, 5))
    #scatter plot of the calculated variance
    plt.scatter(variance_df['phase'], variance_df['calculated_variance'], label='Calculated
                                                            Variance', color='tab:blue')
    #line plot of the theoretical variance
    plt.plot(variance_df['phase'], variance_df['theoretical_variance'], label='Theoretical 
                                                Variance', color='tab:orange')
    plt.xlabel('Phase $\\phi$')
    plt.ylabel('Variance')
    plt.title('Variance vs Phase')
    plt.legend()
    plt.grid(True)
    plt.show()
\end{minted}

Examples of the outputs of these plots for a squeezed vacuum state with real squeezing parameter $r = 1$ are:

\begin{figure}[H]
    \begin{subfigure}{.32\linewidth}
        \centering
        \includegraphics[width=1\linewidth]{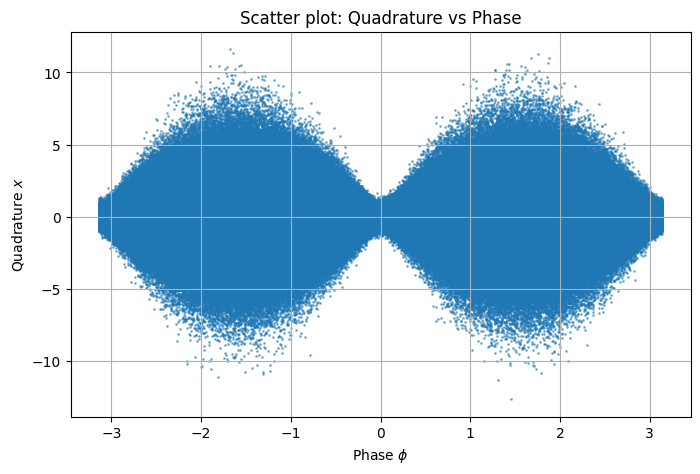}
        \caption{Quadrature vs Phase scatter plot}
        \label{fig:sv_r1_quad_phase}
    \end{subfigure}
    ~
    \begin{subfigure}{.32\linewidth}
        \centering
        \includegraphics[width=1\linewidth]{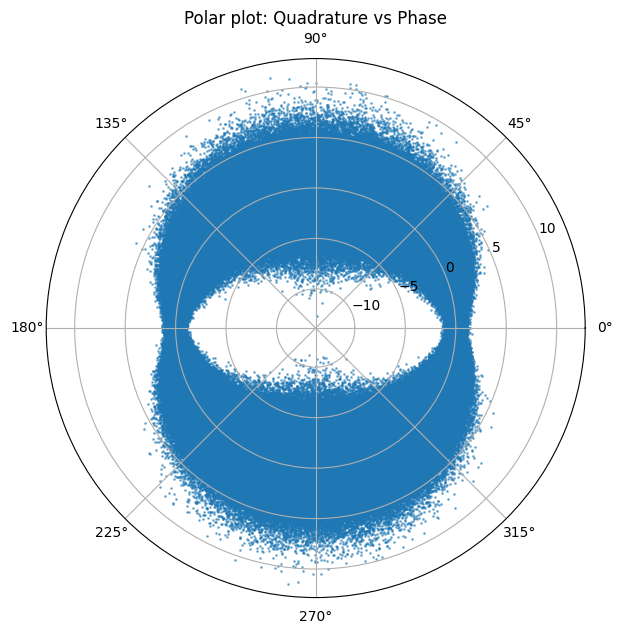}
        \caption{Quadrature vs Phase polar scatter plot}
        \label{fig:sv_r1_quad_phase_polar}
    \end{subfigure}
    ~
    \begin{subfigure}{.32\linewidth}
        \centering
        \includegraphics[width=1\linewidth]{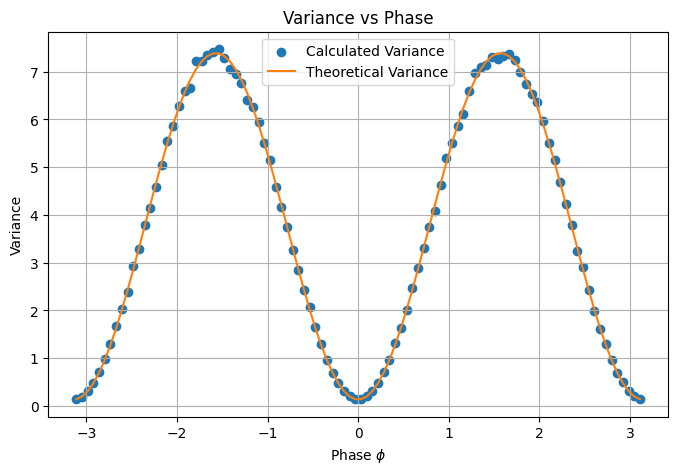}
        \caption{Theoretical and estimated variance against phase}
        \label{fig:sv_r1_variance_phase}
    \end{subfigure}
\end{figure}

\textbf{Task 4.} In the case that the previous analysis indicates squeezing, proof it with the condition given in the introduction.
$$\langle:(\Delta \hat{x}(\phi))^2:\rangle = \langle(\Delta \hat{x}(\phi))^2\rangle - \langle0|(\Delta \hat{x}(\phi))^2|0\rangle < 0$$

\begin{minted}{python}
# To test the condition we need the Vacuum state variance. We can use the squeezed state
# function with r=0
vacuum_variance = squeezed_state_var_th(0, 0)  # This should be 1
print(f'Vacuum state variance: {vacuum_variance}')  

# Generate data for the Vacuum
vacuum_sample_values = cdf_inversion(np.zeros(num_samples), target_cdf_values,
np.zeros(num_samples), squeezed_state_cdf, {'r': 0}, tolerance)

vacuum_variance = np.var(vacuum_sample_values)
print(f'Calculated vacuum state variance from samples: {vacuum_variance}')

# Add another column to the vacuum variance to the variance_df
if len(variance_df) > 1 and 'phase' in variance_df.columns:
    variance_df['vacuum_variance'] = vacuum_variance
    variance_df['normal_ordered_variance'] = (variance_df['calculated_variance']
                                                - vacuum_variance)
    print('Variance with normal ordering:\n', variance_df[['phase', 'calculated_variance', 
        'vacuum_variance', 'normal_ordered_variance']])
    # Plot the normal ordered variance
    plt.figure(figsize=(8, 5))
    plt.scatter(variance_df['phase'], variance_df['normal_ordered_variance'], label='Normally
                Ordered Variance', color='tab:green')
    plt.axhline(0, color='red', linestyle='--', label='Zero Line')
    plt.xlabel('Phase $\\phi$')
    plt.ylabel('Normal Ordered Variance')
    plt.title('Normally Ordered Variance vs Phase')
    plt.legend()
    plt.grid(True)
    plt.show()

\end{minted}

\begin{figure}
    \centering
    \includegraphics[width=0.5\linewidth]{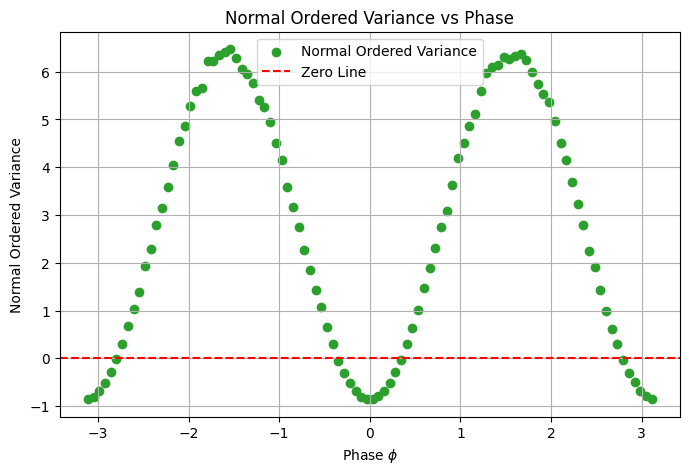}
    \caption{Squeezing condition for the variance of a squeezed vacuum state with real squeezing parameter r=1}
    \label{fig:sv_r1_squeezing_cond}
\end{figure}
    \subsection{Simulation of Gaussian states with Strawberry Fields}

Strawberry Fields is an end-to-end Python library for quantum computing and simulation built by Xanadu that is compatible with continuous-variable (CV) quantum systems. 
It offers tools for designing, optimizing, and simulating quantum optical circuits, including the application of models for Gaussian states and operations, as well as non-Gaussian elements.\cite{Xanadu_2025}

Strawberry Fields uses the convention for defining the quadrature operators ($\hat{x}$ and $\hat{p}$) in terms of the bosonic creation ($\hat{a}^\dagger$) and annihilation ($\hat{a}$) operators as:

\begin{align*}
\hat{x} &= \sqrt{\frac{\hbar}{2}}(\hat{a} + \hat{a}^\dagger) \\
\hat{p} &= -i\sqrt{\frac{\hbar}{2}}(\hat{a} - \hat{a}^\dagger).
\end{align*}

By default, the library sets $\hbar = 2$, but this value can be modified. 
We will adhere to this convention going forward, so the values obtained for the mean vector and covariance matrix will correspond to this choice.

In Strawberry Fields, a multi-mode quantum system is represented by a series of nodes $(q_j)$ in a quantum circuit. 
Each quantum node, such as \texttt{q[j]}, corresponds to a specific mode of the system. 
For example, \texttt{q[0]} represents the first mode, \texttt{q[1]} the second, and so on. 
In the simplest case of a single-mode system, you only need one node in your circuit. 

Let us start simulating some common single-mode states.

\subsubsection{Single mode states}
\label{apx:single_mode}

We first need to install Strawberry Fields. To do it Google Colab as recommended, it's enough to run the following line of code
\begin{minted}{shell}
    !pip install git+https://github.com/XanaduAI/strawberryfields.git
\end{minted}

Then, we import the necessary libraries:

\begin{minted}{python}
import strawberryfields as sf
import matplotlib.pyplot as plt
import numpy as np
\end{minted}

\subsubsection*{Vacuum state} 

In Strawberry Fields, the default state for any un-acted-upon mode is the vacuum state $|0\rangle$. This is the quantum state with the lowest possible energy and no photons. To create a program with a single mode initialized to the vacuum state, you just need to create a quantum circuit with one mode.

\begin{minted}{python}
prog = sf.Program(1) # Single mode program
\end{minted}

Now you can prepare the vacuum state in this mode:

\begin{minted}{python}
with prog.context as q:
    q[0]
\end{minted}

To run the quantum program \texttt{prog} using the Gaussian backend, you use the \texttt{eng.run()} method. This backend is specifically designed for simulating Gaussian states and operations efficiently.

\begin{minted}{python}
eng = sf.Engine('gaussian') # Using the Gaussian backend
state = eng.run(prog).state # Run the program to get the state
\end{minted}

In Strawberry Fields, you can choose from four primary backends to run your quantum programs, with your choice depending on the specific task.

\begin{itemize}
    \item \textbf{Gaussian backend:} This backend is highly efficient for simulating Gaussian states and operations. It is the default for many introductory simulations.
    \item \textbf{Fock backend:} Use this backend when you need to perform simulations in the Fock basis, which is essential for tasks involving photon number-counting and non-Gaussian operations.
    \item \textbf{Bosonic backend:} This backend is an alternative for simulating photonic systems, offering a different approach to computations compared to the Fock basis.
    \item \textbf{TensorFlow (``tf'') backend:} This backend leverages the TensorFlow library to enable the creation of quantum machine learning models.
\end{itemize}

The \texttt{state} object holds all the physical information of your quantum system. One way to access this information is by obtaining the discretized Wigner function using the \texttt{.wigner()} method.

This method takes the quadrature values $(\hat{x},\hat{p})$ as arguments and calculates the Wigner function's value at that specific point in phase space.

For example, to get a full plot of the Wigner function, you would pass in a range of values for $\hat{x}$ and $\hat{p}$, allowing you to visualize the state's quasi-probability distribution.

\begin{minted}{python}
X = np.linspace(-5, 5, 100) # x values for Wigner function
P = np.linspace(-5, 5, 100) # p values for Wigner function
Z = state.wigner(0, X, P) # Wigner function values
\end{minted}

Now you can plot the Wigner function for the vacuum state

\begin{minted}{python}
fig = plt.figure() # Create a new figure
X, P = np.meshgrid(X, P) # Create a meshgrid for X and P
ax = fig.add_subplot(111, projection="3d") # 3D subplot
ax.plot_surface(X, P, Z, cmap="RdYlGn", lw=0.5, rstride=1, cstride=1) # Surface plot
fig.set_size_inches(4.8, 5) # Set figure size
\end{minted}

\begin{figure}[H]
    \centering
    \includegraphics[width=0.5\linewidth]{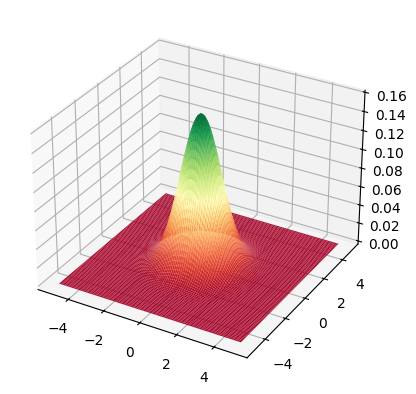}
    \caption{Wigner function for Vaccum state}
\end{figure}

Starting now, we'll use the \texttt{plot\_wigner()} function. It takes the state from a Strawberry Fields simulation and generates both surface and contour plots of the Wigner function for that state.

\begin{minted}{python}
    
def plot_wigner(state):
    """Plots the Wigner function of a given quantum state.

    This function generates and displays two visualizations of the Wigner
    function: a 3D surface plot and a 2D contour plot.

    Args:
        state: A quantum gaussian state object generated with Strawberry Fields.
    """
    # --- Data Generation for Wigner function ---
    X = np.linspace(-5, 5, 100)
    P = np.linspace(-5, 5, 100)
    Z = state.wigner(0, X, P)
    X, P = np.meshgrid(X, P)

    # --- Plotting ---
    # Create a figure to hold the subplots
    fig = plt.figure()
    fig.set_size_inches(10, 5) # Adjusted size for two plots

    # == Subplot 1: Surface Plot ==
    ax1 = fig.add_subplot(1, 2, 1, projection='3d') # 1 row, 2 columns, 1st plot
    ax1.plot_surface(X, P, Z, cmap="RdYlGn", lw=0.5, rstride=1, cstride=1)
    ax1.set_xlabel(r"$\hat{x}$")
    ax1.set_ylabel(r"$\hat{p}$")
    ax1.set_zlabel(r"$W(\hat{x},\hat{p})$")
    ax1.set_title("Surface Plot")

    # == Subplot 2: Contour Plot ==
    ax2 = fig.add_subplot(1, 2, 2) # 1 row, 2 columns, 2nd plot
    contour = ax2.contourf(X, P, Z, cmap="RdYlGn") # Use contourf for filled contours
    fig.colorbar(contour, ax=ax2) # Add a colorbar for the contour plot
    ax2.set_xlabel(r"$\hat{x}$")
    ax2.set_ylabel(r"$\hat{p}$")
    ax2.set_title("Contour Plot")
    ax2.set_aspect('equal') # Ensure the plot is circular, not elliptical

    # Adjust layout and show the plot
    fig.subplots_adjust(wspace=0.4)
    plt.show()
\end{minted}

Applying this function to the vacuum state, we have already simulated

\begin{minted}{python}
plot_wigner(state)
\end{minted}

\begin{figure}[H]
    \centering
    \includegraphics[width=0.8\linewidth]{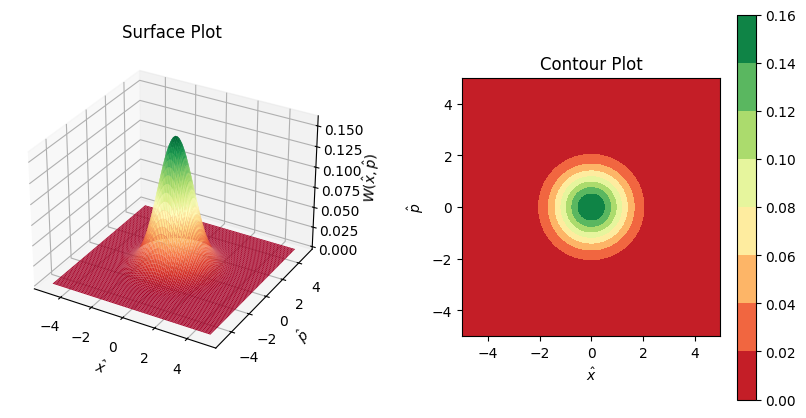}
    \caption{Wigner function for Vaccum state}
\end{figure}

We have already seen in Sec \ref{sec:gaussianiry} that all physical information for Gaussian states is encoded in their covariance matrix and mean vector. You can easily retrieve this data from a Strawberry Fields simulation using the \texttt{.cov()} and \texttt{.means()} functions.

\begin{minted}{python}
# Get the covariance matrix of the vacuum state
cov_vac = state.cov()
cov_vac
\end{minted}

\texttt{\underline{output:}}

\begin{verbatim}
array([[1., 0.],
       [0., 1.]])
\end{verbatim}

\begin{minted}{python}
# Get the mean values of the vacuum state
mean_vac = state.means()
mean_vac
\end{minted}

\begin{verbatim}
array([0., 0.])
\end{verbatim}

For vacuum state:
\[
    \overline{\mathbf{r}}=(0,0)^T,\quad \sigma=\left(\begin{array}{cc}1 & 0 \\ 0 & 1 \end{array}\right). 
\]

\subsubsection*{Coherent state}

A coherent state, written as $|\alpha \rangle$, is essentially a displaced vacuum state $|0\rangle$. You create it by applying the displacement operator, $\hat{D}(\alpha)$. The amount of displacement in the $\hat{x}$ and $\hat{p}$ quadratures is directly related to the real and imaginary parts of $\alpha$. To do this in Strawberry Fields, you simply use the \texttt{sf.ops.Dgate()} on a vacuum mode and provide the magnitude and phase of the complex number $\alpha$ as arguments.

\begin{minted}{python}
prog = sf.Program(1)
with prog.context as q:
    sf.ops.Dgate(1.6,0) | q[0] # Generates a Coherent state with alpha=1.6 and phase=0

eng.reset() #Reset engine to erase information of previous simulated states
state = eng.run(prog).state # Run the program to get the state

plot_wigner(state) # Plot the Wigner function of the coherent state
\end{minted}

\begin{figure}[H]
    \centering
    \includegraphics[width=0.8\linewidth]{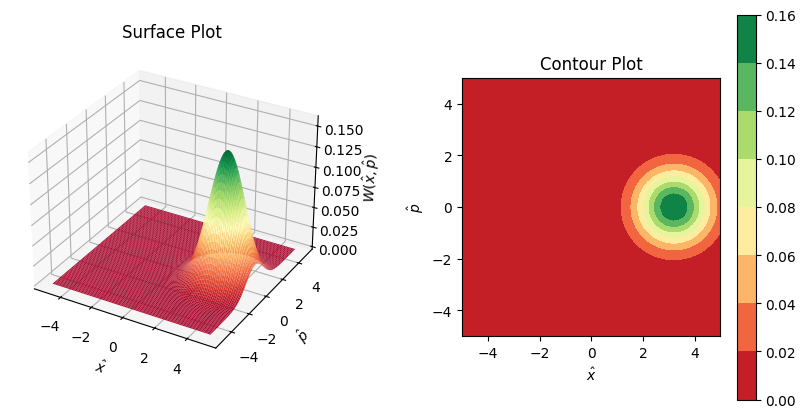}
    \caption{Wigner function for Coherent state with $\alpha = 1.6$} %Puras mentiras
    %jajajajaja
    
\end{figure}

\begin{minted}{python}
# Get the covariance matrix of the coherent state
cov_coh = state.cov()
cov_coh
\end{minted}

\texttt{\underline{output:}}

\begin{verbatim}
array([[1., 0.],
       [0., 1.]])
\end{verbatim}

\begin{minted}{python}
# Get the mean values of the coherent state
mean_coh = state.means()
mean_coh
\end{minted}

\begin{verbatim}
array([1.6, 0. ])
\end{verbatim}

For a coherent state
\[
    \overline{\mathbf{r}}=(3.2,0)^T,\quad \sigma=\left(\begin{array}{cc}1 & 0 \\ 0 & 1 \end{array}\right). 
\]

\subsubsection*{Squeezed vacuum state}

The squeezed state $|\xi\rangle$ is the result of applying the squeeze operator $\hat{S}(\xi)$ to the vacuum state $|0\rangle$. In Strawberry Fields, you can generate this state using the \texttt{sf.ops.Sgate()} on a vacuum qumode. This gate takes the magnitude and phase of the complex squeezing parameter $\xi$ as its arguments.

\begin{minted}{python}
prog = sf.Program(1)
with prog.context as q:
    sf.ops.Sgate(0.5) | q[0] # Squeezing with r=0.5 and phi=0
    
eng.reset() #Reset engine to erase information of previous simulated states
state = eng.run(prog).state # Run the program to get the state
plot_wigner(state) # Plot the Wigner function of the squeezed state
\end{minted}

\begin{figure}[H]
    \centering
    \includegraphics[width=0.8\linewidth]{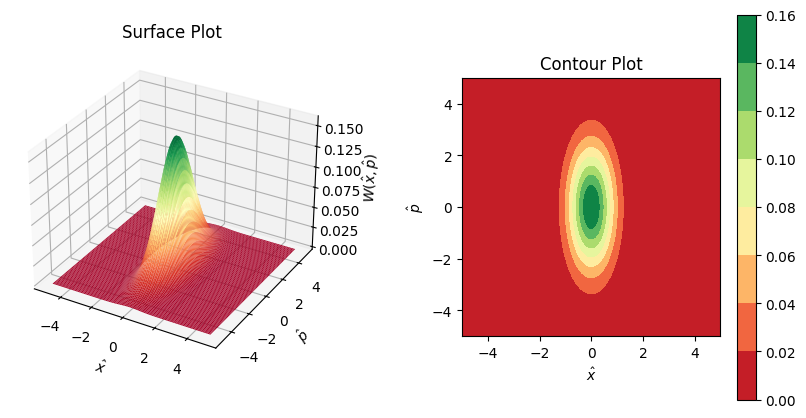}
    \caption{Wigner function for Squeezed state with $\xi = 0.5$} 
    
\end{figure}

\begin{minted}{python}
# Get the covariance matrix of the squeezed state
cov_sqz = state.cov()
cov_sqz
\end{minted}

\texttt{\underline{output:}}

\begin{verbatim}
array([[0.36787944, 0.        ],
       [0.        , 2.71828183]])
\end{verbatim}

\begin{minted}{python}
# Get the mean values of the squeezed state
mean_sqz = state.means()
mean_sqz
\end{minted}

\texttt{\underline{output:}}
\begin{verbatim}
array([0., 0.])
\end{verbatim}

For squeezed state:
\[
    \overline{\mathbf{r}}=(0,0)^T,\quad \sigma=\left(\begin{array}{cc}0.36787944 & 0 \\ 0 & 2.71828183 \end{array}\right). 
\]

\subsubsection*{Displaced squeezed state}

We can displace and squeeze the vacuum state. For instance, we can first apply a displacement with parameter $\alpha = 1.6 e^{i\frac{\pi}{4}}$, then a squeeze operation with parameter of $\xi  = 0.5 e^{i\frac{\pi}{2}}$.

\begin{minted}{python}
prog = sf.Program(1)
with prog.context as q:
    sf.ops.Dgate(1.6,np.pi/4) | q[0] # Displacement with alpha=1.6 and phase=pi/4
    sf.ops.Sgate(0.5,np.pi/2) | q[0] # Squeezing with r=0.5 and phi=pi/2

eng.reset() #Reset engine to erase information of previous simulated states
state = eng.run(prog).state # Run the program to get the state
plot_wigner(state) # Plot the Wigner function of the displaced-squeezed state
\end{minted}

\begin{figure}[H]
    \centering
    \includegraphics[width=0.7\linewidth]{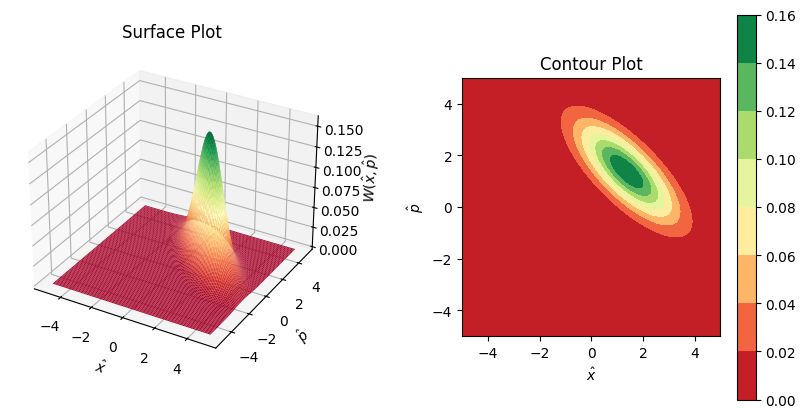}
    \caption{Wigner function for Displaced squeezed state with $\alpha = 1.6 e^{i\frac{\pi}{4}}$ and $\xi  = 0.5 e^{i\frac{\pi}{2}}$.} 
    
\end{figure}

\begin{minted}{python}
# Get the covariance matrix of the displaced-squeezed state
cov_ds = state.cov()
cov_ds
\end{minted}
\texttt{\underline{output:}}
\begin{verbatim}
array([[ 1.54308063, -1.17520119],
       [-1.17520119,  1.54308063]])
\end{verbatim}

\begin{minted}{python}
# Get the means of the displaced-squeezed state
mean_ds = state.means()
mean_ds
\end{minted}
\texttt{\underline{output:}}
\begin{verbatim}
array([1.37242222, 1.37242222])
\end{verbatim}

For displaced squeezed state:
\[
    \overline{\mathbf{r}}=(1.37242222,1.37242222)^T,\quad \sigma=\left(\begin{array}{cc}1.54308063 & -1.17520119 \\ -1.17520119 & 1.54308063 \end{array}\right). 
\]

\subsubsection{Two mode states}
\label{apx:two_mode}

We can also create a two mode vacuum state with Strawberry Fields and print its covariance matrix and means vector. 
Let us create a two mode vaccum state:

\begin{minted}{python}

prog = sf.Program(2)

with prog.context as q:
    
    q[0] 
    q[1]  

eng = sf.Engine("gaussian")
state=eng.run(prog).state
\end{minted}

\begin{minted}{python}
cov_mtx = state.cov()
cov_mtx
\end{minted}

\texttt{\underline{output:}}
\begin{verbatim}
array([[1., 0., 0., 0.],
       [0., 1., 0., 0.],
       [0., 0., 1., 0.],
       [0., 0., 0., 1.]])
\end{verbatim}

\begin{minted}{python}
means=state.means()
means
\end{minted}

\texttt{\underline{output:}}
\begin{verbatim}
array([0., 0., 0., 0.])
\end{verbatim}

Let us generate a two mode coherent state with the first mode in a coherent state with parameter $\alpha = 1.6$ and the second mode in a vacuum state.

\begin{minted}{python}
prog = sf.Program(2)

with prog.context as q:
    
    sf.ops.Dgate(1.6,0) | q[0] 
    # If we don't assign any state on the second mode, it will be a vacuum state by default

eng = sf.Engine("gaussian")
state=eng.run(prog).state

cov_mtx = state.cov()
cov_mtx
\end{minted}

\texttt{\underline{output:}}
\begin{verbatim}
array([[1., 0., 0., 0.],
       [0., 1., 0., 0.],
       [0., 0., 1., 0.],
       [0., 0., 0., 1.]])
\end{verbatim}

\begin{minted}{python}
means=state.means()
means
\end{minted}

\texttt{\underline{output:}}
\begin{verbatim}
array([3.2, 0. , 0. , 0. ])
\end{verbatim}

Here it is worth noting that the library uses representation of the 2N-operator vector, denoted by $\hat{\boldsymbol{r}} = (\hat{x_1},\dots,\hat{x_N} ,\hat{p_1}\dots,\hat{p_N})^T$, which is commonly referred to as the \textit{xp-order}.
In Sec. \ref{sec:gaussianiry} we have worked with the representation $\boldsymbol{\hat{r}}=( \hat{x}_1, \hat{p}_1,\dots,\hat{x}_N, \hat{p}_N)^T$. As a consequence, the order of covariance matrix elements also differs from the ones obtained in Strawberry Fields.
We can apply a basis transformation to go from the \textit{xp-order} to the other one using the permutation matrix ($\Pi$) that flips the elements of $(\hat{x_1},\dots,\hat{x_N} ,\hat{p_1}\dots,\hat{p_N})^T$ to $( \hat{x}_1, \hat{p}_1,\dots,\hat{x}_N, \hat{p}_N)^T$

The change of basis is as follows:
\begin{equation}
\begin{split}
&\boldsymbol{\hat{r}} \rightarrow \Pi \cdot \hat{r} \\
&\sigma \rightarrow \Pi \cdot \sigma \cdot \Pi,
\end{split}
\end{equation}

From now on, we are going to work with the representation in which the $\boldsymbol{\hat{r}}$ vector is written as $\boldsymbol{\hat{r}}=( \hat{x}_1, \hat{p}_1,\dots,\hat{x}_N, \hat{p}_N)^T$. We implemented the functions \texttt{COB\_CovMtx()} and \texttt{COB\_Means()} that implement this change of basis. They receive as an argument the means vector and the covariance matrix that SF returns respectively.

\begin{minted}{python}
def Bj(matrix):
    """Calculate the permuatation matrix associated with the transformation """

    
    size, _ = matrix.shape

    # Create a sample matrix
    matrix = np.eye(size)

    # Split the matrix vertically into two sub-matrices
    matrix1, matrix2 = np.split(matrix, 2, axis=1)

    
    a0=np.vstack(([row[0] for row in matrix1],[row[0] for row in matrix2]))
    # Create a zeros matriz of shape size x size
    #Bj=np.zeros((size,size))
    for i in range((size//2)-1):
        a=np.vstack(([row[i+1] for row in matrix1],[row[i+1] for row in matrix2]))
        a0=np.vstack((a0,a))
    
    Bj=a0.transpose()
       
    return Bj


def COB_CovMtx(cov_mtx):
    """Transform the Covariance Matrix  """
    

    # SF defines vector of quadrature operator as r=(x1,x2,...,xn,p1,p2,...,pn)
    # We want to work with thw quadrature operator as r=(x1,p1,...,xn,pn)
    # Then, we should do a basis transfromation

  
    #Apply transfromation of basis of Covariance Matrix (V) from to (Bj^-1*V*Bj)
    cov_mtx2 = np.linalg.inv(Bj(cov_mtx)) @ cov_mtx @ Bj(cov_mtx)

    return cov_mtx2


def COB_Means(means):
    """Transform the means vector """
    
    # SF defines vector of quadrature operator as r=(x1,x2,...,xn,p1,p2,...,pn)
    # We want to work with the quadrature operator as r=(x1,p1,...,xn,pn)
    # Then, we should do a basis transfromation


    # Create a matrix to use the sizes on Bj() function
    matrix=np.eye(means.size)

    # #Transform vector of means (<r>) to (Bj^-1*<r>) 
    means2 = np.linalg.inv(Bj(matrix)) @ means

    return  means2  
\end{minted}

Now, we can generate a two mode quantum state where the first mode is in a vacuum state and the second mode is a squeezed state with squeezing parameter $\xi = 0$.

\begin{minted}{python}
#Solution

prog = sf.Program(2)

with prog.context as q:
    
    sf.ops.Sgate(0.5,0) | q[0]  #squeezing  along x (phase = 0)  
    # If we don't assign any state on the second mode, it will be a vacuum state by default
    
eng = sf.Engine("gaussian")
state=eng.run(prog).state

cov_mtx = state.cov()
cov_mtx
\end{minted}

\texttt{\underline{output:}}
\begin{verbatim}
array([[0.36787944, 0.        , 0.        , 0.        ],
       [0.        , 1.        , 0.        , 0.        ],
       [0.        , 0.        , 2.71828183, 0.        ],
       [0.        , 0.        , 0.        , 1.        ]])



\end{verbatim}

Then, we can use the function created to express the covariance matrix in the desired basis

\begin{minted}{python}
COB_CovMtx(cov_mtx)
\end{minted}

\texttt{\underline{output:}}
\begin{verbatim}
array([[0.36787944, 0.        , 0.        , 0.        ],
       [0.        , 2.71828183, 0.        , 0.        ],
       [0.        , 0.        , 1.        , 0.        ],
       [0.        , 0.        , 0.        , 1.        ]])
\end{verbatim}

With this representation the covariance matrix for any two mode quantum state is given by the following block matrix:

\begin{equation}
\sigma=\left(\begin{array}{cc}\sigma_{A} & \sigma_{AB} \\ \sigma_{AB}^T & \sigma_{B}\end{array}\right),    
\end{equation}

\begin{itemize}
    \item $\sigma_{A}$ is the local reduced covariance matrix for mode 1. For a two-mode system, it's a 2$\times$2 matrix that describes the variances and correlations within the first mode alone.
    \item $\sigma_{B}$ is the local reduced covariance matrix for mode 2. Similarly, for a two-mode system, it's a 2$\times$2 matrix describing the variances and correlations within the second mode.
    \item $\sigma_{AB}$ is a $2\times2$ matrix that contains the inter-modal correlations between mode 1 and mode 2. It's often referred to as the correlation matrix.
\end{itemize}

This block matrix structure is a standard and powerful way to represent the statistical properties of multi-mode Gaussian states, such as vacuum, coherent, and squeezed states. It cleanly separates the properties of each individual mode from the correlations that may exist between them, which are crucial for phenomena like entanglement.

\subsubsection{Beam splitter transformation}
\label{apx:BS}

You may have noticed that the off-diagonal sub-block matrix, $\sigma_{AB}$, was a zero matrix in our previous examples. This indicates that there are no correlations between the two modes. This result is expected because we generated the modes independently. We created one mode as a squeezed state and the other as a vacuum state, with no interaction or shared history between them. Later, we will see that these inter-modal correlation elements play a crucial role in analyzing the nature of correlations between modes, particularly in the context of entanglement.
A typical way to introduce correlations between two independent quantum modes is by interfering them through a beam splitter as explained on Sec. \ref{sec:BS_transf}. 
The Strawberry Fields library provides a function for this transformation: \texttt{sf.ops.BSgate()}. 
It receives as arguments the $\theta$ and $\phi$ parameters of the beam splitter.
For a typical 50/50 beam splitter, the parameters are $\theta = \frac{\pi}{4}$ and $\phi = 0$.

For example, we can apply a 50/50 beam splitter operation to a squeezed state with $\xi = 0.5$ and a vacuum state. 

\begin{minted}{python}
prog = sf.Program(2)

with prog.context as q:
    

    sf.ops.Sgate(0.5,0) | q[0]  


    #Mix squeezed mode and vaccum q[1] on a Beam-splitter
    sf.ops.BSgate(theta=np.pi/4, phi=0) | (q[0],q[1])
    
eng = sf.Engine("gaussian")
state=eng.run(prog).state

xp_cov_sqz = state.cov()
cov_mtx_sqz = COB_CovMtx(xp_cov_sqz)
cov_mtx_sqz
\end{minted}

\texttt{\underline{output:}}
\begin{verbatim}
array([[ 0.68393972,  0.        , -0.31606028,  0.        ],
       [ 0.        ,  1.85914091,  0.        ,  0.85914091],
       [-0.31606028,  0.        ,  0.68393972,  0.        ],
       [ 0.        ,  0.85914091,  0.        ,  1.85914091]])
\end{verbatim}

\subsubsection{Simon criterion}
\label{apx:simon}

It is not difficult to implement a function that takes a two-mode covariance matrix $\sigma$ in its block form

\begin{equation*}
\sigma=\left(\begin{array}{cc}\sigma_{A} & \sigma_{AB} \\ \sigma_{AB}^T & \sigma_{B}\end{array}\right),    
\end{equation*}

and check for separability of two-mode systems by applying the Simon criterion presented in Eq. \ref{simoncriterion} on Sec. \ref{sec:entanglement}.
\begin{minted}{python}
def split(array, nrows, ncols):
    """Split a matrix into sub-matrices."""

    r, h = array.shape
    return (array.reshape(h//nrows, nrows, -1, ncols)
                 .swapaxes(1, 2)
                 .reshape(-1, nrows, ncols))


def SimonCriterion(covariance_mtx, hbar):
    """Apply Simon criterion to a two-mode system"""
    
    hbar2=hbar**2
    
    A, C, Ct, B =  split(covariance_mtx, 2, 2)
    J=np.array([[0,1],[-1,0]])
    
    left= np.linalg.det(A)*np.linalg.det(B)+ ((hbar2/4)-abs(np.linalg.det(C)))**2 - \
        np.trace(A@J@C@J@B@J@Ct@J)  
    right= (hbar2/4)*(np.linalg.det(A)+np.linalg.det(B)) 
    
    if left >= right:
        print("The bipartite system is separable")
        
    else:
        print("The bipartite system is entangled")
\end{minted}

The entanglement generated by the interference of two independent squeezed vacuum states on a 50:50 beam splitter can be analysed by examining the output correlations using the Simon criterion.
The first input port is injected with a squeezed vacuum state characterised by a squeezing parameter $\xi_1 = 0.5$.
The second input port receives a phase-shifted squeezed vacuum state with a squeezing parameter $\xi_2 = 0.5 e^{i\pi}$.

\begin{minted}{python}
prog = sf.Program(2)

with prog.context as q:


    sf.ops.Sgate(0.5,0) | q[0]
    sf.ops.Sgate(0.5,np.pi) | q[1]


    #Mix squeezed mode and vaccum q[1] on a Beam-splitter
    sf.ops.BSgate(theta=np.pi/4, phi=0) | (q[0],q[1])

eng = sf.Engine("gaussian")
state=eng.run(prog).state

xp_cov_tmsv = state.cov()
cov_mtx_tmsv = COB_CovMtx(xp_cov_tmsv)
\end{minted}

Applying the Simon criterion we check the output modes are entangled

\begin{minted}{python}
SimonCriterion(cov_mtx_tmsv, hbar=2)
\end{minted}

\texttt{\underline{output:}}
\begin{verbatim}
The bipartite system is entangled
\end{verbatim}

\subsubsection{Logarithmic negativity}
\label{apx:log_neg}

The Walrus, a library developed by Xanadu, provides a convenient function to calculate the logarithmic negativity introduced in Eq.~\eqref{eq:log2neg}
on Sec.~\ref{sec:entanglement}.
The \texttt{log\_negativity()} function from the \texttt{thewalrus.quantum} module can compute the logarithmic negativity directly from a covariance matrix.
The result this function returns is calculated using the natural logarithm. 
To convert it to base 2 as we defined it on Eq. \eqref{eq:log2neg}, we should divide it by $\ln2$.

We will calculate the logarithmic negativity for the state generated by mixing  a squeezed vacuum with parameter $\xi = 0.5$ and a squeezed vacuum with $\xi = 0.5 e^{i\pi}$ on a beam splitter.

\begin{minted}{python}
from thewalrus.quantum import log_negativity

En_tsmv=log_negativity(xp_cov_tmsv, split=1)/np.log(2)
print("The logarithmic negativity of the system is:", En_tsmv)
\end{minted}

\texttt{\underline{output:}}
\begin{verbatim}
The logarithmic negativity of the system is: 1.4426950408889623
\end{verbatim}

Because this value is greater than zero, we can conclude that the beam splitter's output modes are entangled.

\subsubsection{Reduced states}
\label{apx:reducedStates}

The covariance matrix and the vector of means of a multimode Gaussian system contain its complete physical information.
To analyze a specific mode or a subset of modes, one must obtain the reduced state of the system.
This process involves tracing out the degrees of freedom of the unwanted modes.
In the \texttt{Strawberry Fields} library, this is implemented via the \texttt{reduced\_gaussian()} function.
This function accepts the indices of the modes of interest as arguments and returns the corresponding reduced means vector and covariance matrix.
We shall analyze the reduced states resulting from the interference of two squeezed vacuum states with squeezing parameters $\xi = 0.5$ and $\xi = 0.5 e^{i\pi}$ on a balanced beam splitter.

The first reduced mode yields the following means vector:

\begin{minted}{python}
red_mean1, red_cov1 = state.reduced_gaussian([0])
print(red_mean1)
\end{minted}

\textbf{Output:}
\begin{verbatim}
array([0., 0.])
\end{verbatim}

And the corresponding covariance matrix:

\begin{minted}{python}
print(red_cov1)
\end{minted}

\textbf{Output:}
\begin{verbatim}
array([[ 1.54308063, 0.0],
       [ 0.0, 1.54308063]])
\end{verbatim}

Computing the covariance matrix and means vector for the second mode yields identical results.
The symmetry of the covariance matrix, characterised by equal diagonal elements and vanishing correlations, indicates that the reduced states of this squeezed mixture are thermal states.

\subsubsection{Optical network}
\label{apx:network}

Now, we can simulate and characterise quantum correlations in optical networks using the computational tools studied.
Here we construct an optical network consisting of a four-mode Gaussian state and three 50-50 beam splitters as depicted in Figure \ref{fig:3BS}. 
Then, we mixed the resulting output modes of the first beam splitter with two vacuum modes using the other two beam splitters. 

\begin{figure}[htbp]
  \centering
  \includegraphics[width=0.35\textwidth]{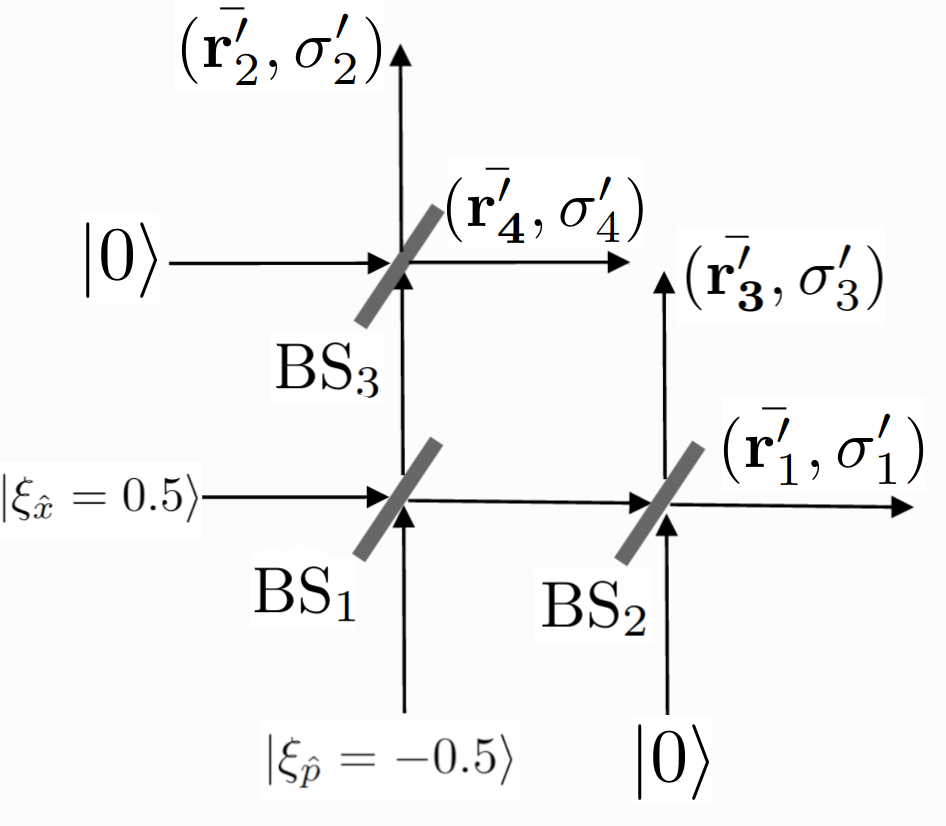}
  \caption{Optical network with three beam splitters}
  \label{fig:3BS}
\end{figure}

\textbf{Step by step}

1. Mix two initial squeezed modes.

\begin{minted}{python}
prog = sf.Program(4)

with prog.context as q:

    sf.ops.Squeezed(r=0.5, p=0) | q[0]  #squeezing  along x (phase = 0)
    sf.ops.Squeezed(r=0.5, p=np.pi) | q[1]  #squeezing  along p (phase = pi)

    #Mix squeezed mode and vaccum q[1] on a Beam-splitter
    sf.ops.BSgate(theta=np.pi/4, phi=0) | (q[0],q[1])
\end{minted}

2. Mix the resulting output modes of the first beam splitter with two vacuum modes using the other two beam splitters.

\begin{minted}{python}
with prog.context as q:

    #Mix the first ouput mode with vaccum q[2] on a Beam-splitter
    sf.ops.BSgate(theta=np.pi/4, phi=0) | (q[0],q[2])
    #Mix the second ouput mode with vaccum q[3] on a Beam-splitter
    sf.ops.BSgate(theta=np.pi/4, phi=0) | (q[1],q[3])
\end{minted}

3. Store the final covariance matrix for the four-mode Gaussian state obtained after the action of the three beam splitters.

\begin{minted}{python}
eng = sf.Engine("gaussian")
state=eng.run(prog).state

cov_mtx=state.cov() #Covariance matrix V on StrawberryFields basis 

#Applying transfromation of basis from SF (r=[x1,x2,...,x1,...,xn]) to (r=[x1,p1,...,xn,pn])
cov_mtx2=COB_CovMtx(cov_mtx) 
cov_mtx2[abs(cov_mtx2) < 1e-11] = 0 #Defining a threshold value
cov_mtx2
\end{minted}

\texttt{\underline{output:}}
\begin{verbatim}
array([[ 1.27154  ,  0.       , -0.587601 ,  0.       ,  0.27154  ,  0.       , -0.587601 ,  0.       ],
       [ 0.       ,  1.27154  ,  0.       ,  0.587601 ,  0.       ,  0.27154  ,  0.       ,  0.587601 ],
       [-0.587601 ,  0.       ,  1.27154  ,  0.       , -0.587601 ,  0.       ,  0.27154  ,  0.       ],
       [ 0.       ,  0.587601 ,  0.       ,  1.27154  ,  0.       ,  0.587601 ,  0.       ,  0.27154  ],
       [ 0.27154  ,  0.       , -0.587601 ,  0.       ,  1.27154  ,  0.       , -0.587601 ,  0.       ],
       [ 0.       ,  0.27154  ,  0.       ,  0.587601 ,  0.       ,  1.27154  ,  0.       ,  0.587601 ],
       [-0.587601 ,  0.       ,  0.27154  ,  0.       , -0.587601 ,  0.       ,  1.27154  ,  0.       ],
       [ 0.       ,  0.587601 ,  0.       ,  0.27154  ,  0.       ,  0.587601 ,  0.       ,  1.27154  ]])
\end{verbatim}

Once we have the covariance matrix, we can characterise entanglement between different parts of the network.

To characterise the entanglement between the output modes after the second beam splitter, i.e, the modes with reduced covariance matrix $\sigma_1'$ and $\sigma_3'$.
To analyze a single mode or a subset of modes, we need to obtain the reduced state. 
This process involves tracing out the unwanted modes, which can be done using the \texttt{reduced\_gaussian()} function in the Strawberry Fields library. 
Let us analyse the entanglement in the following bipartitions. \\

1. Between output modes of the second beam splitter ($\sigma_1'$ and $\sigma_3'$).

\begin{minted}{python}
reduced=state.reduced_gaussian([0,2])
reducedCov1=reduced[1]
reducedCov1[abs(reducedCov1) < 1e-10] = 0 #Defining a threshold value 
reducedCov2=COB_CovMtx(reducedCov1)
\end{minted}

Applying Simon criterion

\begin{minted}{python}
SimonCriterion(reducedCov2,hbar=sf.hbar)
\end{minted}

\texttt{\underline{output:}}
\begin{verbatim}
The bipartite system is separable
\end{verbatim}

Using Logarithmic negativity.
\begin{minted}{python}
#using default function of thewalrus for logarithmic negativity (it uses log in base e)
En_13 = log_negativity(reducedCov1, split=1)*np.log2(np.e)
print("The logarithmic negativity of the system is:", En_13)
\end{minted}

\texttt{\underline{output:}}
\begin{verbatim}
The logarithmic negativity of the system is: 0.0
\end{verbatim}

2. Between output modes of the second beam splitter ($\sigma_2'$ and $\sigma_4'$).

\begin{minted}{python}
reduced=state.reduced_gaussian([1,3])
reducedCov3=reduced[1]
reducedCov3[abs(reducedCov3) < 1e-10] = 0 #Defining a threshold value 
reducedCov4=COB_CovMtx(reducedCov3)
\end{minted}

Applying Simon criterion

\begin{minted}{python}
SimonCriterion(reducedCov4,hbar=sf.hbar)
\end{minted}

\texttt{\underline{output:}}
\begin{verbatim}
The bipartite system is separable
\end{verbatim}

Using Logarithmic negativity.
\begin{minted}{python}
#using default function of thewalrus for logarithmic negativity (it uses log in base e)
En=log_negativity(reducedCov3, split=1)*np.log2(np.e) 
print("The logarithmic negativity of the system is:", En)
\end{minted}

\texttt{\underline{output:}}
\begin{verbatim}
The logarithmic negativity of the system is: 0.0
\end{verbatim}

3. Between one output mode of the second beam splitter ($\sigma_1'$) and one output mode of the third beam splitter ($\sigma_4'$).

\begin{minted}{python}
reduced=state.reduced_gaussian([0,3])
reducedCov5=reduced[1]
reducedCov5[abs(reducedCov1) < 1e-10] = 0 #Defining a threshold value 
reducedCov6=COB_CovMtx(reducedCov5)
\end{minted}

Applying Simon's criterion

\begin{minted}{python}
SimonCriterion(reducedCov6,hbar=sf.hbar)
\end{minted}

\texttt{\underline{output:}}
\begin{verbatim}
The bipartite system is entangled
\end{verbatim}

Using Logarithmic negativity.
\begin{minted}{python}
En=log_negativity(reducedCov5, split=1)*np.log2(np.e) 
print("The logarithmic negativity of the system is:", En)
\end{minted}

\texttt{\underline{output:}}
\begin{verbatim}
The logarithmic negativity of the system is: 0.5480589169169516
\end{verbatim}

\end{widetext}

\section{Final remarks}

In these lecture notes, we have developed a coherent introduction to the foundations of quantum optics with a focus on nonclassical states of light and their role in quantum information. 
Beginning with the quantization of the electromagnetic field and the Fock-space bosonic formalism, we established the basic language needed to describe optical quantum systems. 
This framework allowed us to introduce key families of states --- coherent, thermal, and squeezed --- and to analyze how their properties depart from classical expectations through fluctuations, uncertainty relations, and phase-space representations.

A central theme throughout the notes has been the distinction between classicality and nonclassicality. 
By framing classical states as convex mixtures of coherent states and by introducing quasiprobability distributions, in particular the Glauber–Sudarshan $P$-function, we provided an operational criterion for identifying genuinely quantum features of light. 
This perspective highlights how negativity or singularity in phase-space representations serves as a clear signature of nonclassical behavior and complements more familiar operator-based or variance-based criteria.

We then extended these ideas to composite systems and continuous-variable entanglement, emphasizing the close connection between local nonclassicality and the generation of quantum correlations. 
In this context, nonclassicality emerges not merely as a formal property of single-mode states, but also as a physical resource that can be transformed into entanglement through linear-optical operations such as beam splitters. 
This link provides a unifying viewpoint on how quantum features at the level of states translate into operational advantages for information processing, communication, and metrology.

Altogether, the study of nonclassicality and its interplay with entanglement continues to motivate the development of new theoretical tools and experimental strategies. 
While no single approach provides a complete characterization of quantumness, the coexistence and partial unification of moment-based criteria, phase-space formalisms, and resource-theoretic perspectives form a rich and versatile framework for understanding and exploiting quantum optical systems. 
This plurality of methods reflects the multifaceted nature of quantum correlations and underscores the importance of combining complementary viewpoints when analyzing complex quantum states.

Finally, by incorporating simulation-based laboratories and data analysis tools, these notes aim to bridge conceptual understanding with practical implementation. 
Numerical experiments using homodyne detection models and Gaussian-state simulators illustrate how abstract theoretical concepts can be explored in realistic settings. 
Together, the material presented here is intended to equip students and early researchers with both the conceptual foundations and the technical tools needed to engage with modern quantum optics and quantum information science. 
We hope that these notes serve not only as an introduction to nonclassical light and quantum correlations, but also as a starting point for deeper study and experimentation in this rapidly evolving field.

\section{Acknowledgments}
The authors are grateful to the organizers of the Colombian Escuela-Congreso ``100 años de revolución cuántica: Transformando el conocimiento, la industria y la sociedad" for the invitation to teach at this event.
These lecture notes are the result of our course ``Óptica cuántica para la era de la información", taught in Medellín, Colombia, from September 30 to October 3, 2025.
E.A. acknowledges financial support from the Austrian Science Fund (FWF) through the Elise Richter project No. 10.55776/V1037.

\bibliographystyle{apsrev-new}
\bibliography{LectureNotes}

\appendix
\begin{widetext}
    \section{Quadratic Hamiltonians and Gaussian Wigner functions}
\label{apx:gaussian}

For a multimode system, it is convenient to group the quadrature operators for each $N$ modes into a single $2N$ dimensional vector  $\mathbf{\hat{r}}=( \hat{x}_1, \hat{p}_1,\dots,\hat{x}_N, \hat{p}_N)^T$. 
The canonical commutation relations can then be compactly expressed as:
\begin{align}
    [\mathbf{\hat{r}}, \mathbf{\hat{r}}^T] &= \mathbf{\hat{r}} \mathbf{\hat{r}}^T - (\mathbf{\hat{r}} \mathbf{\hat{r}}^T)^T \nonumber %\\ &
    =  i \Omega
\end{align}
with the symplectic form $\Omega = \bigoplus_{k=1}^{N} \begin{pmatrix}
0 & 1 \\
-1 & 0
\end{pmatrix}.$

A general quadratic Hamiltonian in the quadrature operators can be written as:
\begin{equation}\label{eq:quadratic_H}
    \hat{H} =\frac{1}{2}(\mathbf{\hat{r}}-\mathbf{\bar{r}})^{\text{T}}H(\mathbf{\hat{r}}-\mathbf{\bar{r}}),
\end{equation}
where $H$ is a $2N \times 2N$ symmetric, and positive-definite matrix, and $\mathbf{\overline{r}}$ is a real displacement vector \cite{Serafini_2023}.
Here, it is useful to define the joint displacement operator for $N$ modes, which is a generalization of the single-mode displacement operator defined in Sec. \ref{sec:coherent}:
\begin{equation}
    \hat{D}(\mathbf{\alpha}) = \hat{D}_1(\alpha_1) \otimes \cdots \otimes \hat{D}_N(\alpha_N),
\end{equation}
with $\mathbf{\alpha} = (\alpha_1,...,\alpha_N)^T$ and $\alpha_k = \frac{ x_k+ i p_k}{\sqrt{2}}$. 
In terms of the quadrature operators, the displacement operator can also be written as:  
\begin{equation}
    \hat{D}(\mathbf{r}) = \exp(i\mathbf{r}^T \Omega \mathbf{\hat{r}}).
\end{equation}

This implies $\hat{D}({-\mathbf{\bar{r}}})\mathbf{\hat{r}}\hat{D}({\mathbf{\bar{r}}}) = \mathbf{\hat{r}}-\mathbf{\bar{r}}$. 
Substituting this transformation into the Hamiltonian (Eq. \ref{eq:quadratic_H}) allows us to write it as:
\begin{equation}
    \hat{H} = \frac{1}{2}\hat{D}({-\mathbf{\bar{r}}})\mathbf{\hat{r}}^{\text{T}}H\mathbf{\hat{r}}\hat{D}({\mathbf{\bar{r}}}) \label{eq:quadratic_H_D}
\end{equation}

The positive-definite matrix $H$ can be diagonalized by a symplectic transformation $S$ by virtue of the Williamson's theorem, such that $H = S^T \mathcal{D} S$ \cite{Williamson1936}. 
Here, $S$ is a symplectic matrix ($S \Omega S^T = \Omega$) and $\mathcal{D}$ is a diagonal matrix of the form:
\begin{equation*}
   \mathcal{D} = \bigoplus_{k=1}^{N} \begin{pmatrix}
\omega_k & 0 \\
0 & \omega_k
\end{pmatrix},
\end{equation*}
where $\omega_k$ are the symplectic eigenvalues, corresponding to the frequencies of the system's normal modes. Then the Hamiltonian can be written as

\begin{equation*}
    \hat{H} =\frac{1}{2}\hat{D}({-\mathbf{\bar{r}}})\mathbf{\hat{r}}^{\text{T}}S^T \mathcal{D} S\mathbf{\hat{r}}\hat{D}({\mathbf{\bar{r}}}).
\end{equation*}

A symplectic transformation on the operators is represented by a unitary operator $\hat{S}$ such that $\hat{S}\mathbf{\hat{r}}\hat{S}^{\dagger} = S\mathbf{\hat{r}}$. 
This allows the Hamiltonian to be expressed in terms of these normal modes:
\begin{align}
    \hat{H} &=\frac{1}{2}\hat{D}({-\mathbf{\bar{r}}})(\hat{S}\mathbf{\hat{r}}\hat{S}^{\dagger})^T \mathcal{D}  \hat{S}\mathbf{\hat{r}}\hat{S}^{\dagger}\hat{D}({\mathbf{\bar{r}}}) %\\ &
=\frac{1}{2}\hat{D}({-\mathbf{\bar{r}}})\hat{S}\mathbf{\hat{r}}^T \mathcal{D} \mathbf{\hat{r}}\hat{S}^{\dagger}\hat{D}({\mathbf{\bar{r}}}).
\end{align}

The term quadratic in $\mathbf{\hat{r}}$ simplifies to a sum of single-mode Hamiltonians:
\begin{align*}
    \mathbf{\hat r}^{T} \mathcal{D} \mathbf{\hat r} &= \sum_{k=1}^N \omega_k(\hat x_k^2+\hat p_k^2) %\\    &
    = \sum_{k=1}^N \hat H_{\omega_k}
\end{align*}
Therefore, the total Hamiltonian can be elegantly expressed as a displaced and symplectically transformed sum of independent harmonic oscillators:
\begin{equation}\label{eq:H_quad}
    \hat{H} = \hat{D}({-\mathbf{\bar{r}}})\hat{S}\left(\sum_{k=1}^N \hat{H}_{\omega_k}\right)\hat{S}^{\dagger}\hat{D}({\mathbf{\bar{r}}})
\end{equation}

The density matrix for a system in thermal equilibrium is given by:
\begin{equation}
    \hat \rho = \frac{e^{-\beta \hat{H}}}{\text{Tr}\left[e^{-\beta \hat{H}}\right]}.
\end{equation}
Substituting the structure of the quadratic Hamiltonian from Eq. \ref{eq:H_quad} yields:
\begin{align}
    \hat \rho &= \frac{e^{-\beta \left( \hat{D}({-\mathbf{\bar{r}}})\hat{S}\left(\sum_{k=1}^N \hat{H}_{\omega_k}\right)\hat{S}^{\dagger}\hat{D}({\mathbf{\bar{r}}})\right)}}{\text{Tr}\left[e^{-\beta \left( \hat{D}({-\mathbf{\bar{r}}})\hat{S}\left(\sum_{k=1}^N \hat{H}_{\omega_k}\right)\hat{S}^{\dagger}\hat{D}({\mathbf{\bar{r}}})\right)}\right]} 
    %\\ & 
    =\frac{\hat{D}({-\mathbf{\bar{r}}})\hat{S}\left(\bigotimes_{k=1}^N e^{-\beta \hat{H}_{\omega_k}}\right)\hat{S}^{\dagger}\hat{D}({\mathbf{\bar{r}}})}{\prod_{k=1}^N \text{Tr} \left[e^{-\beta \hat{H}_{\omega_k}}\right]}
\end{align}
Expressing this in the eigenbasis of the harmonic oscillators gives (with $\beta_k = \beta$ for all modes):
\begin{equation*}
    \hat{\rho} = \left[\prod_{k=1}^N (1-e^{-\beta_k \omega_k})\right]\hat{D}({-\mathbf{\bar{r}}})\hat{S} \left(\bigotimes_{k=1}^N \left(\sum_{m=0}^\infty e^{-\beta_k \omega_k m}|m\rangle\langle m| _k \right)\right)
    \hat{S}^{\dagger}\hat{D}({\mathbf{\bar{r}}})
\end{equation*}

The characteristic function $\chi(\alpha)$ of the state $\hat{\rho}$ can be defined as the expectation value of the displacement operator, $\chi(\alpha) = \text{Tr}[\hat{D}(\alpha)\hat{\rho}]$. 
Using the definition $\hat{D}(\mathbf{r}) = \exp(i\mathbf{\hat{r}}^T \Omega \mathbf{r})$
\begin{align*}
    \chi(\mathbf{r}) &= \text{Tr}\left[ \hat{D}(-\mathbf{r})\hat{\rho} \right]%\\ &
    = \left[\prod_{k=1}^N (1-e^{-\beta_k \omega_k})\right] \text{Tr}\left[ \hat D(-\mathbf{r}) \hat{D}({-\mathbf{\bar{r}}})\hat{S}  \left(\bigotimes_{k=1}^N \left(\sum_{m=0}^\infty e^{-\beta_k \omega_k m}|m\rangle\langle m| _k \right)\right)
    \hat{S}^{\dagger}\hat{D}({\mathbf{\bar{r}}})\right].
\end{align*}
Using the cyclic property of the trace 

\begin{equation}
	\chi(\mathbf{r})= \left[\prod_{k=1}^N (1-e^{-\beta_k \omega_k})\right]\text{Tr}\left[\hat{S}^{\dagger}\hat{D}({\mathbf{\bar{r}}}) \hat D(-\mathbf{r}) \hat{D}({-\mathbf{\bar{r}}})\hat{S}
	\left(\bigotimes_{k=1}^N \left(\sum_{m=0}^\infty e^{-\beta_k \omega_k m}|m\rangle\langle m| _k \right)\right)
	\right]. \label{eq:ant}
\end{equation}

Using the relation  $\hat{D}(\mathbf{r_1})\hat{D}(\mathbf{r_2}) = \hat{D}(\mathbf{r_1}+\mathbf{r_2}) e^{-i\mathbf{r_1}^T \Omega \mathbf{r_2}/2}$, we can prove that $\hat D(-\mathbf{r})\hat{D}({-\mathbf{\bar{r}}}) = \hat{D}({-\mathbf{\bar{r}}})\hat D(-\mathbf{r})e^{i \mathbf{\bar{r}}^T \Omega \mathbf{r} }$.
Then Eq. \ref{eq:ant} becomes
\begin{align*}
	\chi(\mathbf{r}) =& \left[\prod_{k=1}^N (1-e^{-\beta_k \omega_k})\right] e^{i \mathbf{\bar{r}}^T \Omega \mathbf{r}} \text{Tr}\left[\hat{S}^{\dagger}\hat{D}({\mathbf{\bar{r}}}) \hat{D}({-\mathbf{\bar{r}}})\hat D(-\mathbf{r})\hat{S}  \left(\bigotimes_{k=1}^N \left(\sum_{m=0}^\infty e^{-\beta_k \omega_k m}|m\rangle\langle m| _k \right)\right)
	\right] \\
	=& \left[\prod_{k=1}^N (1-e^{-\beta_k \omega_k})\right] e^{i \mathbf{\bar{r}}^T \Omega \mathbf{r}} \text{Tr}\left[\hat{S}^{\dagger}\hat D(-\mathbf{r})\hat{S} \left(\bigotimes_{k=1}^N \left(\sum_{m=0}^\infty e^{-\beta_k \omega_k m}|m\rangle\langle m| _k \right)\right)
	\right].
\end{align*}
It can be proven that  $\hat{S}^{\dagger}\hat D(-\mathbf{r})\hat{S} = \hat D(S^{-1}\mathbf{r})$. Hence:
\begin{align}
	\chi(\mathbf{r}) =& \left[\prod_{k=1}^N (1-e^{-\beta_k \omega_k})\right] e^{i \mathbf{\bar{r}}^T \Omega \mathbf{r}} \text{Tr}\left[\hat D(S^{-1}\mathbf{r})
	\left(\bigotimes_{k=1}^N \left(\sum_{m=0}^\infty e^{-\beta_k \omega_k m}|m\rangle\langle m| _k \right)\right)
	\right] %\\ &
	= \left[\prod_{k=1}^N (1-e^{-\beta_k \omega_k})\right] e^{i \mathbf{\bar{r}}^T \Omega \mathbf{r}}\chi(S^{-1}\mathbf{r})  \label{eq:charF_gauss}
\end{align}
The problem now reduces to calculating $\chi(S^{-1}\mathbf{r})$. 
Since the state is a tensor product of single-mode states, its characteristic function is the product of individual characteristic functions. 
For a single-mode  state with $\xi_k = \beta_k \omega_k$, the characteristic function is:
\begin{align*}
	\chi_k(\alpha) =& \text{Tr} \left[\hat{D}(\alpha)\sum_{m}^{\infty} e^{-\xi_k m} \ket{m}\bra{m}  \right] %\\ &
	= \frac{1}{\pi}\int d^2 \beta\bra{\beta}\sum_{m}^{\infty} e^{-\xi_k m} \ket{m}\bra{m} \hat{D}(\alpha) \ket{\beta} %\\ &
	= \frac{1}{\pi} \int d^2 \beta \sum_{m}^{\infty} e^{-\xi_k m} \braket{\beta | m}\bra{m}\hat{D}(\alpha)\ket{\beta}
\end{align*}
We know that 
\begin{align*}
	\braket{\beta | m} &= e^{\frac{-|\beta|^2}{2}} \sum_{n=0}^{\infty} \frac{\beta^{*n}}{\sqrt{n!}}\braket{n|m} %\\ &
    = e^{\frac{-|\beta|^2}{2}} \sum_{n=0}^{\infty} \frac{\beta^{*n}}{\sqrt{n!}}\delta_{nm}
    %\\ &
    = e^{\frac{-|\beta|^2}{2}} \frac{\beta^{*m}}{\sqrt{m!}}
\end{align*}
and 
\begin{align*}
	\bra{m}\hat{D}(\alpha)\ket{\beta} &= e^{\frac{1}{2}(\alpha \beta^* - \alpha^* \beta)} \braket{m|\alpha + \beta}
    %\\   &
    = e^{\frac{1}{2}(\alpha \beta^* - \alpha^* \beta)}e^{\frac{-|\alpha + \beta|^2}{2}}\frac{(\alpha+\beta)^m}{\sqrt{m!}},
\end{align*}
then
\begin{align*}
	\chi(\alpha) =& \frac{1}{\pi} \int d^2 \beta \sum_{m}^{\infty} e^{\xi_k m} e^{\frac{-|\beta|^2}{2}}\frac{\beta^{*m}}{\sqrt{m!}} e^{\frac{1}{2}\left[\alpha^* \beta - \alpha \beta^* \right]}e^{\frac{-|\alpha + \beta|^2}{2}}\frac{(\alpha+\beta)^2}{\sqrt{m!}}\\
	=& \frac{1}{\pi} \int d^2 \beta e^{\frac{-1}{2} \left( |\beta|^2 + |\alpha+\beta|^2 + \alpha^* \beta - \alpha \beta^*\right)}e^{e^{-\xi_k}\beta^* (\alpha+\beta)}\\
	=& \frac{1}{\pi} \int d^2 \beta e^{\frac{-1}{2} \left( |\beta|^2 + |\alpha|^2+|\beta|^2 + \alpha \beta^* + \alpha^* \beta + \alpha^* \beta - \alpha \beta^*\right)}e^{e^{-\xi_k}\beta^* (\alpha+\beta)}\\
	=& \frac{1}{\pi} \int d^2 \beta e^{-\frac{1}{2} \left(2 |\beta|^2 +|\alpha|^2 + 2\alpha^*\beta \right)} e^{e^{-\xi_k} \beta^* \alpha} e^{-\xi_k |\beta|^2}   
\end{align*}
\begin{align}
	\chi_k(\alpha) &= \frac{e^{-\frac{|\alpha|^2}{2}}}{\pi} \int d^2 \beta e^{-\frac{|\beta|^2}{2}\left( 1- e^{-\xi_k}\right)} e^{\alpha \beta^* e^{-\xi_k}-\alpha^* \beta}
    %\\ &
    = \frac{e^{-\frac{|\alpha|^2}{2}}}{\pi} I \label{eq:charf_sg}
\end{align}

Now let us solve the integral 
\begin{equation*}
	I = \int d^2 \beta e^{-\frac{|\beta|^2}{2}\left( 1- e^{-\xi_k}\right)} e^{\alpha \beta^* e^{-\xi_k}-\alpha^* \beta}
\end{equation*}
with $A = 1 - e^{-\xi_k}$, $B = \alpha e^{-\xi_k}$ and $C = -\alpha^*$
\begin{equation*}
	I = \int d^2 \beta e^{\left(-A|\beta|^2 + B \beta^* + C \beta\right)}
\end{equation*}
replacing $\beta =x+iy$ 
\begin{align*}
	I =& \int dxdy e^{\left(-A(x^2 + y^2)  + B(x-iy) + C(x+iy)\right)}
    %\\	=&  
    = \int dx e^{-Ax^2 + (B+C)x} \int dy e^{-Ay^2 + i(B-C)y}\\
	=& e^{\frac{(B+C)^2}{4A}} e^{\frac{-(B-C)^2}{4A}} \int dx e^{-A\left[x^2 -  \frac{B+C}{A}x +\left(\frac{B+C}{2A}\right)^2\right]}  \int dy e^{-A\left[y^2 - i \frac{B-C}{A}y +\left(\frac{i(B-C)}{2A}\right)^2\right]}\\
	=& e^{\frac{(B+C)^2}{4A}} e^{\frac{-(B-C)^2}{4A}} \int dx e^{-A\left[x -\frac{B+C}{2A}\right]^2}\int dy e^{-A\left[y -i\frac{B-C}{2A}\right]^2}\\
	=& e^{\frac{1}{4A}\left((B+C)^2 -(B-C)^2\right)} \sqrt{\frac{\pi}{A}}\sqrt{\frac{\pi}{A}}\\
	I =& \frac{\pi}{A} e^{\frac{BC}{A}}.
\end{align*}
Replacing the values of $A$, $B$ and $C$
\begin{equation}\label{eq:eq_I}
	I = \frac{\pi}{A}\frac{e^{\frac{e^{-\xi_k}|\alpha|^2}{1 - e^{-\xi_k}}}}{1 - e^{-\xi_k}},
\end{equation}
then replacing Eq. \ref{eq:eq_I} on Eq. \ref{eq:charf_sg}
\begin{equation*}
	\chi_k(\alpha) = \frac{e^{-\frac{|\alpha|^2}{2}\left(\frac{1+e^{-\xi_k}}{1-e^{-\xi_k}}\right)}}{1-e^{-\xi_k}}
\end{equation*}

If we define $ \nu_k = \left(\frac{1+e^{-\xi_k}}{1-e^{-\xi_k}}\right)$ and replacing $\alpha = \frac{x_k+ip_k}{\sqrt{2}}$
\begin{equation}
	\chi_k(\alpha) =\frac{e^{-\frac{|\alpha|^2}{2}\nu_k}}{1-e^{-\xi_k}}
% \end{equation}
% \begin{equation}
% 	\chi_k(\alpha) 
    =\frac{e^{-\frac{1}{4}(x_k^2 +p_k^2)\nu_k}}{1-e^{-\xi_k}}
% \end{equation}.
% \begin{equation*}
%     \chi_k(\mathbf{r}_k) 
     = \frac{1}{1-e^{-\xi_k}}e^{-\frac{1}{4}(x_k^2 +p_k^2)\nu_k}
\end{equation}

Using vector notation 

\begin{equation}
    \chi_k(\mathbf{r}_k) = \frac{1}{1-e^{-\xi_k}}e^{-\frac{1}{4}\mathbf{r}_k^T  \begin{pmatrix}
\nu_k & 0 \\
0 & \nu_k
\end{pmatrix} \mathbf{r}_k }
\end{equation}
We have seen that for the multimode system, the characteristic function will be the product:
\begin{align}
	\chi(\mathbf{r}) =& \bigotimes_{k=1}^N \chi_k(\mathbf{r_k})
    %\\ &
    = \bigotimes_{k=1}^N\frac{1}{1-e^{-\xi_k}} e^{-\frac{1}{4}\mathbf{r_k}^T  \begin{pmatrix}
			\nu_k & 0 \\
			0 & \nu_k
		\end{pmatrix} \mathbf{r_k}}
        %\\	\chi(\mathbf{r}) =& 
        = \left[\prod_{k=1}^N (1-e^{-\xi_k})^{-1}\right] e^{-\frac{1}{4}\mathbf{r}^T \bigoplus_{k=1}^N\begin{pmatrix}
			\nu_k & 0 \\
			0 & \nu_k
		\end{pmatrix} \mathbf{r}}\label{eq:X_r}
\end{align}

With Eq. \ref{eq:X_r} now we can calculate the function $\chi(S^{-1}\mathbf{r})$ needed on Eq. \ref{eq:charF_gauss}
\begin{equation*}
	\chi(S^{-1}\mathbf{r}) = \left[\prod_{k=1}^N (1-e^{-\xi_k})^{-1}\right] e^{-\frac{1}{4}(S^{-1}\mathbf{r})^T \bigoplus_{k=1}^N\begin{pmatrix}
			\nu_k & 0 \\
			0 & \nu_k
		\end{pmatrix} (S^{-1}\mathbf{r})}
\end{equation*}

Defining $V = \bigoplus_{k=1}^N \begin{pmatrix}
	\nu_k & 0 \\
	0 & \nu_k
\end{pmatrix}$
\begin{align}
	\chi(S^{-1}\mathbf{r}) =& \left[\prod_{k=1}^N (1-e^{-\xi_k})^{-1}\right] e^{-\frac{1}{4}\mathbf{r}^T S^{-1T} V S^{-1}\mathbf{r}}
    %\\ &
	= \left[\prod_{k=1}^N (1-e^{-\xi_k})^{-1}\right] e^{-\frac{1}{4}\mathbf{r}^T S^{-1T} \Omega^T V \Omega S^{-1}\mathbf{r}}\\
	=& \left[\prod_{k=1}^N (1-e^{-\xi_k})^{-1}\right] e^{-\frac{1}{4}\mathbf{r}^T \Omega^T S  V S^{T} \Omega\mathbf{r}}
    %\\	\chi(S^{-1}\mathbf{r}) &
    = \left[\prod_{k=1}^N (1-e^{-\xi_k})^{-1}\right] e^{-\frac{1}{4}\mathbf{r}^T \Omega^T \sigma \Omega\mathbf{r}}, \label{eq:charF_Sr}
\end{align}
with 
\begin{equation*}
	\sigma = S  V S^{T} = S \left[ \bigoplus_{k=1}^N \begin{pmatrix}
		\nu_k & 0 \\
		0 & \nu_k
	\end{pmatrix} \right] S^{T}.
\end{equation*}

Then, replacing Eq. \ref{eq:charF_Sr} in Eq. \ref{eq:charF_gauss}, we finally obtain the characteristic function for the state governed by the quadratic Hamiltonian.
\begin{equation}
	\chi(\mathbf{r}) = e^{-\frac{1}{4}\mathbf{r}^T \Omega^T \sigma \Omega\mathbf{r}} e^{i \mathbf{\bar{r}}^T \Omega \mathbf{r}}.
\end{equation}

The Wigner function is obtained by taking the Fourier transform of the characteristic function:
$\chi(\mathbf{r})$, in this case
\begin{align*}
	W(\mathbf{r}) =& \frac{1}{\left(2 \pi\right)^{2N}} \int d ^{2N} \mathbf{r'} e^{i\mathbf{r'}^T\Omega \mathbf{r}} \chi(\mathbf{r'})  
    %\\ &
	= \frac{1}{\left(2 \pi\right)^{2N}} \int d ^{2N} \mathbf{r'} e^{-\frac{1}{4}\mathbf{r'}^T \Omega^T \sigma \Omega\mathbf{r'}}e^{i \mathbf{\bar{r}}^T \Omega \mathbf{r'}}e^{i\mathbf{r'}^T\Omega \mathbf{r}}    
    \\
	=& \frac{1}{\left(2 \pi\right)^{2N}} \int d ^{2N} \mathbf{r'} e^{-\frac{1}{4}\mathbf{r'}^T \Omega^T \sigma \Omega\mathbf{r'}} e^{i \mathbf{r'}^T\Omega^T (\mathbf{\bar{r}}-\mathbf{r})}.
\end{align*}
This is a standard multidimensional Gaussian integral, defining $A = \frac{1}{4}\Omega^T \sigma \Omega$ and $b= i \Omega^T(\mathbf{\bar{r}}-\mathbf{r})$ we have
\begin{align*}
	W(\mathbf{r}) =& \frac{1}{\left(2 \pi\right)^{2N}} \int d ^{2N} \mathbf{r'}e^{-\mathbf{r'}^T A \mathbf{r'}+\mathbf{r'}b}
    %\\ &
	= \frac{1}{\left(2 \pi\right)^{2N}} \frac{\pi^n}{\sqrt{\text{Det}(A)}}e^{b^T A^{-1}b}\\
	=& \frac{1}{\left(2 \pi\right)^{2N}} \frac{\pi^n}{\sqrt{\text{Det}(\frac{1}{4}\Omega^T \sigma \Omega)}} e^{(i \Omega^T(\mathbf{\bar{r}}-\mathbf{r}))^T (\frac{1}{4}\Omega^T \sigma \Omega)^{-1}(i \Omega^T(\mathbf{\bar{r}}-\mathbf{r}))}
    %\\ &
	= \frac{1}{\pi^N \sqrt{\text{Det}(\sigma)}} e^{-(\mathbf{r}-\mathbf{\bar{r}})^T \sigma^{-1}(\mathbf{r}-\mathbf{\bar{r}})}
\end{align*}

redefining $\sigma = \frac{\sigma}{2}$ the last equation becomes
\begin{equation}
	W(\mathbf{r}) = \frac{1}{ \sqrt{(2\pi)^{2N}\text{Det}(\sigma)}} e^{-\frac{1}{2}(\mathbf{r}-\mathbf{\bar{r}})^T \sigma^{-1}(\mathbf{r}-\mathbf{\bar{r}})}.
\end{equation}
The resulting Wigner function is a multivariate Gaussian distribution with mean vector $\mathbf{\bar{r}}$ and covariance matrix $\sigma$. 
This confirms that states prepared via quadratic Hamiltonians in quadrature operators are indeed Gaussian states.
\end{widetext}

\end{document}